\documentclass{article}
\usepackage[utf8]{inputenc}
\usepackage[T1]{fontenc}
\usepackage[english]{babel}
\usepackage{amssymb}
\usepackage{amsmath}
\usepackage{amsfonts} 
\usepackage{amssymb}
\usepackage{mathrsfs}
\usepackage{graphicx}
\usepackage{enumerate}
\usepackage{epstopdf}
\usepackage{caption}
\usepackage{subcaption} 	
\usepackage{float}
\usepackage{hyperref}

\usepackage{geometry}
\title{Construction of a new $(3+1)$-dimensional KdV equation and its closed-form solutions with solitary wave behaviour and conserved
vectors}

\author{{Nardjess Benoudina$^{1}$\footnote{Corresponding author}, Chaudry Massood Khalique$^{2,3}$, Ji Lin$^{1}$}\\
{\small \it\  $^1$Department of physics, Zhejiang Normal University,} \\
{\small \it\ Jinhua, 321004, PR China }\\
{\small \it\ $^2$Material Science, Innovation and Modelling Research Focus Area, }\\
{\small \it\	Department of Mathematics and Applied Mathematics, North-West University, }  \\
{\small \it  Mafikeng Campus,  Private Bag X2046, Mmabatho 2735, Republic of South Africa}\\
{\small \it\ $^3$Department of Mathematics and Informatics, Azerbaijan University,}\\
{\small \it\	  Jeyhun Hajibeyli str., 71, AZ1007, Baku, Azerbaijan}}

\begin{document}

\allowdisplaybreaks

\maketitle

\begin{abstract}
	This paper discusses the construction of a new $(3+1)$-dimensional Korteweg-de Vries (KdV) equation. 
	By employing the KdV's recursion operator, we extract two equations, and with elemental computation steps, the obtained result is
  $
  3u_{xyt}+3u_{xzt}-(u_{t}-6uu_{x}+u_{xxx})_{yz}-2\left(u_{x}\partial_{x}^{-1}u_{y}\right)_{xz}-2\left(u_{x}\partial_{x}^{-1}u_{z}\right)_{xy}=0.
  $
   We then transform the new equation to a simpler one to avoid the appearance of the integral in the equation. Thereafter, we apply the Lie symmetry technique and gain a $7$-dimensional Lie algebra $L_7$ of point symmetries. The one-dimensional optimal system of Lie subalgebras is then computed and   used in the reduction process to achieve seven exact solutions. These obtained solutions are  graphically illustrated as 3D and 2D plots that show different propagations of solitary wave solutions such as breather, periodic, bell shape, and others.
   Finally, the conserved vectors are computed by invoking Ibragimov's method.
  \end{abstract}


\section{Introduction}
The Korteweg-de Vries (KdV) equation is a classic wave equation given by
\begin{equation} \label{TKdV}
  u_{t}-6 u u_{x}+u_{xxx}=0,
\end{equation}
which was first introduced in de Vries's dissertation (1894) \cite{Miles1981}. Later in 1895, Diederik Johannes Korteweg (1848-1941) and Gustav de Vries (1866-1934) gave an explanation to the Eq.\eqref{TKdV} that it describes "long waves in a rectangular canal must necessarily change their form as they advance, becoming steeper in front and less steep behind" \cite{Korteweg1895}. In addition, the KdV equation has wide applications for shallow water waves due to its description of long waves traveling in canals \cite{Miura1976}. Moreover, the KdV equation has many generalizations and extensions that are derived to describe the sophisticated nonlinear phenomena; among them are the KdV equations in three, five, seven, or more order forms \cite{Qu2007,Wazwaz2006,Wen-xiuMa1993}, the K(m,n) equation \cite{Retz2012}, the modified KdV (mKdV) equation \cite{Miura1968}, the potential KdV equation \cite{Wang2014a}, the Gardner equation (the combined KdV-mKdV equations) \cite{Fu2004}, the generalized KdV equation with Fisher-type nonlinearity \cite{Kocak2021}, and so on.

The recursion operator plays an essential role in the theory of integrable equations. It helps to find the families of equations that can be integrated by a given spectral problem in a compact form. In addition, the Hamiltonian treatment of integrable equations is another vital role of recursion operators \cite{Landi1994,Yamaguchi1989}. Based on the notion of a recursion operator, the generalized symmetries of a system can be generated. Finally, it can be used in the construction of high-dimensional integrable equations \cite{Lou1997}. In the current study, we   use  the KdV's recursion operator to derive a new $(3+1)$-dimenional KdV equation
\begin{equation}\label{NewKdV}
  3u_{xyt}+3u_{xzt}-(u_{t}-6uu_{x}+u_{xxx})_{yz}-2\left(u_{x}\partial_{x}^{-1}u_{y}\right)_{xz}-2\left(u_{x}\partial_{x}^{-1}u_{z}\right)_{xy}=0,
   \end{equation}
   where $u=u(x,y,z,t)$, and $\partial_{x}^{-1}=\left(\partial_{x}^{-1} f\right)(x)=\int_{-\infty}^{x} f(t) d t$. 
   This model is complicated to study in this form. Therefore, we use the transformation $u=q_x$  and obtain a simpler equation 
   \begin{equation}\label{NewKdVTren}
      3q_{xxyt}+3q_{xxzt}-(q_{xt}-6q_{x}q_{xx}+q_{xxxx})_{yz}-2\left(q_{xx}q_{y}\right)_{xz}-2\left(q_{xx}q_{z}\right)_{xy}=0,
       \end{equation}
which does not have the integral sign in it.

The Lie symmetry method is an invaluable approach for tackling complex nonlinear partial differential equations (NPDEs). It is characterized as an indirect method, as it revolves around the reduction of the independent variables in NPDEs to a lower-dimensional form. This transformation facilitates the exploration of a more manageable equation, which can be readily solved or adapted for use with alternative techniques \cite{Olver1993,Bluman2004,Hydon2000,Benoudina2023a,Khalique2020a,Benoudina2022,Kumar2019a}. 
We apply the Lie symmetry technique to Eq.\eqref{NewKdVTren} and find the $7$-dimensional vector field of symmetries. Thereafter, we construct the optimal system  following the literature \cite{Hu2015,Olver1993,Benoudina2020} and then  use the obtained subalgebras in the reduction process. As a result, Eq.\eqref{NewKdVTren}  reduces to simpler equations to solve. We obtain seven similarity solutions  of 
Eq.\eqref{NewKdVTren}  and then use the transformation $u(x,y,z,t)=q(x,y,z,t)_x$ to obtain exact solutions of our new equation \eqref{NewKdV}. 
Concerning the physical behaviour, we provide a graphical illustration as $3$D and $2$D plots, which we discuss  to show many solitary wave dynamics among them: periodic, bright and dark solitons, breather, lump, bright-dark solitons, and their interactions. This model represents a novel research avenue that requires further investigation to gain a comprehensive understanding of its solution behaviour and its potential applications in the natural world. 

The structure of this paper is outlined as follows: In Section~\ref{sec2}, we provide the derivation of the new $(3+1)$-KdV equation. The investigation of Lie symmetries and the exploration of the optimal system are discussed in Section~\ref{sec3}. Furthermore, Section~\ref{sec4} is dedicated to the examination of reductions and the exploration of similarity solutions. We delve into their relevance to the propagation of solitary waves in Section~\ref{sec5}. Conserved vectors are derived in Section~\ref{sec6}.  Lastly, in Section~\ref{sec7}, we present the conclusions drawn from our study.

\section{Construction of new $(3+1)$-KdV equation} \label{sec2}

In this part, we consider a KdV recursion operator with  parameters, namely
\begin{equation}\label{recursionKdV}
    \mathcal{D}=\frac{1}{5}\partial_{xx}-\frac{16}{15}u+\frac{2}{3}u_{x}\partial^{-1}.
\end{equation}
By considering the theory of recursion operators for developing higher-dimensional integrable equations \cite{Lou1997} and some of the applications of this technique \cite{Benoudina2023}, the  hierarchy equation can be given as
\begin{equation} \label{heirarchy}
u_{t}=\sum_{i=1}^{n} \alpha_{i} \mathfrak{R}^{s_{i}} u_{x_{i}} 
\end{equation}
with $\alpha_{i}$ $(i=1,...,n)$ being arbitrary constants, and $s_{i}=0, \pm 1, \pm 2, \ldots$ for invertible $\mathfrak{R}(u)$ and $s_{i}=0, 1, 2, \ldots$ for non-invertible $\mathfrak{R}(u)$. Setting $\alpha_{i}=1, i=1,...,n$, $n=1$ and $s_{i}=1$ into Eq.\eqref{heirarchy}  we get 
\begin{equation} \label{KdV1}
    u_{t}= \mathcal{D}(u_{x}) 
  \end{equation}
and $u=u(x,y,z,t)$.  Computing  Eq.\eqref{KdV1} using the assumed recursion operator \eqref{recursionKdV} we get
\begin{equation}
u_{t}+\frac{6}{15}uu_{x}-\frac{1}{5}u_{xxx}=0.
\end{equation}
Differentiating  the above equation with respect to $y$ and $z$, we have 
\begin{equation}\label{KdVx}
u_{tyz}+\frac{6}{15}(uu_{x})_{yz}-\frac{1}{5}u_{xxxyz}=0.
\end{equation}
Now, let's consider the heirarchy
\begin{equation}
u_{t}= \mathcal{D}(u_{y}).
\end{equation}
This gives
\begin{equation}\label{KdVy1}
u_{t}+\frac{16}{15}uu_{y}-\frac{1}{5}u_{xxy}-\frac{2}{3}u_{x}\partial^{-1}_{x}u_{y}=0.
\end{equation}
Differentiating  Eq.\eqref{KdVy1} with respect to the variables $x$ and $z$ and then  multiplying by $3$ we achieve

\begin{equation}\label{KdVy}
3u_{xzt}+\frac{16}{5}(uu_{y})_{xz}-\frac{3}{5}u_{xxxyz}-2(u_{x}\partial^{-1}_{x}u_{y})_{xz}=0.
\end{equation}
Likewise, for the hierarchy $u_{t}= \mathcal{D}(u_{z})$, one obtains
\begin{equation} \label{KdVz}
3u_{xyt}+\frac{16}{5}(uu_{z})_{xy}-\frac{3}{5}u_{xxxyz}-2(u_{x}\partial^{-1}_{x}u_{z})_{xy}=0.
\end{equation}
Noting that $(uu_{x})_{yz}=(uu_{y})_{xz}=(uu_{z})_{xy}$, from equations \eqref{KdVy},  \eqref{KdVz} and \eqref{KdVx} we obtain Eq.\eqref{NewKdV}.

\section{Lie symmetry analysis} \label{sec3}

We assume that the one-parameter Lie group of point transformations for Eq.\eqref{NewKdVTren} are given by
\begin{equation}
   \begin{array}{lllll}
    \tilde{x}=x+\epsilon \xi_{1}(x, y, z, t, u)+O\left(\epsilon^{2}\right), \\
    \tilde{y}=y+\epsilon \xi_{2}(x, y, z, t, u)+O\left(\epsilon^{2}\right), \\
    \tilde{z}=z+\epsilon \xi_{3}(x, y, z, t, u)+O\left(\epsilon^{2}\right), \\
    \tilde{t}=t+\epsilon \tau(x, y, z, t, u)+O\left(\epsilon^{2}\right), \\
    \tilde{q}=q+\epsilon \eta(x, y, z, t, u)+O\left(\epsilon^{2}\right),
   \end{array}
   \end{equation}
where $\epsilon $ is a small parameter, and  $\xi_{1},\, \xi_{2}, \, \xi_{3}, \tau$, and $\eta$ define the infinitesimal generator 
\begin{equation}
   \mathfrak{S} =\xi_{1} \frac{\partial}{\partial x}+\xi_{2} \frac{\partial}{\partial y}+\xi_{3} \frac{\partial}{\partial z}+\tau \frac{\partial}{\partial t}+\eta \frac{\partial}{\partial q}.
 \end{equation}
 In order to establish the symmetries, it is necessary to acquire the infinitesimals, which are derived from the invariance condition
 \begin{equation} \label{winvcon1}
   \left.p r^{(6)} \mathfrak{S}(\Delta)\right|_{\Delta=0}=0,
 \end{equation}
 where $pr^{(6)} \mathfrak{S}$ is the sixth prolongation of  $\mathfrak{S}$ and
 $
 \Delta  \equiv	3q_{xxyt}+3q_{xxzt}-(q_{xt}-6q_{x}q_{xx}+q_{xxxx})_{yz}-2\left(q_{xx}q_{y}\right)_{xz}-2\left(q_{xx}q_{z}\right)_{xy} 
$.
Consequently, the following system of deterministic linear PDEs is obtained after performing symbolic computations with Maple:
 \begin{equation}
   \begin{aligned}
    &\left(\eta\right)_q=\left(\xi_2\right)_y,\left(\eta\right)_x=\frac{\left(\xi_1\right)_t}{6},\left(\eta\right)_y=-\frac{3\left(\xi_1\right)_t}{2},\left(\eta\right)_z=-\frac{3\left(\xi_1\right)_t}{2},\left(\tau \right)_t=3\left(\xi_2\right)_y \\ 
   & \left(\tau \right)_q=0,\left(\tau \right)_x=0,\left(\tau \right)_y=0,\left(\tau \right)_z=0,\left(\xi_1\right)_q=0,\left(\xi_1\right)_x=\left(\xi_2\right)_y,\left(\xi_1\right)_y=0,\left(\xi_1\right)_z=0,\left(\xi_2\right)_t=0 \text {, } \\
   &\left(\xi_2\right)_q=0,\left(\xi_2\right)_x=0,\left(\xi_2\right)_z=0,\left(\xi_2\right)_{y, y}=0,\left(\xi_3\right)_t=0,\left(\xi_3\right)_q=0,\left(\xi_3\right)_x=0,\left(\xi_3\right)_y=0,\left(\xi_3\right)_z=\left(\xi_2\right)_y.
   \end{aligned}
   \end{equation}
Solving the above system we get
\begin{equation}
   \xi_1=C_1 x+F_1(t),\,\xi_2=C_1 y+C_2,\,\xi_3=C_1 z+C_4,\, \tau=3C_1 t+C_3,\,\eta=-\frac{1}{6}(x+9 y+9 z)F_1'(t) -C_1 q+F_2(t),
   \end{equation}
 where $F_1(t)$ and $F_2(t)$ are functions of $t$. For simplification, we assume  $F_1(t)=C_5t+C_6$ and $F_2(t)=C_{7}$. Then, through the constants $C_{i}, i=1,...,7$, the   seven Lie point symmetries 
      \begin{equation} \label{NewSymkdv}
      \begin{aligned}
      &\mathfrak{S}_1 = \frac{\partial}{\partial x}, \, \mathfrak{S}_2  =\frac{\partial}{\partial y}, \,
       \mathfrak{S}_3 = \frac{\partial}{\partial z}, \, \mathfrak{S}_4= \frac{\partial}{\partial t}, \,
      \mathfrak{S}_5= t\frac{\partial}{\partial q} , \,\mathfrak{S}_6= t\frac{\partial}{\partial x}-\frac{1}{2}\left(\frac{1}{3}x+3y+3z\right) \frac{\partial}{\partial q},\\
  & \mathfrak{S}_7= x\frac{\partial}{\partial x} +y\frac{\partial}{\partial y} +z\frac{\partial}{\partial z} f+3 t\frac{\partial}{\partial t} -q\frac{\partial}{\partial q},
      \end{aligned}      
\end{equation}
are obtained, which form a basis for the Lie algebra $L_7$ of symmetries   of Eq.\eqref{NewKdVTren}.

      Additionally, the Lie brackets $[\mathfrak{S}_{i},  \mathfrak{S}_{j}]=  \mathfrak{S}_{i}  \mathfrak{S}_{j}-  \mathfrak{S}_{j}  \mathfrak{S}_{i}$ for the Lie algebra of symmetries $L_7$ are provided to generate the Table \ref{KDVCommutator}.  We also examine the adjoint relationships by utilizing the Lie brackets
      \begin{equation} \label{KdVadjoint}
         \begin{aligned}
         A d_{e^ {\epsilon \mathbb{W}}}(\mathfrak{S}) &=e^{-\epsilon \mathbb{W}} \mathfrak{S} e^{\epsilon \mathbb{W}} \\
         &=\mathfrak{S}-\epsilon[\mathbb{W}, \mathfrak{S}]+\frac{1}{2 !} \epsilon^{2}[\mathbb{W},[\mathbb{W}, \mathfrak{S}]]-\cdots .
         \end{aligned}
       \end{equation}
       The construction of Table~\ref{tab:KdVadjoint} is based on the adjoint relation (\ref{KdVadjoint}) that exists between the Lie algebra elements $\left\{\mathfrak{S}_{1}, \mathfrak{S}_{2},..., \mathfrak{S}_{7} \right\}$.

       \begin{table}
         \caption{Commutation relations of Lie algebra of symmetries} \label{KDVCommutator}
         \begin{center}
         \begin{tabular}{cccccccc}
         \hline
         \hline
          $[ \mathfrak{S}_{i}, \mathfrak{S}_{j}]$ & $ \mathfrak{S}_{1}$ & $ \mathfrak{S}_{2}$ & $ \mathfrak{S}_{3}$ & $ \mathfrak{S}_{4}$ & $ \mathfrak{S}_{5}$ & $ \mathfrak{S}_{6}$ & $ \mathfrak{S}_{7}$  \\
          \hline
          $ \mathfrak{S}_{1}$ &$0$ & $0$ & $0$ & $0$ & $ 0$ & $-\frac{1}{6}\mathfrak{S}_{5}$ & $\mathfrak{S}_{1}$ \\
          \hline
          $ \mathfrak{S}_{2}$ & $0$ & $0$ & $0$ & $0$ & $ 0$ & $ -\frac{2}{3} \mathfrak{S}_{5}$ & $\mathfrak{S}_{2}$   \\
          \hline
          $ \mathfrak{S}_{3}$ & $0$ & $0$ & $0$ & $0$ &  $0$ & $-\frac{2}{3} \mathfrak{S}_{5}$ & $\mathfrak{S}_{3}$   \\
          \hline
          $ \mathfrak{S}_{4}$ &  $0$ & $0$ & $0$ & $0$ & $0$  & $\mathfrak{S}_{1}$ & $3 \mathfrak{S}_{4}$\\
          \hline
          $ \mathfrak{S}_{5}$ &$0$ & $0$ & $0$ & $0$ & $ 0$ & $ 0$ & $-\mathfrak{S}_{5}$   \\
          \hline
          $ \mathfrak{S}_{6}$ & $\frac{1}{6}\mathfrak{S}_{5}$ & $\frac{2}{3}\mathfrak{S}_{5}$ & $-\frac{2}{3}\mathfrak{S}_{5}$ & $-\mathfrak{S}_{1}$  & $0$ & $0$& $-2\mathfrak{S}_{6}$ \\
          \hline
          $ \mathfrak{S}_{7}$ & $-\mathfrak{S}_{1}$ & $ -\mathfrak{S}_{2}$ & $-\mathfrak{S}_{3}$ & $- 3\mathfrak{S}_{4}$ & $\mathfrak{S}_{5}$ & $2\mathfrak{S}_{6}$ & $0$  \\
          \hline
          \hline
          \end{tabular}
          \end{center}
         \end{table}

         \begin{table}
            \caption{Adjoint table of Lie algebra of symmetries} \label{tab:KdVadjoint}
            \begin{center}
            \begin{tabular}{cccccccc}
            \hline
            \hline
             $ A d_{e^ {\epsilon \mathbb{W}_{i}}}(\mathfrak{S}_{j})$ & $ \mathfrak{S}_{1}$ & $ \mathfrak{S}_{2}$ & $ \mathfrak{S}_{3}$ & $ \mathfrak{S}_{4}$ & $ \mathfrak{S}_{5}$ & $ \mathfrak{S}_{6}$ & $ \mathfrak{S}_{7}$  \\
             \hline
             $ \mathfrak{S}_{1}$ &$ \mathfrak{S}_{1}$ & $ \mathfrak{S}_{2}$ & $ \mathfrak{S}_{3}$ & $ \mathfrak{S}_{4}$ & $ \mathfrak{S}_{5}$ &  $ \mathfrak{S}_{6}+\frac{\epsilon_1}{s}\mathfrak{S}_{5}$ & $ \mathfrak{S}_{7}-\epsilon_1\mathfrak{S}_{1}$  \\
             $ \mathfrak{S}_{2}$ &$ \mathfrak{S}_{1}$ & $ \mathfrak{S}_{2}$ & $ \mathfrak{S}_{3}$ & $ \mathfrak{S}_{4}$ & $ \mathfrak{S}_{5}$ & $ \mathfrak{S}_{6}+\frac{2\epsilon_2}{3}\mathfrak{S}_{5}$ & $ \mathfrak{S}_{7}-\epsilon_3\mathfrak{S}_{2}$ \\
             $ \mathfrak{S}_{3}$ &$ \mathfrak{S}_{1}$ & $ \mathfrak{S}_{2}$ & $ \mathfrak{S}_{3}$ & $ \mathfrak{S}_{4}$ & $ \mathfrak{S}_{5}$& $  \mathfrak{S}_{6}-\frac{2\epsilon_{3}}{3}\mathfrak{S}_{5}$ &$\mathfrak{S}_{7}-\epsilon_3\mathfrak{S}_{3}$ \\
             $ \mathfrak{S}_{4}$ & $ \mathfrak{S}_{1}$ & $ \mathfrak{S}_{2}$ & $ \mathfrak{S}_{3}$ & $ \mathfrak{S}_{4}$ & $ \mathfrak{S}_{5}$ & $ \mathfrak{S}_{6}-\epsilon_4\mathfrak{S}_{1}$ & $ \mathfrak{S}_{7}-3\epsilon_{4}\mathfrak{S}_{4}$ \\
             $ \mathfrak{S}_{5}$ &$ \mathfrak{S}_{1}$ & $ \mathfrak{S}_{2}$ & $ \mathfrak{S}_{3}$ & $ \mathfrak{S}_{4}$ & $ \mathfrak{S}_{5}$ & $ \mathfrak{S}_{6}$ & $ \mathfrak{S}_{7}+\epsilon_5\mathfrak{S}_{5}$  \\
             $ \mathfrak{S}_{6}$ &$ \mathfrak{S}_{1}-\frac{\epsilon_6}{6}\mathfrak{S}_{5}$ & $ \mathfrak{S}_{2}-\frac{2\epsilon_6}{3}\mathfrak{S}_{5}$ & $\mathfrak{S}_{3}-\frac{2\epsilon_{6}}{3}\mathfrak{S}_{5}$ &  $ \mathfrak{S}_{4}+\epsilon_6\mathfrak{S}_{1}$ & $ \mathfrak{S}_{5}$ & $ \mathfrak{S}_{6}$ & $ \mathfrak{S}_{7}+2\epsilon_6\mathfrak{S}_{6}$ \\
             $ \mathfrak{S}_{7}$ &$ e^{\epsilon_{7}}\mathfrak{S}_{1}$ & $ e^{\epsilon_7}\mathfrak{S}_{2}$ & $ e^{\epsilon_7}\mathfrak{S}_{3}$  & $ e^{3\epsilon_7}\mathfrak{S}_{4}$ &$ e^{-\epsilon_7}\mathfrak{S}_{5}$ & $ e^{-2\epsilon_7}\mathfrak{S}_{6}$ & $ \mathfrak{S}_{7}$   \\
             \hline
             \hline
             \end{tabular}
             \end{center}
            \end{table}

   \subsection{Optimal system}
   
   This section aims at the construction of the one-dimensional optimal system of  Lie subalgebras \cite{Hu2015} spanned by the set of symmetries \eqref{NewSymkdv}. 
   As a first step, we search for the real invariant function $\Phi$ that satisfies $\Phi\left(Ad_{g}(\mathfrak{S})\right)=\Phi(\mathfrak{S})$, for  any subgroup $g$ and $\mathfrak{S} \in \mathfrak{Q}$.
\begin{equation} \label{KdVgenAd}
   \begin{split}
   Ad_{exp(\epsilon \mathbf{V})}(\mathfrak{S})&=e^{-\epsilon \mathbf{V}}\mathfrak{S} e^{\epsilon \mathbf{V}} \\
                          &=\mathfrak{S}-\epsilon [\mathbf{V},\mathfrak{S}]+\frac{1}{2!}\epsilon^{2}[\mathbf{V},[\mathbf{V},\mathfrak{S}]]-...     \\
                         &=\left(a_{1}\mathfrak{S}_{1}+...+a_{7}\mathfrak{S}_{7} \right)-\epsilon [b_{1}\mathfrak{S}_{1}+...+b_{7}\mathfrak{S}_{7},a_{1}\mathfrak{S}_{1}+...+a_{7}\mathfrak{S}_{7}]+O\left(\epsilon^{2} \right) \\
                         &=\left(a_{1}\mathfrak{S}_{1}+...+a_{7}\mathfrak{S}_{7} \right)-\epsilon \left(\Theta_{1}\mathfrak{S}_{1}+...+\Theta_{7}\mathfrak{S}_{7} \right)+O\left(\epsilon^{2} \right),
   \end{split}
   \end{equation}
   where $\Theta_{i}\equiv \Theta_{i} \left(a_{1},...,a_{7},b_{1},...,b_{7} \right), \; i=1,...,7$ are to be determined. By employing Table \ref{KDVCommutator}, we find the following system:
   \begin{equation} \label{Theta}
      \begin{aligned}
      &\Theta_{1}=  a_1 b_7+a_4 b_6-a_6 b_4-b_1 a_7, \quad
      \Theta_{2}=  a_2 b_7-b_2 a_7, \quad
      \Theta_{3}=a_3 b_7-b_3 a_7,\quad
      \Theta_{4}=3 a_4 b_7-3b_4a_7, \\
      &\Theta_{5}=\frac{b_1 a_6}{6}+\frac{2 b_2 a_6}{3}+\frac{2 b_3 a_6}{3}+b_5 a_7+\left(-\frac{a_1}{6}-\frac{2 a_2}{3}-\frac{2 a_3}{3}\right) b_6-a_5 b_7, \,
      \Theta_{6}=-2 a_6 b_7+2 b_6 a_7, \,
      \Theta_{7}=0.
      \end{aligned}
    \end{equation}
In addition, the invariance of the function $\Phi\left(a_{{1}},a_{{2}}, \dots,a_{{7}} \right)$ is verified only for the following condition with any $b_{j},1\leq j \leq 5$:
\begin{equation} \label{KdVinvCon}
   \Theta_{1}\frac{\partial {\Phi}}{\partial{a_{1}}}+\Theta_{2}\frac{\partial {\Phi}}{\partial{a_{2}}}+...+\Theta_{7}\frac{\partial{\Phi}}{\partial{a_{7}}}=0.
   \end{equation}   
Substituting the values of $\Theta_{i}$'s from Eqs.(\ref{Theta}) into Eq.(\ref{KdVinvCon}), and collecting the coefficients $b_{j},1\leq j \leq 7$, we obtain
\begin{equation} \label{phisys}
      \begin{aligned}
      &-\Phi_{a_1} a_6-3 \Phi_{a_4} a_7=0,-\Phi_{a_5} a_5=0, -\Phi_{a_1} a_7+\frac{\Phi_{a_5} a_6}{6}=0, \\
      & \Phi_{a_1} a_4+\Phi_{a_5}\left(-\frac{a_1}{6}-\frac{2 a_2}{3}-\frac{2 a_3}{3}\right)+2 \Phi_{a_6} a_7=0,-\Phi_{a_2} a_7+\frac{2 \Phi_{a_5} a_6}{3}=0,\\
      & \Phi_{a_1} a_1+\Phi_{a_2} a_2+\Phi_{a_3} a_3+3 \Phi_{a_4} a_4-\Phi_{a_5} a_5  -2 \Phi_{a_6} a_6=0, -\Phi_{a_3} a_7+\frac{2 \Phi_{a_5} a_6}{3}=0.
      \end{aligned}
      \end{equation}
Solving the above system we get the invariant function 
\begin{equation}
   \Phi\left(a_{{1}},a_{{2}}, \dots,a_{{7}} \right) ={F} \left( a_{{7}} \right).
   \end{equation}
   Now, we proceed with the computation of the general adjoint matrix  $A=A_1A_2,...,A_7$.  The matrices $A_{i},\,i=1\ldots7$ can be derived with the aid of the adjoint Table \ref{tab:KdVadjoint}. For instance,
   \begin{equation}
   \begin{split}
   Ad_{exp(\epsilon_{1}\mathfrak{S}_{1})}(\mathfrak{S}) &= a_{1}Ad_{exp(\epsilon_{1}\mathfrak{S}_{1})}(\mathfrak{S}_{1})+a_{2}Ad_{exp(\epsilon_{1}\mathfrak{S}_{1})}(\mathfrak{S}_{2})+...+a_{7}Ad_{exp(\epsilon_{1}\mathfrak{S}_{1})}(\mathfrak{S}_{7})\\
                                  &= a_{1}\mathfrak{S}_{1}+a_{2}\mathfrak{S}_{2}+a_{3}\mathfrak{S}_{4}+a_{5}\mathfrak{S}_{5}+\left(\frac{\epsilon_{1}}{6}\mathfrak{S}_{5}+\mathfrak{S}_{6}\right)a_{6}+(\mathfrak{S}_{7}-\epsilon_1\mathfrak{S}_{1})a_{7}\\
                                  &=(a_{1},a_{2},...,a_{7})A_{1}(\mathfrak{S}_{1},\mathfrak{S}_{2},...,\mathfrak{S}_{7})^{T}.
   \end{split}
   \end{equation}
Therefore, we get 
\begin{equation}
   A_1=\left(\begin{array}{ccccccc}
   1 & 0 & 0 & 0 & 0 & 0 & 0 \\
   0 & 1 & 0 & 0 & 0 & 0 & 0 \\
   0 & 0 & 1 & 0 & 0 & 0 & 0 \\
   0 & 0 & 0 & 1 & 0 & 0 & 0 \\
   0 & 0 & 0 & 0 & 1 & 0 & 0 \\
   0 & 0 & 0 & 0 & \frac{\epsilon_1}{6} & 1 & 0 \\
   -\epsilon_1 & 0 & 0 & 0 & 0 & 0 & 1
   \end{array}\right).
   \end{equation}

Similarly, we calculate the matrices $A_{i}, i=2,...,7$ and obtain
\begin{equation*}
   A_2=\left(\begin{array}{ccccccc}
   1 & 0 & 0 & 0 & 0 & 0 & 0 \\
   0 & 1 & 0 & 0 & 0 & 0 & 0 \\
   0 & 0 & 1 & 0 & 0 & 0 & 0 \\
   0 & 0 & 0 & 1 & 0 & 0 & 0 \\
   0 & 0 & 0 & 0 & 1 & 0 & 0 \\
   0 & 0 & 0 & 0 & \frac{2 \epsilon_2}{3} & 1 & 0 \\
   0 & -\epsilon_2 & 0 & 0 & 0 & 0 & 1
   \end{array}\right),  \,\,
   A_3=\left(\begin{array}{ccccccc}
      1 & 0 & 0 & 0 & 0 & 0 & 0 \\
      0 & 1 & 0 & 0 & 0 & 0 & 0 \\
      0 & 0 & 1 & 0 & 0 & 0 & 0 \\
      0 & 0 & 0 & 1 & 0 & 0 & 0 \\
      0 & 0 & 0 & 0 & 1 & 0 & 0 \\
      0 & 0 & 0 & 0 & \frac{2 \epsilon_3}{3} & 1 & 0 \\
      0 & 0 & -\epsilon_3 & 0 & 0 & 0 & 1
      \end{array}\right), 
   \end{equation*}
   \begin{equation*}
      A_4=\left(\begin{array}{ccccccc}
         1 & 0 & 0 & 0 & 0 & 0 & 0 \\
         0 & 1 & 0 & 0 & 0 & 0 & 0 \\
         0 & 0 & 1 & 0 & 0 & 0 & 0 \\
         0 & 0 & 0 & 1 & 0 & 0 & 0 \\
         0 & 0 & 0 & 0 & 1 & 0 & 0 \\
         -\epsilon_4 & 0 & 0 & 0 & 0 & 1 & 0 \\
         0 & 0 & 0 & -3 \epsilon_4 & 0 & 0 & 1
         \end{array}\right), \,\,
   A_5=\left(\begin{array}{ccccccc}
      1 & 0 & 0 & 0 & 0 & 0 & 0 \\
      0 & 1 & 0 & 0 & 0 & 0 & 0 \\
      0 & 0 & 1 & 0 & 0 & 0 & 0 \\
      0 & 0 & 0 & 1 & 0 & 0 & 0 \\
      0 & 0 & 0 & 0 & 1 & 0 & 0 \\
      0 & 0 & 0 & 0 & 0 & 1 & 0 \\
      0 & 0 & 0 & 0 & \epsilon_5 & 0 & 1
      \end{array}\right),
   \end{equation*}
      
   \begin{equation*}
      A_6=\left(\begin{array}{ccccccc}
         1 & 0 & 0 & 0 & -\frac{\epsilon_6}{6} & 0 & 0 \\
         0 & 1 & 0 & 0 & -\frac{2 \epsilon_6}{3} & 0 & 0 \\
         0 & 0 & 1 & 0 & -\frac{2 \epsilon_6}{3} & 0 & 0 \\
         \epsilon_6 & 0 & 0 & 1 & 0 & 0 & 0 \\
         0 & 0 & 0 & 0 & 1 & 0 & 0 \\
         0 & 0 & 0 & 0 & 0 & 1 & 0 \\
         0 & 0 & 0 & 0 & 0 & 2 \epsilon_6 & 1
         \end{array}\right), \,\,
         A_7=\left(\begin{array}{ccccccc}
            e^\epsilon_7 & 0 & 0 & 0 & 0 & 0 & 0 \\
            0 & e^\epsilon_7 & 0 & 0 & 0 & 0 & 0 \\
            0 & 0 & e^\epsilon_7 & 0 & 0 & 0 & 0 \\
            0 & 0 & 0 & e^{3 \epsilon_7} & 0 & 0 & 0 \\
            0 & 0 & 0 & 0 & e^{-\epsilon_7} & 0 & 0 \\
            0 & 0 & 0 & 0 & 0 & e^{-2} \epsilon_7 & 0 \\
            0 & 0 & 0 & 0 & 0 & 0 & 1
            \end{array}\right).
\end{equation*}
Now, we proceed to the classification discussion with the help of the sign of the invariant function $\Phi$ to select the appropriate representative element $\mathbf{W}=\sum_{i=1}^{5}{\tilde{a}}\mathfrak{S}_{i}$, which should be equivalent to $\mathfrak{S}=\sum_{i=1}^{5}a_{i}\mathfrak{S}_{i}$,
\begin{equation} \label{sysoptsys}
   (\tilde{a}_{1},\tilde{a}_{2},...,\tilde{a}_{5})=(a_{1},a_{2},...,a_{5})A.
   \end{equation}
   For the invariant function $\Phi=F(a_7)$,  we choose the simple case $F(a_7)=a_7$, in which we can distinguish two cases $a_7=1$ and $a_{7}=0$. For $a_{7}=1$, we select the representative element $\mathbf{W} = \mathfrak{S}_{7}$. Then, we solve the system \eqref{sysoptsys} to get
   \begin{equation}
   \epsilon_1=-\frac{a_4 a_6}{3}+a_1, \epsilon_2=a_2, \epsilon_3=a_3, \epsilon_4=\frac{a_4}{3}, \epsilon_5=-\frac{1}{18 }(-a_4 a_6^2+3 a_1 a_6+12 a_2 a_6+12 a_3 a_6+18 a_5), {\epsilon}_6=-\frac{a_6}{2}.
   \end{equation}
   For $a_{7}=0$, we substitute the latter into the system \eqref{phisys} to reach a new system, which we solve and obtain the new invariant function $\Phi=F\left({a_3}/{a_2}, {a_4}/{a_2^3}, a_6 a_2^2\right)$. Following the similar reasoning, we obtain the one-dimensional optimal system of subalgebras that  is spanned by
   \begin{equation}
   \{ \mathfrak{S}_{1},\, \mathfrak{S}_{7},\, \mathfrak{S}_{1}+\mathfrak{S}_{2}, \, \alpha \mathfrak{S}_{2}+\mathfrak{S}_{3} ,\, \alpha \mathfrak{S}_{2}+\mathfrak{S}_{4} ,\, \alpha \mathfrak{S}_{2}+\mathfrak{S}_{6} ,\, \alpha \mathfrak{S}_{2}+\mathfrak{S}_{3}+\mathfrak{S}_{4} ,\, \alpha \mathfrak{S}_{2}+\mathfrak{S}_{3}+\mathfrak{S}_{6} ,\, \alpha \mathfrak{S}_{2}+\mathfrak{S}_{4}+\mathfrak{S}_{6}, \, \alpha \mathfrak{S}_{2}+\mathfrak{S}_{3}+\mathfrak{S}_{4}+\mathfrak{S}_{6} \},
\end{equation}
where $\alpha=-1$  or $\alpha=1$. In the following analysis, we consider $\alpha=1$.

\section{Similarity reductions and new solutions} \label{sec4}

\subsection{Subalgebra: $\mathfrak{S}_{2}=\frac{\partial}{\partial{y}}$}
The characteristic equations for the symmetry subalgebra $\mathfrak{S}_{2}=\frac{\partial}{\partial{y}}$ are
\begin{equation} \label{Scharact}
   \frac{dx}{0}=\frac{dy}{1}=\frac{dz}{0}=\frac{dt}{0}=\frac{dq}{0},
   \end{equation} 
   which yields  the group-invariant solution  $q(x,y,z,t)=f(\xi,\zeta,\eta)$,  with similarity variables   $\xi=x$, $\zeta=z$, and $\eta=t$. Substituting the group-invariant solution into Eq.\eqref{NewKdVTren} we get   $f_{\xi \xi \zeta \tau}=0$, whose solution is
   \begin{equation}
   q(x,y,z,t)=F_1(z,t)x+F_{2}(z,t)+F_{3}(x,t)+F_{4}(x,z),
\end{equation}
where $F_1(z,t)$, $F_{2}(z,t)$, $F_{3}(x,t)$, and $F_{4}(x,z)$ are arbitrary functions of their arguments. 
Therefore, the solution of Eq.\eqref{NewKdV} is
\begin{equation}\label{kdvsol1}
   u(x,y,z,t)=F_1(z,t)+\frac{\partial}{\partial x} F_{3}(x,t)+\frac{\partial}{\partial x} F_{4}(x,z).
\end{equation}

\begin{figure}
  \centering
    \begin{subfigure}[b]{0.24\textwidth}
      \centering
        \includegraphics[width=\textwidth]{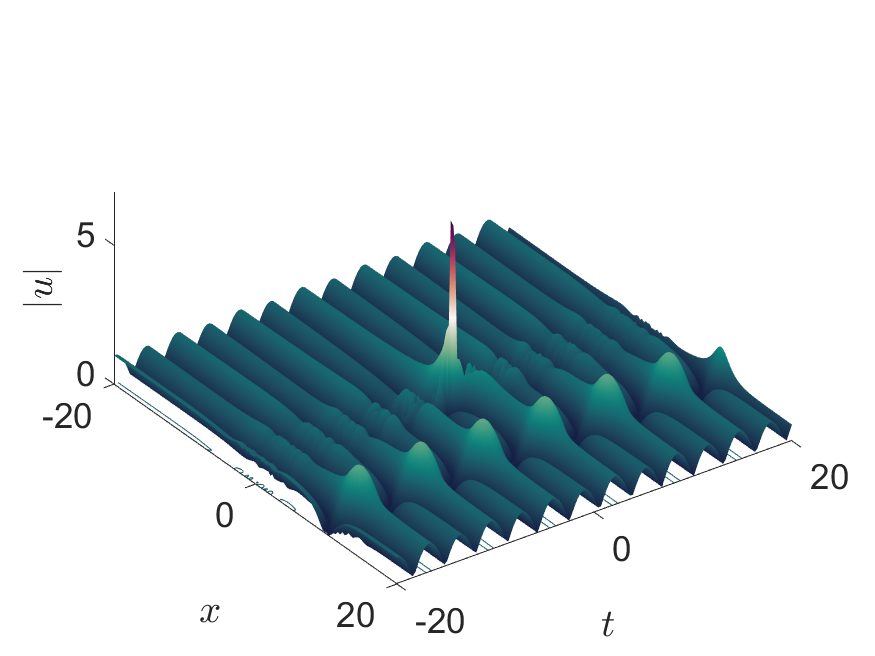}
        \caption{}
        \label{kdvfig1_1}
    \end{subfigure}
    \centering
    \begin{subfigure}[b]{0.24\textwidth}
      \centering
        \includegraphics[width=\textwidth]{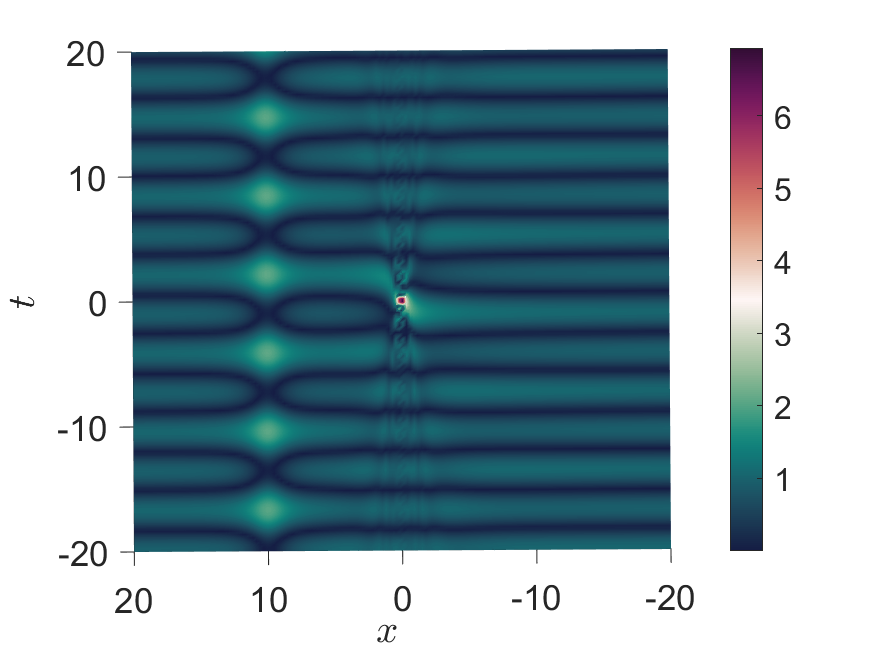}
        \caption{}
        \label{kdvfig1_1v}
    \end{subfigure}
    \centering
    \begin{subfigure}[b]{0.24\textwidth}
      \centering
        \includegraphics[width=\textwidth]{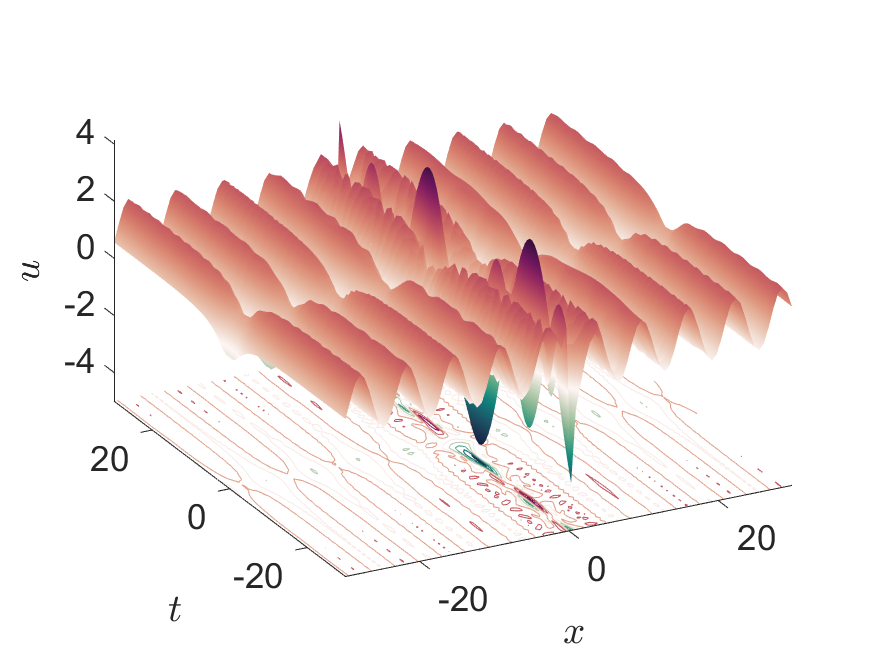}
        \caption{}
        \label{kdvfig1_2}
    \end{subfigure}
    \centering
    \begin{subfigure}[b]{0.24\textwidth}
      \centering
        \includegraphics[width=\textwidth]{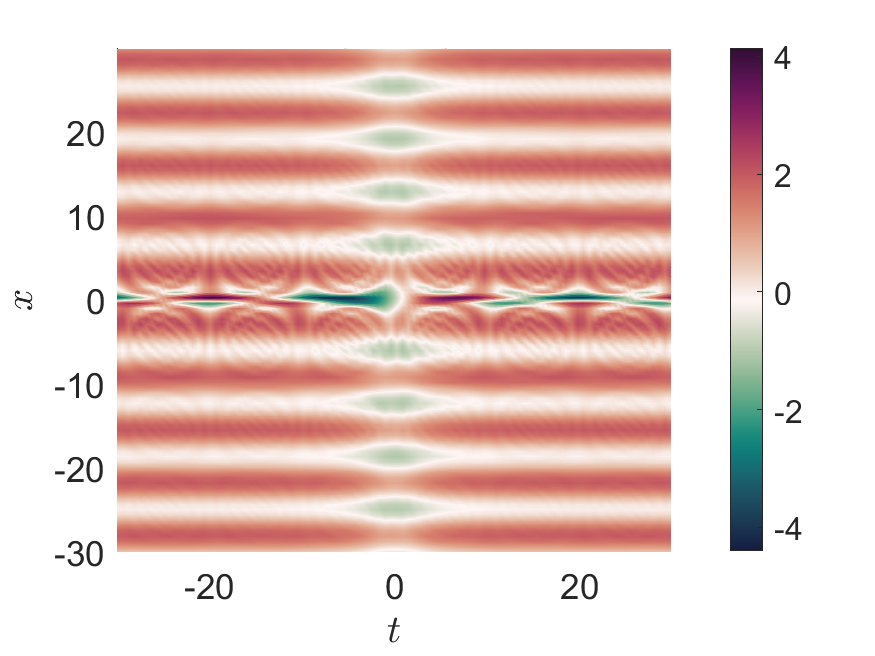}
        \caption{}
        \label{kdvfig1_2v}
    \end{subfigure}
    \centering
    \begin{subfigure}[b]{0.24\textwidth}
      \centering
        \includegraphics[width=\textwidth]{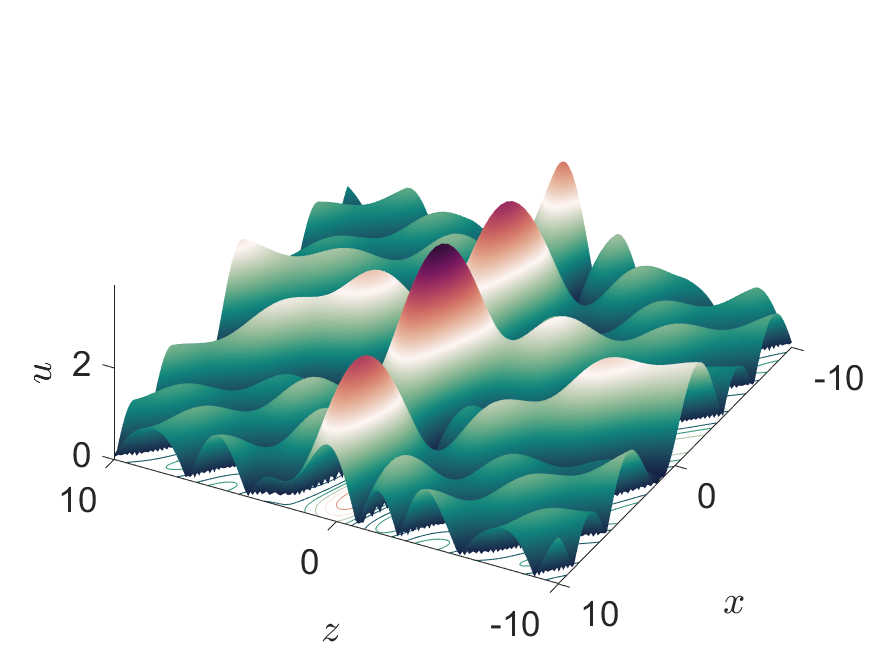}
        \caption{}
        \label{kdvfig1_3}
    \end{subfigure}
    \centering
    \begin{subfigure}[b]{0.24\textwidth}
      \centering
        \includegraphics[width=\textwidth]{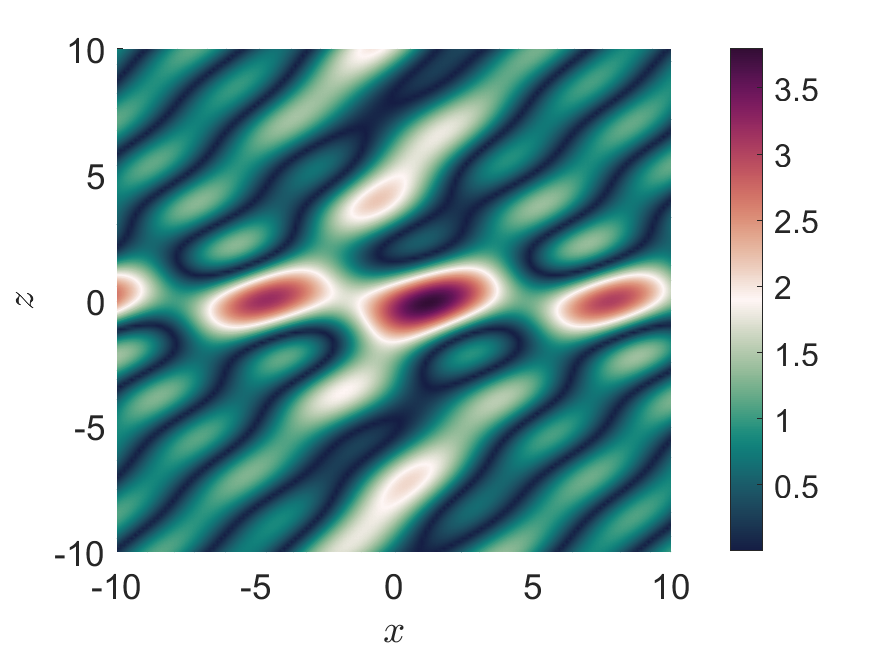}
        \caption{}
        \label{kdvfig1_3v}
    \end{subfigure}
    \centering
    \begin{subfigure}[b]{0.24\textwidth}
      \centering
        \includegraphics[width=\textwidth]{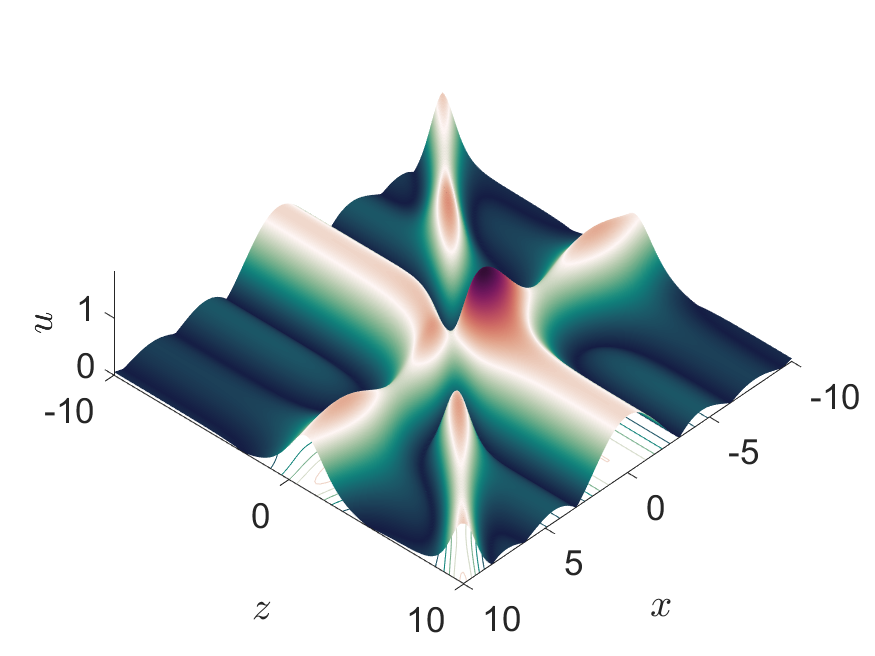}
        \caption{}
        \label{kdvfig1_4}
    \end{subfigure}
    \centering
    \begin{subfigure}[b]{0.24\textwidth}
      \centering
        \includegraphics[width=\textwidth]{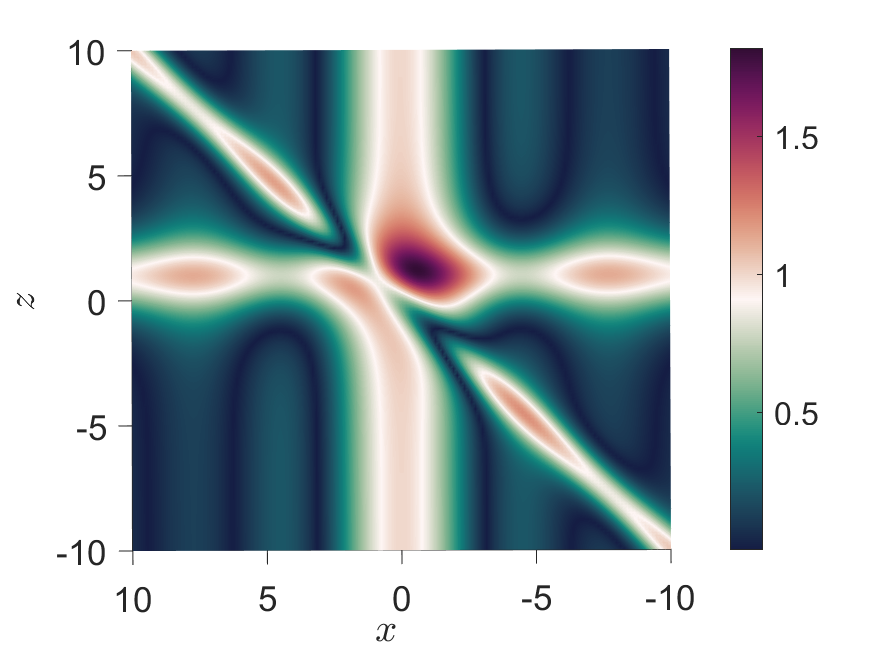}
        \caption{}
        \label{kdvfig1_4v}
    \end{subfigure}
    \centering
    \begin{subfigure}[b]{0.24\textwidth}
      \centering
        \includegraphics[width=\textwidth]{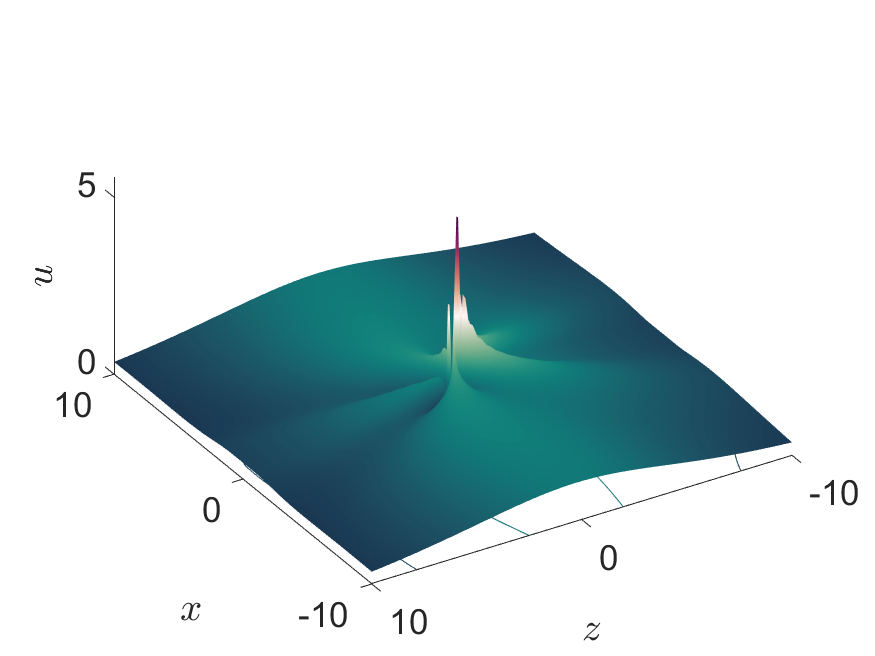}
        \caption{}
        \label{kdvfig1_5}
    \end{subfigure}
    \centering
    \begin{subfigure}[b]{0.24\textwidth}
      \centering
        \includegraphics[width=\textwidth]{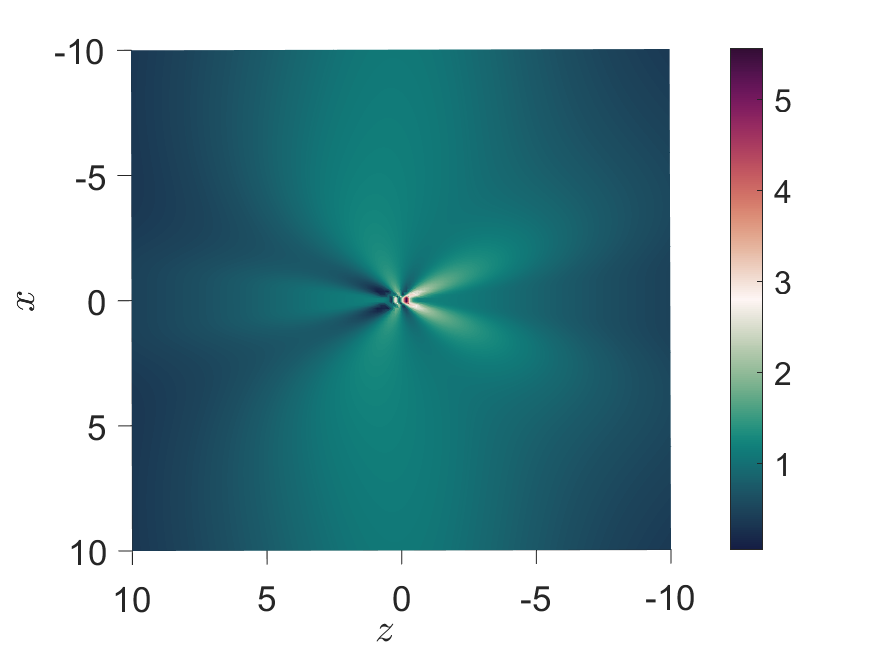}
        \caption{}
        \label{kdvfig1_5v}
    \end{subfigure}
    \centering
    \begin{subfigure}[b]{0.24\textwidth}
      \centering
        \includegraphics[width=\textwidth]{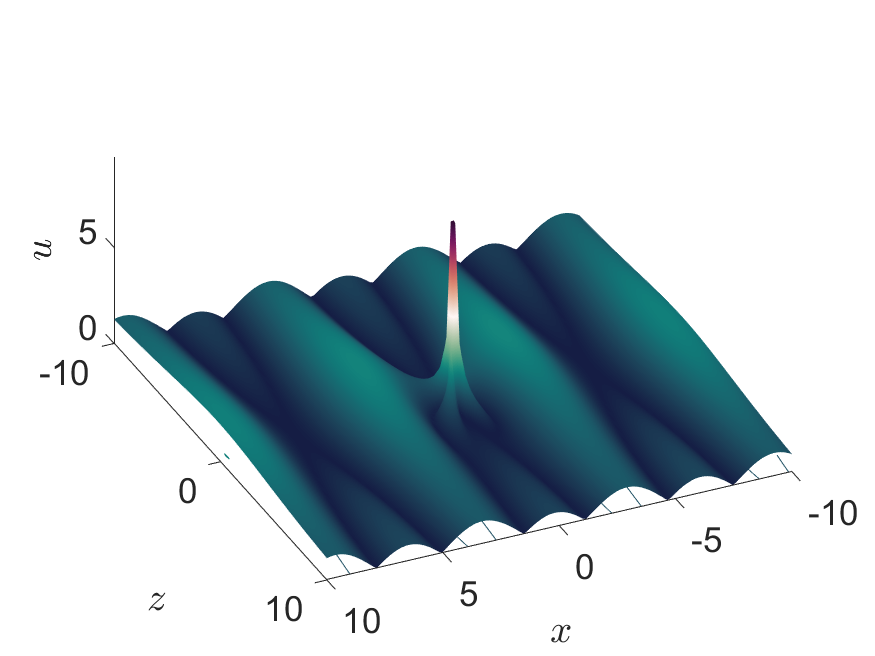}
        \caption{}
        \label{kdvfig1_6}
    \end{subfigure}
    \centering
    \begin{subfigure}[b]{0.24\textwidth}
      \centering
        \includegraphics[width=\textwidth]{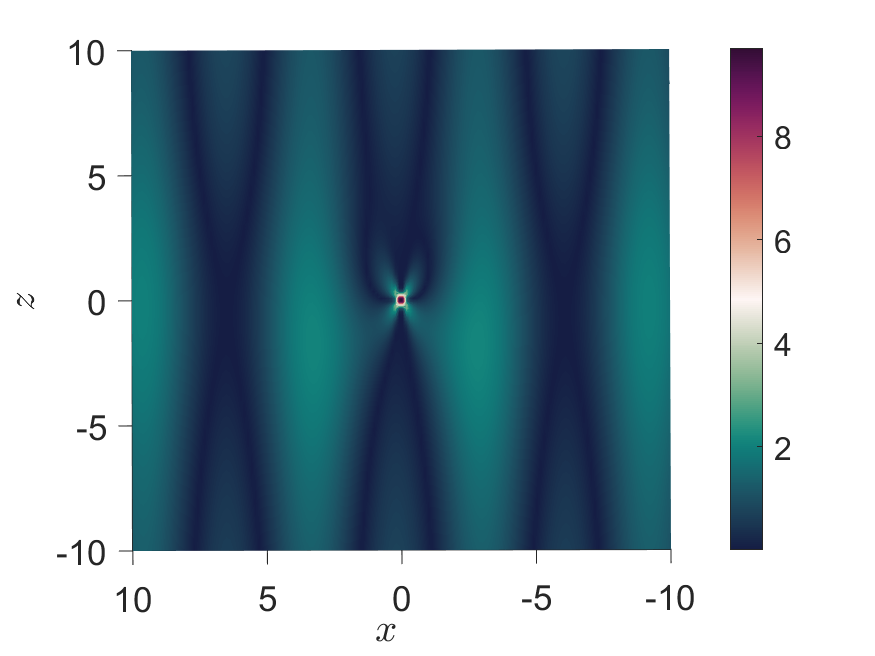}
        \caption{}
        \label{kdvfig1_6v}
    \end{subfigure}
    \caption{Profiles of the dynamical behaviour of the solution \eqref{kdvsol1} in 3D and contour plots} \label{kdvfig1}
  \end{figure}

\subsection{Subalgebra: $\mathfrak{S}_{8}=\mathfrak{S}_{1}+\mathfrak{S}_{2}+\mathfrak{S}_{3}+\mathfrak{S}_{4}=\frac{\partial}{\partial{x}}+\frac{\partial}{\partial{y}}+\frac{\partial}{\partial{z}}+\frac{\partial}{\partial{t}}$}

The similarity variables for $\mathfrak{S}_{8}$ are $\xi=y-x$, $\zeta=z-x$,   $\eta=t-x$, and the group-invariant solution is $q(x,y,z,t)=f(\xi,\zeta,\eta)$. Substituting this  into Eq.\eqref{NewKdVTren}, we get
\begin{equation}\label{Redeq1}
  \begin{aligned}
  & 2 f_\zeta f_{\xi \xi \xi \xi}+4 f_{\xi \zeta} f_{\eta \eta \eta}+2 f_{\xi} f_{\zeta \zeta \zeta \zeta}-6 f_{\eta \eta \xi \zeta \zeta \zeta}-12 f_{\eta \xi \xi \zeta \zeta \zeta}-12 f_{\eta \eta \xi \xi \zeta \zeta} +\left(-6 f_\eta+13\right) f_{\eta \eta \xi \zeta}-4 f_{\eta \eta \eta \xi \xi \zeta}\\
  & +\left(2 f_\zeta+3\right) f_{\eta \eta \eta \xi}+\left(6 f_\zeta+6\right) f_{\eta \eta \xi \xi}+\left(-4 f_{\xi}-6 f_\eta\right) f_{\xi \xi \xi \zeta}+\left(6 f_\zeta+3\right) f_{\eta \xi \xi \xi}+\left(-2 f_{\eta \eta}-4 f_{\xi \zeta}-6 f_{\zeta \zeta}-10 f_{\xi \xi}\right. \\
  &\left. -8 f_{\eta \zeta}-12 f_{\eta \xi}\right) f_{\xi \zeta \zeta}+\left(-2 f_{\eta \eta}+4 f_{\xi \zeta}-12 f_{\eta \zeta}-10 f_{\zeta \zeta}-10 f_{\xi \xi}-12 f_{\eta \xi}\right) f_{\eta \xi \zeta}+\left(-2 f_{\eta \eta}-4 f_{\xi \zeta}-12 f_{\eta \zeta}-10 f_{\zeta \zeta}\right. \\
  & \left.-6 f_{\xi \xi}-8 f_{\eta \xi}\right) f_{\xi \xi \zeta}+\left(4 f_{\xi \zeta}-8 f_{\xi \xi}-8 f_{\eta \xi}\right) f_{\eta \zeta \zeta}+\left(8 f_{\xi \zeta}-4 f_{\xi \xi}-4 f_{\eta \xi}\right) f_{\eta \eta \zeta}+\left(8 f_{\xi \zeta}-4 f_{\eta \zeta}-4 f_{\zeta \zeta}\right) f_{\eta \eta \xi} \\
  & +\left(4 f_{\xi \zeta}-8 f_{\eta \zeta}-8 f_{\zeta \zeta}\right) f_{\eta \xi \xi}+\left(-4 f_{\xi \xi}-4 f_{\eta \xi}\right) f_{\zeta \zeta \zeta}+\left(-4 f_{\eta \zeta}-4 f_{\zeta \zeta}\right) f_{\xi \xi \xi}+\left(-6 f_\eta-4 f_\zeta\right) f_{\xi \zeta \zeta \zeta}-6 f_{\xi \xi \xi \zeta \zeta \zeta} \\
  &+\left(2 f_{\xi}+3\right) f_{\eta \eta \eta \zeta} -4 f_{\xi \xi \zeta \zeta \zeta \zeta}-4 f_{\eta \xi \xi \xi \xi \zeta} -6 f_{\eta \eta \xi \xi \xi \zeta}-4 f_{\eta \eta \eta \xi \zeta \zeta}-f_{\xi \zeta \zeta \zeta \zeta \zeta}-f_{\eta \eta \eta \eta \xi \zeta}-4 f_{\eta \xi \zeta \zeta \zeta \zeta} -f_{\xi \xi \xi \xi \xi \zeta}\\
  & -12 f_{\eta \xi \xi \xi \zeta \zeta}+\left(6 f_{\xi}+6\right) f_{\eta \eta \zeta \zeta}+\left(10-6 f_\zeta-12 f_\eta\right) f_{\eta \xi \zeta \zeta}+\left(-6 f_{\xi}-6 f_\zeta-12 f_\eta\right) f_{\xi \xi \zeta \zeta} -4 f_{\xi \xi \xi \xi \zeta \zeta}\\
  &\left(-6 f_{\xi}-12 f_\eta+10\right) f_{\eta \xi \xi \zeta} +\left(6 f_{\xi}+3\right) f_{\eta \zeta \zeta \zeta}  =0.
  \end{aligned}
  \end{equation}
We now use the symmetry subalgebra $\frac{\partial}{\partial \xi}-\frac{\partial}{\partial \zeta}-\frac{\partial}{\partial \eta}$ 
of  Eq.\eqref{Redeq1} to  reduce Eq.\eqref{Redeq1}. 
Thus we obtain
\begin{equation} \label{Redeq1_2}
   \begin{aligned}
   & -16 \,g_{\theta \theta \theta \theta \theta \theta}+50\, g_{\theta \vartheta \vartheta \vartheta}-160\, g_{\theta \theta \theta \theta \vartheta \vartheta}-80\, g_{\theta \theta \theta \theta \theta \vartheta}-16\, g_{\theta \vartheta \vartheta \vartheta \vartheta \vartheta} -48\, g_{\vartheta \vartheta} g_{\theta \theta \theta}-96\, g_{\vartheta \vartheta} g_{\theta \theta \vartheta}-48\, g_{\vartheta \vartheta} g_{\theta \vartheta \vartheta} \\
   &-32\, g_\theta g_{\theta \theta \theta \vartheta} +32\, g_\theta g_{\theta \vartheta \vartheta \vartheta}+16\, g_\theta g_{\vartheta \vartheta \vartheta \vartheta}-32\, g_{\vartheta} g_{\theta \theta \theta \theta}-96\, g_{\vartheta} g_{\theta \theta \theta \vartheta}-96\, g_{\vartheta} g_{\theta \theta \vartheta \vartheta}-32\, g_{\vartheta} g_{\theta \vartheta \vartheta \vartheta}-48\, g_{\theta \theta} g_{\theta \theta \theta}\\
   & -96\, g_{\theta \theta} g_{\theta \theta \vartheta}-192\, g_{\theta \vartheta} g_{\theta \theta \vartheta} -48\, g_{\theta \theta} g_{\theta \vartheta \vartheta}-96\, g_{\theta \vartheta} g_{\theta \vartheta \vartheta}+64\, g_{\theta \theta \vartheta \vartheta}+12\, g_{\vartheta \vartheta \vartheta \vartheta}-80\, g_{\theta \theta \vartheta \vartheta \vartheta \vartheta}+26 \,g_{\theta \theta \theta \vartheta} \\
   &-16\, g_\theta g_{\theta \theta \theta \theta}-96\, g_{\theta \vartheta} g_{\theta \theta \theta} -160\, g_{\theta \theta \theta \vartheta \vartheta \vartheta}=0,
   \end{aligned}
   \end{equation}
   where the similarity variables are defined as $\theta=\xi+\zeta$ and $\vartheta=\xi+\eta$, and the group-invariant solution is $f(\xi,\zeta,\eta)=g(\theta,\vartheta)$. 
   Using the symmetry subalgebra 
   $\frac{\partial}{\partial \theta}-\frac{\partial}{\partial \vartheta}$ 
   of Eq.\eqref{Redeq1_2}, we reduce \eqref{Redeq1_2} to the ODE
\begin{equation} \label{Redeq1_3}
   -768 h_{\rho\rho} h_{\rho\rho\rho}-256 h_\rho h_{\rho\rho\rho\rho}+152 h_{\rho\rho\rho\rho}-512 h_{\rho\rho\rho\rho\rho\rho}=0,
   \end{equation}
where $g(\theta,\vartheta)=h(\rho)$ is the group-invariant solution with $\rho=\theta+\vartheta$. 
Integrating Eq.\eqref{Redeq1_3} four times gives
   \begin{equation}\label{Redeq1_4}
   	192 h_{\rho\rho}^2+ 32 h_\rho^3  -57 h_\rho^2=0.
   \end{equation}
Solving the ODE \eqref{Redeq1_4}  yields
\begin{equation}
   q(x,y,z,t)=-\frac{3 \sqrt{19}}{1 + {e}^{\frac{\sqrt{19}\left\lbrace \ln \left(\tanh \left(c_4 \rho+c_3\right)+1\right)-\ln \left(1-\tanh \left(c_4 \rho +c_3\right)\right)+2c_1\right\rbrace  }{16 c_4}}} +c_2,
   \end{equation}
   and consequently, we obtain the solution of  Eq.\eqref{NewKdV} as
   \begin{equation}\label{kdvsol3}
      u \left( x,y,z,t \right) =-\frac{
         57 e^{\frac{\sqrt{19}\left\lbrace \ln \left(\tanh \left(c_4       		\rho+c_3\right)+1\right)-\ln \left(1-\tanh \left(c_4 \rho +c_3\right)\right)+2 c_1\right\rbrace }{16 c_4}}}{2\left(1 + e^{\frac{\sqrt{19}\left\lbrace \ln \left(\tanh \left(c_4 \rho +c_3\right)+1\right)-\ln \left(1-\tanh \left(c_4 \rho +c_3\right)\right)+2 c_1\right\rbrace}{16 c_4}}\right)^2},
   \end{equation}
where $\rho = -4 x+2 y+z+t$.

   \subsection{Subalgebra: $\mathfrak{S}_{4}=\frac{\partial}{\partial{t}}$}
In this case, the group-invariant solution for $\mathfrak{S}_{4}$ is $q(x,y,z,t)=f(X,Y,Z)$, with the similarity variables $X=x$, $Y=y$, and $Z=z$, which reduces Eq.\eqref{NewKdVTren} to 
\begin{equation}\label{Redeq2}
   2 f_{X Y Z} f_{X X}+4 f_{X Y} f_{X X Z}+4 f_{X Z} f_{X X Y}+6 f_X f_{X X Y Z}-f_{X X X X Y Z}-2 f_{X X X Z} f_Y -4 f_{X X X} f_{Y Z}-2 f_{X X X Y} f_Z=0.
\end{equation}
As before, we apply the Lie symmetry approach twice to get an ODE, which on solving gives
\begin{equation}
   q(x,y,z,t)=\frac{6}{\alpha_1-x+\sin (y)+z}+\alpha_2.
\end{equation}
As a consequence, the solution of Eq.\eqref{NewKdV} is
\begin{equation} \label{kdvsol2}
  u(x,y,z,t)=\frac{6}{\left(\alpha_1-x+\sin (y)+z\right)^2},
\end{equation}
which is a steady-state solution of \eqref{NewKdV}.

\begin{figure}
  \centering
    \begin{subfigure}[b]{0.33\textwidth}
      \centering
        \includegraphics[width=\textwidth]{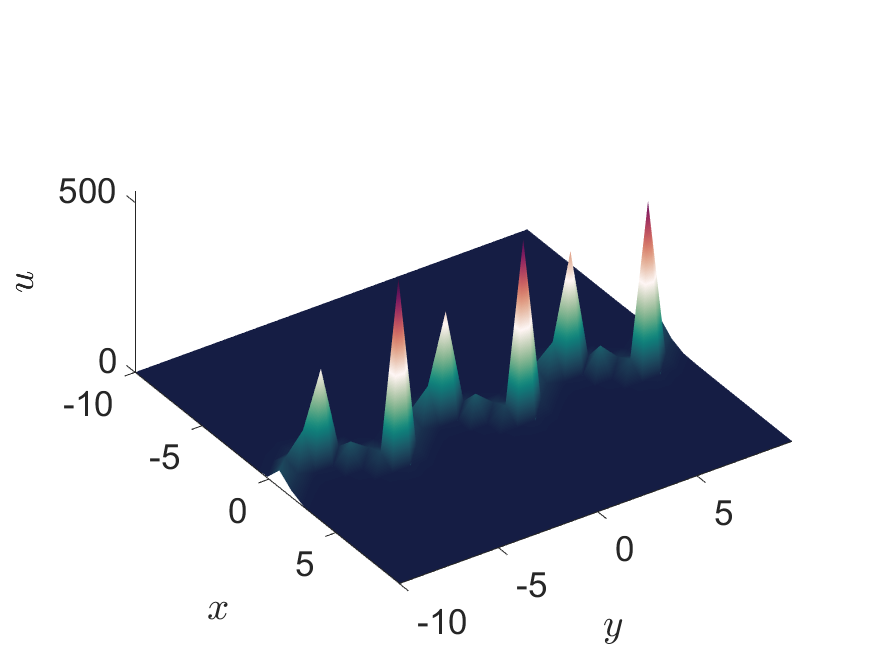}
        \caption{}
        \label{kdvfig22}
    \end{subfigure}
    \centering
    \begin{subfigure}[b]{0.28\textwidth}
      \centering
        \includegraphics[width=\textwidth]{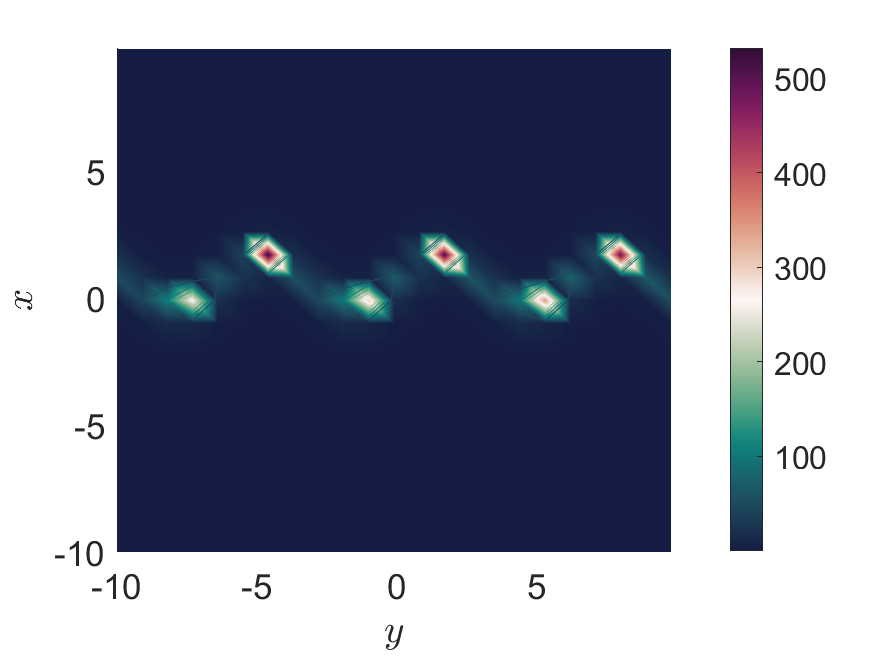}
        \caption{}
        \label{kdvfig2v}
    \end{subfigure}
    \caption{Profiles of the dynamical behaviour of the solution \eqref{kdvsol2} in 3D and contour plots} \label{kdvfig2}
  \end{figure}

\subsection{Subalgebra: $\mathfrak{S}_{9}=\mathfrak{S}_{1}+\mathfrak{S}_{2}=\frac{\partial}{\partial{x}}+\frac{\partial}{\partial{y}}$}

We substitute  the group-invariant solution $q(x,y,z,t)=f(\xi, Z, T)$ with similarity variables $\xi=y-x$, $Z=z$,  $T=t$ associated to the symmetry subalgebra $\mathfrak{S}_{9}$ into 
Eq.\eqref{NewKdVTren} and obtain the NLODE
\begin{equation} \label{Redeq3}
   3 f_{T \xi \xi \xi}+4 f_{T \xi \xi Z}-6 f_{\xi \xi Z} f_{\xi \xi}-4 f_{\xi} f_{\xi \xi \xi Z}-f_{\xi \xi \xi \xi \xi Z}+2 f_{\xi \xi \xi \xi} f_Z=0.
   \end{equation}
  Utilizing the Lie symmetry approach as before, Eq.\eqref{Redeq3} can be reduced twice and finally we obtain the solutions are
   \begin{equation}
  q(x,y,z,t)= -6 \int^{\tanh \left(c_5(x-y+z)+c_4\right)} \frac{\operatorname{ WeierstrassP }\left(-\operatorname{arctanh}\left(a\right)+ c_1,0, c_2\right) c_5}{a^2-1}da+c_3.
   \end{equation}
   Consequently,
   \begin{equation}\label{kdvsol4}
        u(x,y,z,t)= 6 \operatorname{ WeierstrassP }\left(-c_5 (x-c_5  y+c_5 z)-c_4+c_1,0,c_2 \right) c_5^2
   \end{equation}
is the solution of  Eq.\eqref{NewKdV}.
   
   \begin{figure}
    \centering
      \begin{subfigure}[b]{0.4\textwidth}
        \centering
          \includegraphics[width=\textwidth]{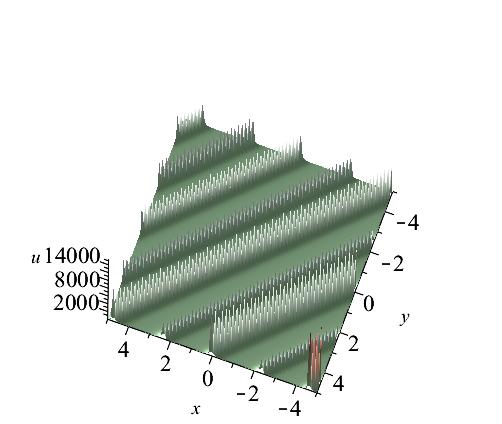}
          \caption{}
          \label{kdvfig44}
      \end{subfigure}
      \centering
      \begin{subfigure}[b]{0.28\textwidth}
        \centering
          \includegraphics[width=\textwidth]{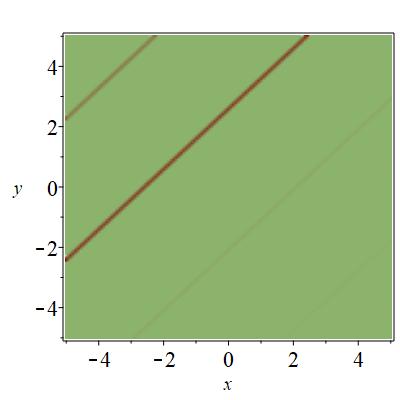}
          \caption{}
          \label{kdvfig4v}
      \end{subfigure}
      \caption{Profiles of the dynamical behaviour of the solution \eqref{kdvsol4} in 3D and contour plots} \label{kdvfig4}
    \end{figure}

\subsection{Subalgebra: $\mathfrak{S}_{10}=\mathfrak{S}_{2}+\mathfrak{S}_{3}=\frac{\partial}{\partial{y}}+\frac{\partial}{\partial{z}}$}
   In this case, we consider the symmetry subalgebra $\mathfrak{S}_{10}$ which gives the invariants $X=x$, $\xi=z-y$, $T=t$,  $q(x,,y,z,t)=f(X,\xi,T)$ and using these Eq.\eqref{NewKdVTren} transforms to
   \begin{equation}\label{Redeq4}
      f_{X X X X \xi \xi}+4 f_{X X X \xi} f_{\xi} +4 f_{X X X} f_{\xi \xi}-2 f_{X \xi \xi} f_{X x}-8 f_{X \xi} f_{X X \xi}-6 f_X f_{X X \xi \xi}+f_{T X \xi \xi}=0.
         \end{equation}
         We reduce  Eq.\eqref{Redeq4} via the Lie symmetry technique, after which we solve the gained PDE. Consequently, the invariants lead to  the solution of Eq.\eqref{NewKdVTren} as
        \begin{equation}
   \begin{aligned}
     q(x,y,z,t)= \, & c_2 \ln \left(\sin \left(-c_6 \cos (t)+c_5(-y+z)-c_6 x+c_4\right)+{i}\left|\cos \left(-c_6 \cos (t)+c_5(-y+z)-c_6 x+c_4\right)\right|\right)\\
      & +c_3\left(\int^{\sin \left(-c_6 \cos (t)+c_5(-y+z)-c_6 x+c_4\right)} \frac{\ln \left(a+\sqrt{a^2-1}\right)}{\sqrt{a^2-1}} da\right) 
      -\frac{1}{12} (\cos (t)+2 x)\left|\sin (t)\right|
      -\frac{ t}{12}-\frac{\pi}{24}+c_1
   \end{aligned}
   \end{equation}
and hence the solution to  Eq.\eqref{NewKdV} is
   \begin{equation}\label{kdvsol5}
      \begin{aligned}
     u(x,y,z,t)= & -6\, c_3 \ln \left(\sin \left(-c_6 \cos (t)+ c_5(-y+z)-c_6 x+c_4\right)+i \mid \cos \left(-c_6 \cos (t)+ c_5(-y+z)- c_6 x\right.\right. \\
      & \left.\left.+ c_4\right) \mid\right) c_6 \cos \left(-c_6 \cos (t)+ c_5(-y+z)-c_6 x+ c_4\right)-6\, c_2 c_6 \cos \left(-c_6 \cos (t)+c_5(-y+z)- c_6 x+ c_4\right)\\
      &-\frac{1}{6}\, |\sin (t)|.
      \end{aligned}
      \end{equation}

      \begin{figure}
        \centering
          \begin{subfigure}[b]{0.3\textwidth}
            \centering
              \includegraphics[width=\textwidth]{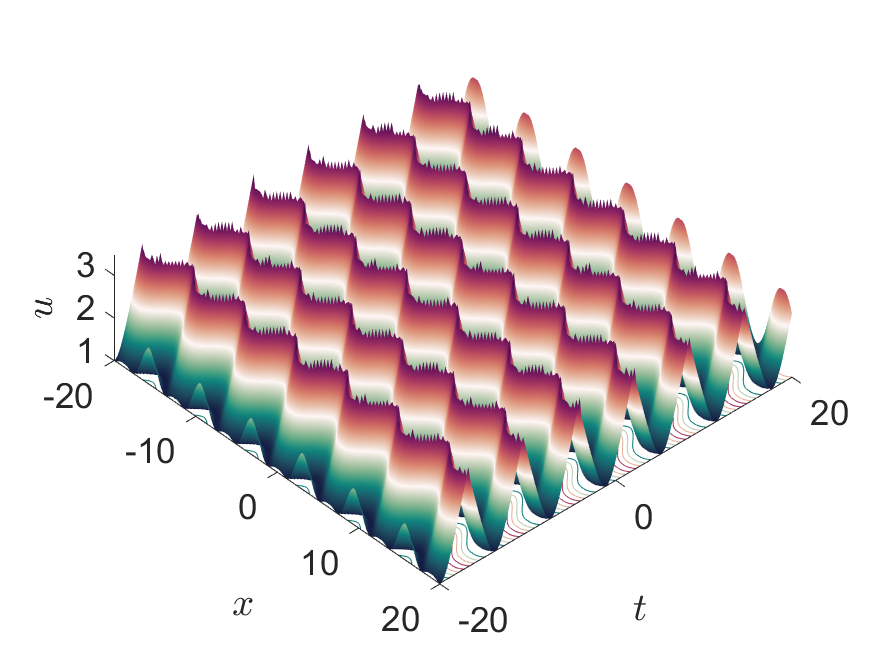}
              \caption{}
              \label{kdvfig5_1}
          \end{subfigure}
          \centering
          \begin{subfigure}[b]{0.3\textwidth}
            \centering
              \includegraphics[width=\textwidth]{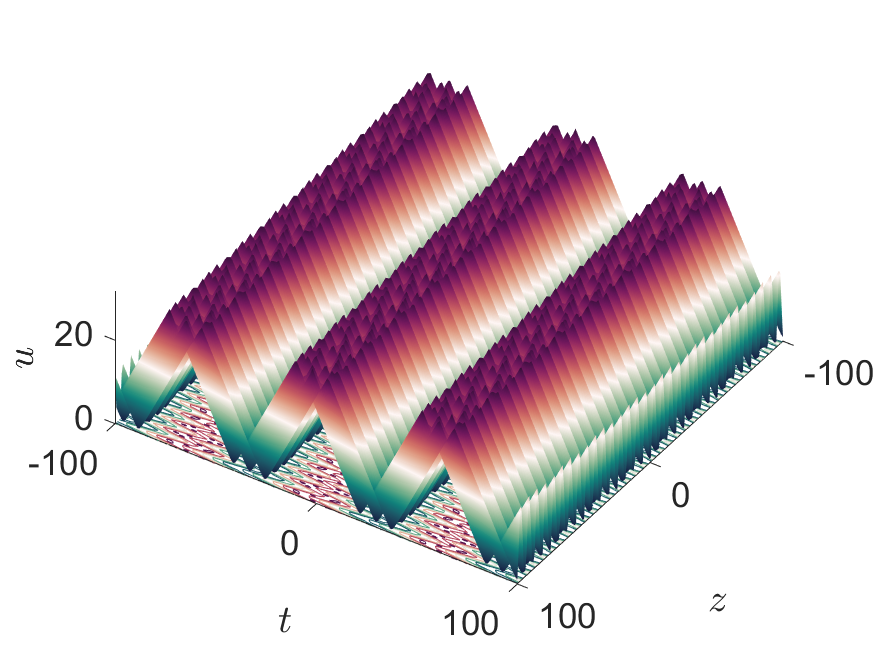}
              \caption{}
              \label{kdvfig5_2}
          \end{subfigure}
          \centering
          \begin{subfigure}[b]{0.3\textwidth}
            \centering
              \includegraphics[width=\textwidth]{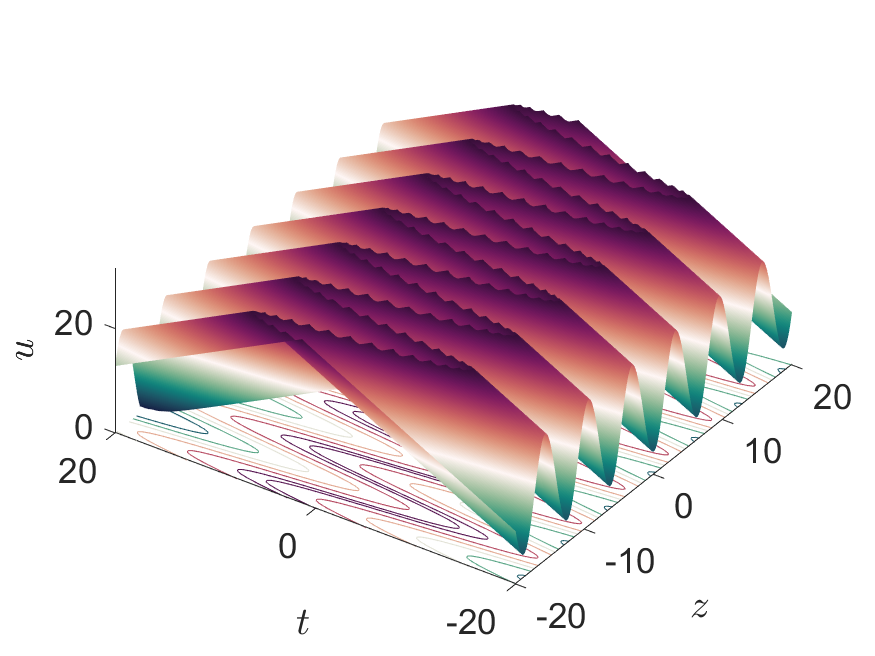}
              \caption{}
              \label{kdvfig5_3}
          \end{subfigure}
          \centering
          \begin{subfigure}[b]{0.28\textwidth}
            \centering
              \includegraphics[width=\textwidth]{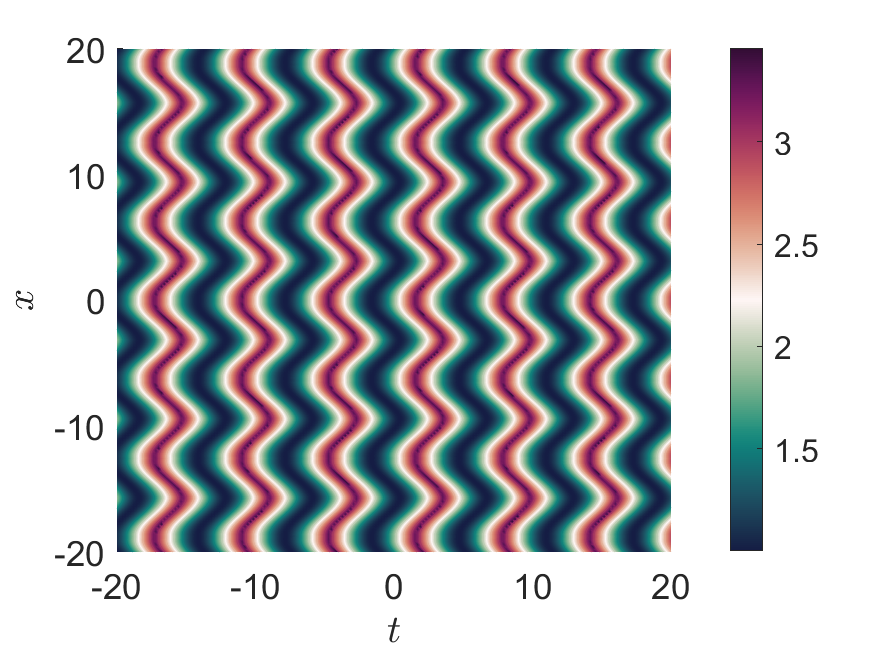}
              \caption{}
              \label{kdvfig5_1v}
          \end{subfigure}
          \centering
          \begin{subfigure}[b]{0.28\textwidth}
            \centering
              \includegraphics[width=\textwidth]{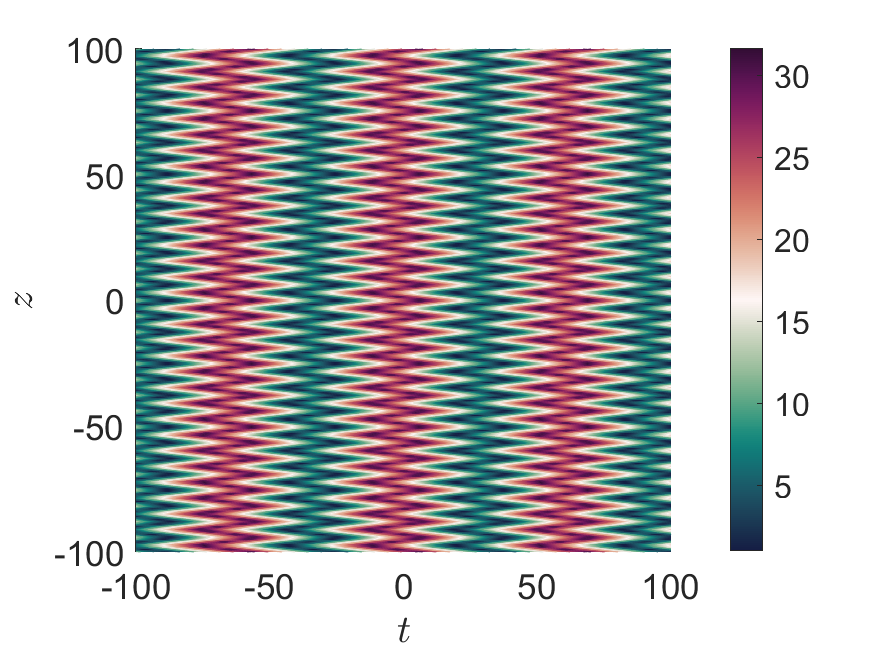}
              \caption{}
              \label{kdvfig5_2v}
          \end{subfigure}
          \centering
          \begin{subfigure}[b]{0.28\textwidth}
            \centering
              \includegraphics[width=\textwidth]{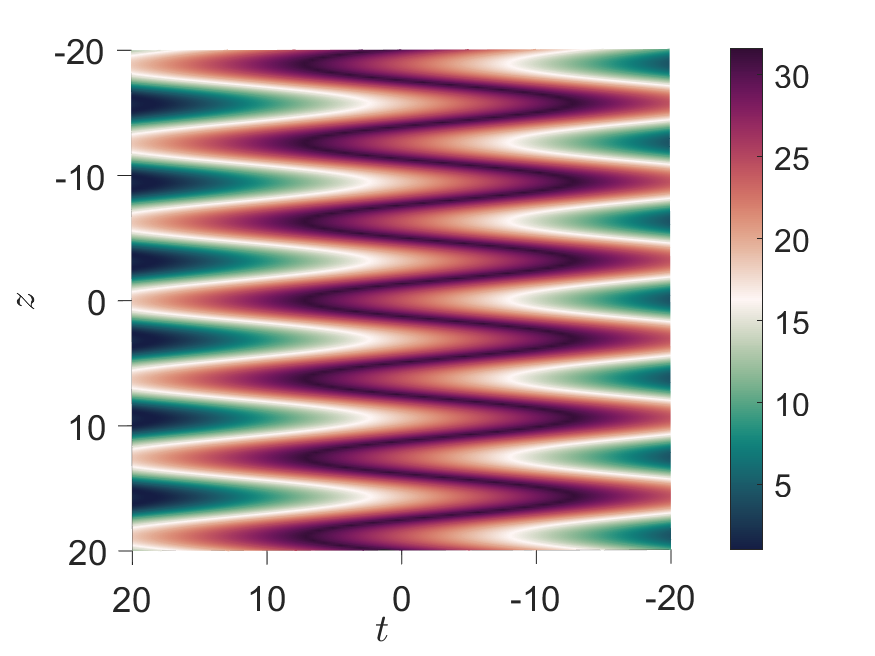}
              \caption{}
              \label{kdvfig5_3v}
          \end{subfigure}
          \caption{Profiles of the dynamical behaviour of the solution \eqref{kdvsol5} in 3D and contour plots} \label{kdvfig5}
        \end{figure}

\subsection{Subalgebra: $\mathfrak{S}_{11}=\mathfrak{S}_{2}+\mathfrak{S}_{6}=t\frac{\partial}{\partial{x}}+\frac{\partial}{\partial{y}}-\frac{1}{2}\left(\frac{1}{3}x+3y+3z\right) \frac{\partial}{\partial q}$}
The symmetry subalgebra $\mathfrak{S}_{11}$ provides us with the invariants $q(x,y,z,t)=f(Z,T,\xi)-\{((x+18 y+18 z) t-9 x) x\}/\{12 t^2\}$, $Z=z$, $T=t$,   $\xi=(ty-x)/t$ and utilizing them  Eq.\eqref{NewKdVTren}  transforms to  
\begin{equation}\label{Redeq5}
      \begin{aligned}
      & T^2f_{T \xi \xi Z} +T\left(3 f_{T \xi \xi Z}-\frac{2 }{3}f_{\xi \xi Z}+3 f_{T \xi \xi \xi}\right) +2 f_Z f_{\xi \xi \xi \xi}-4 f_{\xi} f_{\xi \xi \xi Z}  -9( \xi+Z) f_{\xi \xi \xi Z} -6 f_{\xi \xi Z} f_{\xi \xi}\\
      &-15 f_{\xi \xi Z}-12 f_{\xi \xi \xi} -\frac{1}{T}f_{\xi \xi \xi \xi \xi Z}=0.
      \end{aligned}
      \end{equation}
      Repeated symmetry reductions of Eq.\eqref{Redeq5} results in its  solution as
\begin{equation}
   \begin{aligned}
  q(x,y,z,t) = \, & F_2(t)+ F_3(t) \xi+ F_4(t) \xi^2+\frac{t^4}{2}\left(\int \xi^2 F_1\left(\xi t^\frac{3}{2}\right) d \xi\right)-\xi t^4\left(\int \xi F_1\left(\xi t^\frac{3}{2}\right) d \xi\right) \\
   & +\frac{\xi^2 t^\frac{5}{2} }{2}F_1\left(\xi t^\frac{3}{2}\right)-\frac{9 \xi z}{4}.
   \end{aligned}
   \end{equation}
By taking $F_1\left(\xi t^{3 / 2}\right) =1/ (\xi t^{3 / 2})$ and reverting to variable $u$, we achieve the solution of Eq.\eqref{NewKdV} as
\begin{equation}\label{kdvsol6}
   u(x, y, z, t)=\frac{1}{12 t^2}\left(-18 t^\frac{5}{2} x+18 t^\frac{7}{2} y+(-24 t y+24 x) F_4(t)-12 F_3(t) t-6 t^2+(-2 x-18 y+9 z) t+18 x\right).
\end{equation}

\begin{figure}
  \centering
    \begin{subfigure}[b]{0.33\textwidth}
      \centering
        \includegraphics[width=\textwidth]{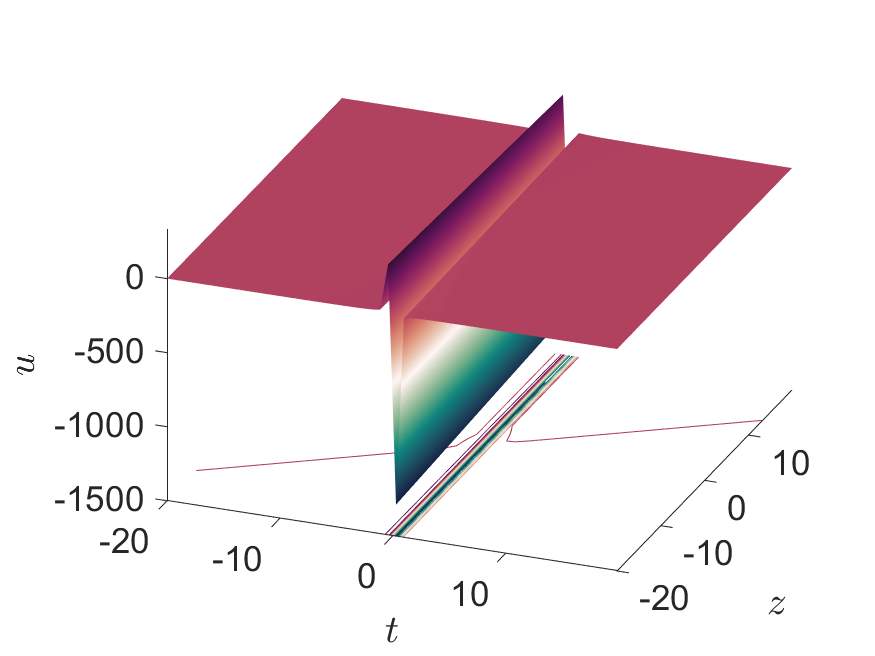}
        \caption{}
        \label{kdvfig66}
    \end{subfigure}
    \centering
    \begin{subfigure}[b]{0.28\textwidth}
      \centering
        \includegraphics[width=\textwidth]{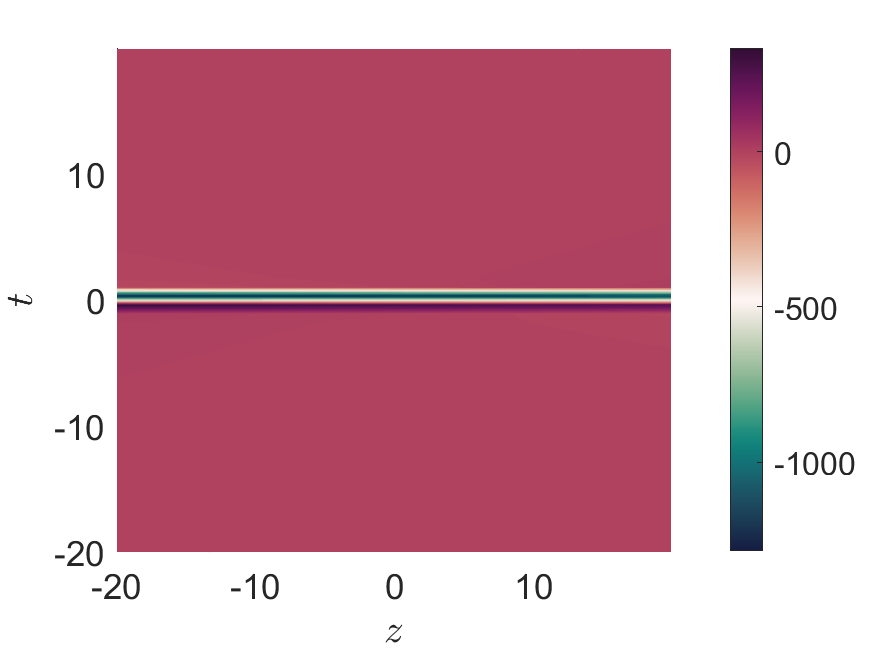}
        \caption{}
        \label{kdvfig6v}
    \end{subfigure}
    \caption{Profiles of the dynamical behaviour of the solution \eqref{kdvsol6} in 3D and contour plots} \label{kdvfig6}
  \end{figure}

\subsection{Subalgebra: $\mathfrak{S}_{12}=\mathfrak{S}_{2}+\mathfrak{S}_{3}+\mathfrak{S}_{4}=\frac{\partial}{\partial{y}}+\frac{\partial}{\partial{z}}+ \frac{\partial}{\partial t}$}
The associated group-invariant solution of $\mathfrak{S}_{12}$ is $q(x,y,z,t)=f(X,\xi, \zeta)$ with similarity variables as $X=x$, $\xi=z-y$, and $\zeta=t-y$. Making use of these invariants, 
Eq.\eqref{NewKdVTren}  reduces to  
\begin{equation}\label{Redeq6}
   \begin{aligned}
   & f_{X X X X \xi \xi} +f_{X X X X \xi \zeta}-6 f_X f_{X X \xi \zeta} -6 f_X f_{X X \xi \xi}-2 f_{X X} f_{X \xi \zeta}-8 f_{X X \xi} f_{X \xi}-4 f_{X X \xi} f_{X \zeta}  -4 f_{X \xi} f_{X X \zeta}\\
   &-2 f_{X X} f_{X \xi \xi}+4 f_{X X X} f_{\xi \xi}+4 f_{X X X} f_{\xi \zeta}+4 f_{X X X \xi} f_{\xi} +2 f_{X X X \xi} f_\zeta+2 f_{\xi} f_{X X X \zeta}-3 f_{X X \zeta \zeta}+f_{X \xi \xi \zeta}+f_{X \xi \zeta \zeta}=0.
   \end{aligned}
   \end{equation}
  As before, repeated use of Lie symmetry technique on  Eq.\eqref{Redeq6} leads us to the solution
\begin{equation}\label{Trensol7}
   \begin{aligned}
   q(x,y,z,t) = \, & -\frac{3 \sqrt{2} }{2}\tanh \left(\frac{\sqrt{2}}{4}\left(c_1+x-2 y+z+t\right)\right)-\frac{3 \sqrt{2} }{4}\ln \left\{\tanh \left(\frac{\sqrt{2}}{4}\left(c_1+x-2 y+z+t\right)\right)-1\right\} \\
   & +\frac{3 \sqrt{2} }{4}\ln \left\{\tanh \left(\frac{\sqrt{2}}{4}\left(c_1+x-2 y+z+t\right)\right)+1\right\}-\frac{3 x}{4}+\frac{3 y}{2}-\frac{3 z}{4}-\frac{3 t}{4}+c_2.
   \end{aligned}
   \end{equation}
  Differentiation of Eq.\eqref{Trensol7} with respect to $x$, gives the  solution of Eq.\eqref{NewKdV} in the form
   \begin{equation}\label{kdvsol7}
   u(x, y, z, t)=-\frac{3}{4 \cosh \left(\frac{\sqrt{2}}{4}\left(c_1+x-2 y+z+t\right)\right)^2}.
   \end{equation}

   \begin{figure}
    \centering
      \begin{subfigure}[b]{0.3\textwidth}
        \centering
          \includegraphics[width=\textwidth]{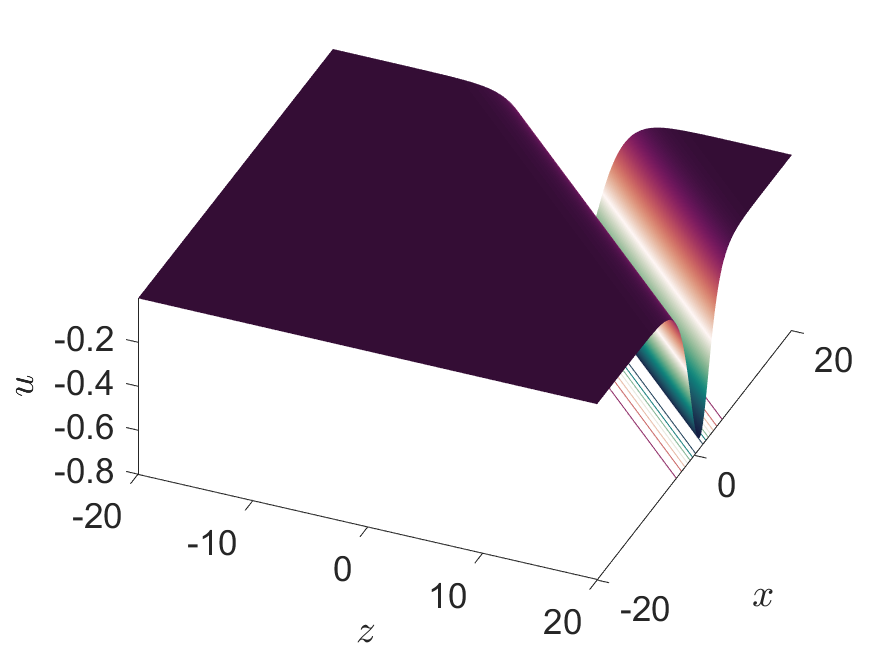}
          \caption{$t=-20$}
          \label{kdvfig7_1}
      \end{subfigure}
      \centering
      \begin{subfigure}[b]{0.3\textwidth}
        \centering
          \includegraphics[width=\textwidth]{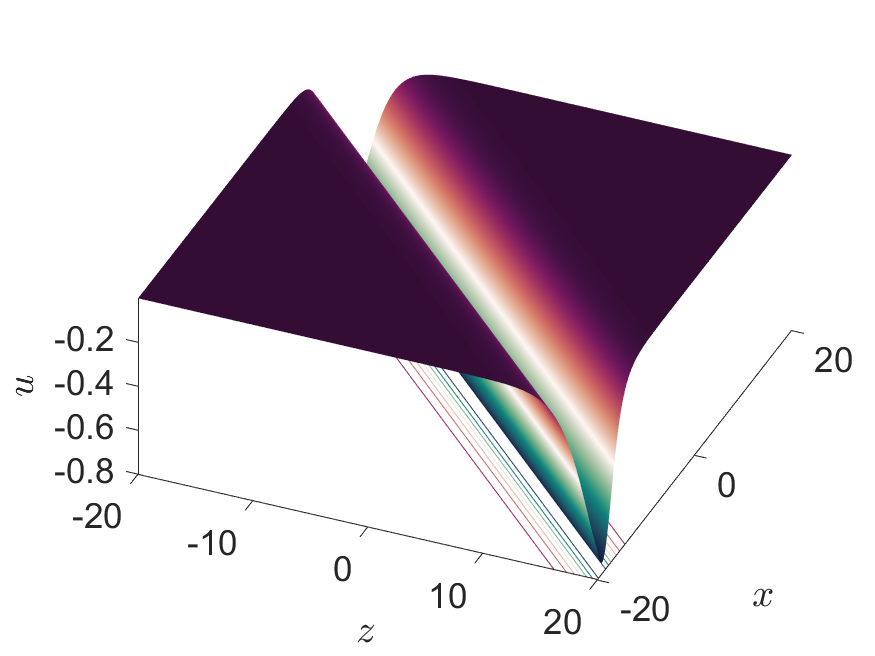}
          \caption{$t=0$}
          \label{kdvfig7_2}
      \end{subfigure}
      \centering
      \begin{subfigure}[b]{0.3\textwidth}
        \centering
          \includegraphics[width=\textwidth]{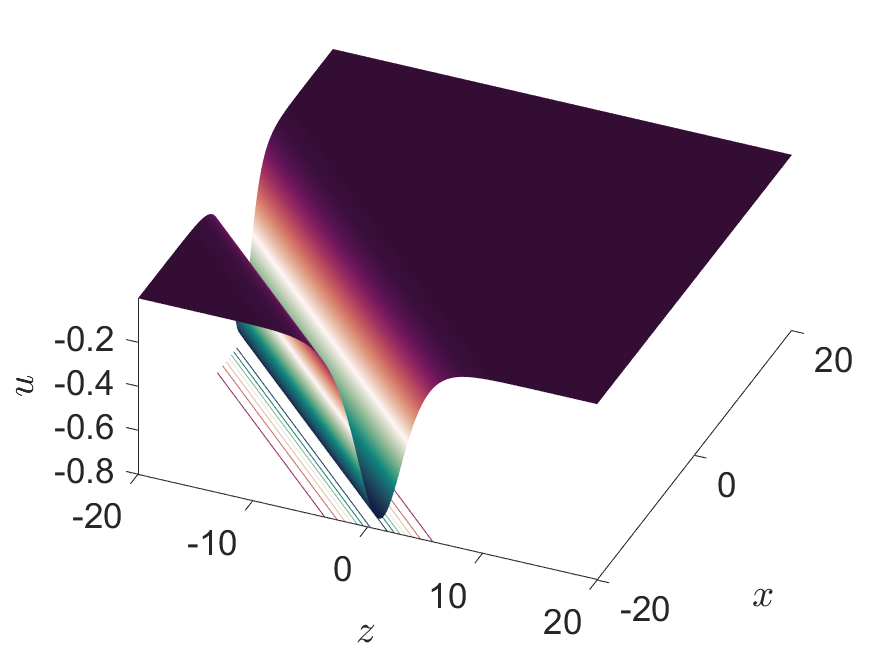}
          \caption{$t=20$}
          \label{kdvfig7_3}
      \end{subfigure}
      \centering
      \begin{subfigure}[b]{0.3\textwidth}
        \centering
          \includegraphics[width=\textwidth]{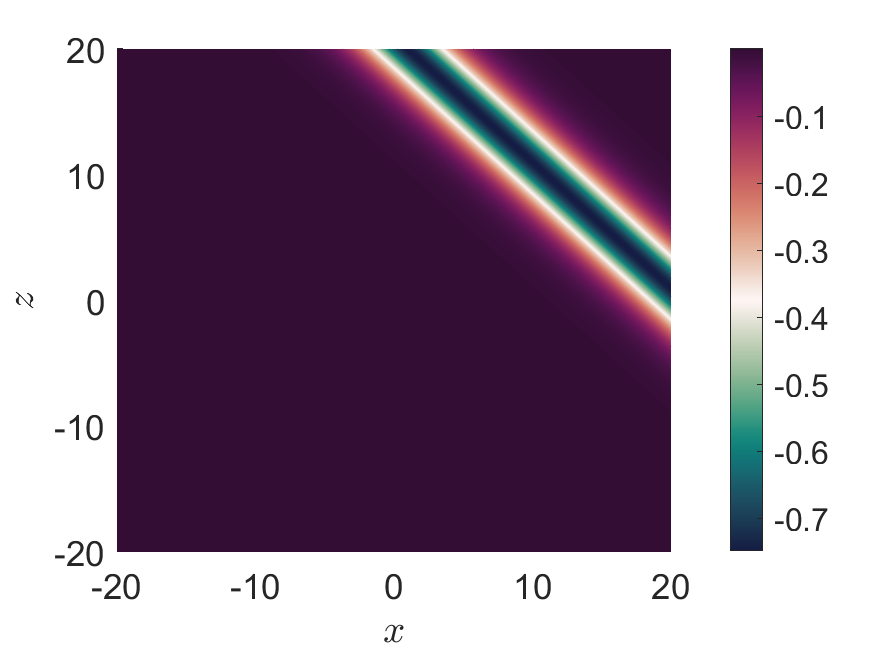}
          \caption{$t=-20$}
          \label{kdvfig7_1v}
      \end{subfigure}
      \centering
      \begin{subfigure}[b]{0.3\textwidth}
        \centering
          \includegraphics[width=\textwidth]{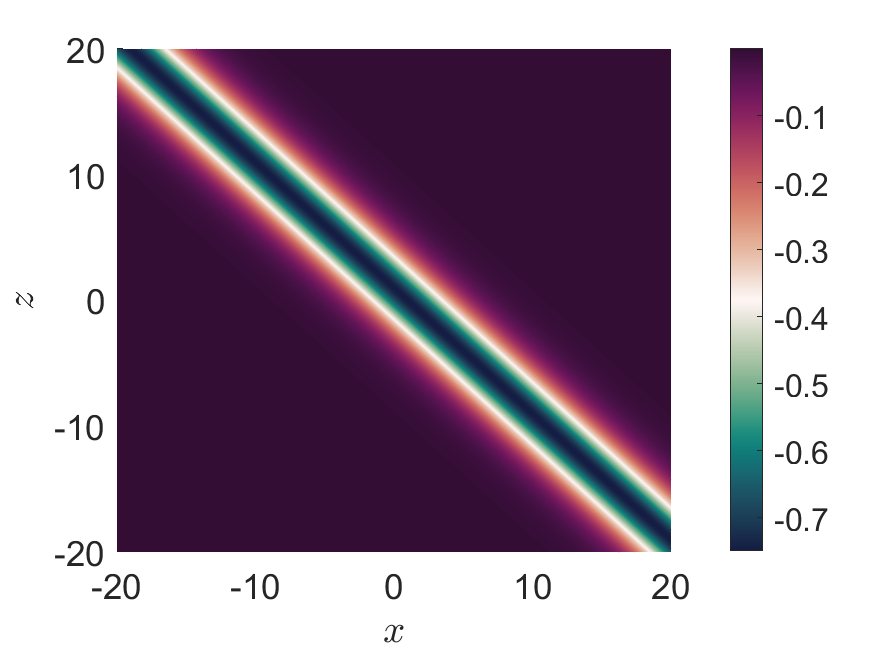}
          \caption{$t=0$}
          \label{kdvfig7_2v}
      \end{subfigure}
      \centering
      \begin{subfigure}[b]{0.3\textwidth}
        \centering
          \includegraphics[width=\textwidth]{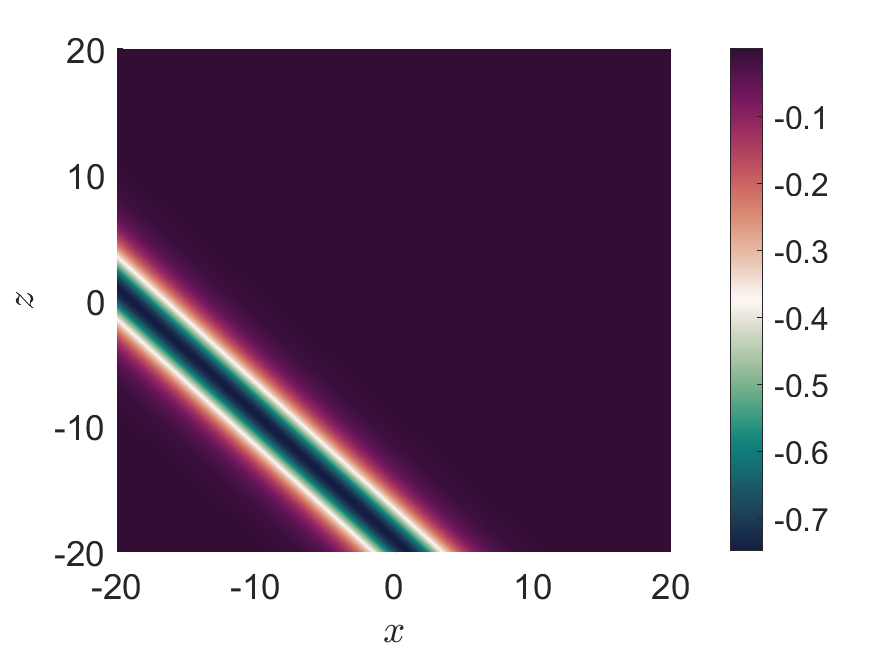}
          \caption{$t=20$}
          \label{kdvfig7_3v}
      \end{subfigure}
      \caption{Profiles of the dynamical behaviour of the solution \eqref{kdvsol7} in 3D and contour plots} \label{kdvfig7}
    \end{figure}

\section{Graphical discussion} \label{sec5}

In this section, we discuss the dynamic behaviour of the obtained solutions of the new (3+1)-dimenional KdV equation. The achieved solutions have been drawn in $3$D and a density plot to
illustrate solitary wave propagation. First, Figure \ref{kdvfig1} shows how the path of solitary waves changes when the arbitrary function and parameters are changed in the solution \eqref{kdvsol1}. Therefore, when we select $F_{1}(z,t)=\operatorname{sech}(z-t)$, $F_{3}(x,t)=\operatorname{Si}(x/z)$, $F_{4}(x,z)=\cos(z-x)$, and $z=10$, we observe an interaction of a lump solution with a breather on a periodic background in Figures~\ref{kdvfig1_1}, \ref{kdvfig1_1v}. However, Figures~\ref{kdvfig1_2}, \ref{kdvfig1_2v}
show a vertical interaction of a breather with dark soliton on a multi-periodic background for the adjustment of $F_1(z,t)=\operatorname{sech}(z/t)$, $F_{3}(x,t)=\operatorname{Si}(xt)$, $F_{4}(x,z)=\cos(z-x)$, and $z=5$.  Furthermore, we  see an interaction of a soliton with a doubly periodic solution in Figures~\ref{kdvfig1_3}, \ref{kdvfig1_3v} for $F_{1}(z,t)=\operatorname{sech}(z/t)$, $F_{3}(x,t)=\operatorname{Si}(xt)$, $F_{4}(x,z)=\cos(z-x)$, and $t=2$.  Figures~\ref{kdvfig1_4}, \ref{kdvfig1_4v} reveal three soliton interactions on a periodic background for the  functions $F_{1}(z,t)=\operatorname{sech}(z-t)$, $F_{3}(x,t)=\operatorname{Si}(xt)$, $F_{4}(x,z)=\operatorname{tanh}(z-x)$, and  $t=1$. A lump solution appears at the point $(0,0)$ in Figures~\ref{kdvfig1_5}, \ref{kdvfig1_5v} for $F_{1}(z,t)=\operatorname{sech}(zt)$, $F_{3}(x,t)=\operatorname{Si}(xt)$, $F_{4}(x,z)=\operatorname{sech}(z/x)\operatorname{tanh}(z/x)$, and $t=0.2$. 
For the same function $F_{1}(z,t)$ and value of $t$,  we take $F_{3}(x,t)=\sin(t-x)$, $F_4(x,z)=\arctan(\sinh(z/x))$  and draw a lump on a periodic background in Figures~\ref{kdvfig1_6}, \ref{kdvfig1_6v}.
Now, we proceed to the graphical illustration of solution \eqref{kdvsol2}. We set $\alpha_1=0.5$ and $z=0.1$ to get a breather solution passing on a periodic path in Figures~\ref{kdvfig22}, \ref{kdvfig2v}.
Taking the arbitrary constants as $c_1=1$, $c_2=5$, $c_3=0$, $c_4=19$, $c_5=1$, and $z=1$ in the solution \eqref{kdvsol4}, nine breathers are produced that are located parallel to each other as seen in the Figures~\ref{kdvfig44}, \ref{kdvfig4v}. 
In Figures~\ref{kdvfig5_1}, \ref{kdvfig5_1v}, a doubly periodic
  solution \eqref{kdvsol5} is shown with the selection of $c_1=0.5$, $c_i=1,\, i=2,...,6$, and $y=z=10$. Besides, when we change the constants in solution \eqref{kdvsol5} to $c_1=5$, $c_2=1$, $c_3=10$, $c_4=5$, $c_5=0.1$, $c_6=1$, $y=x=0$, collision of vertical two periodic appears as shown in Figures~\ref{kdvfig5_2}, \ref{kdvfig5_2v} in the domain $[-100,100]^2$.
  The second periodic propagation over the $z$-axis is vertically located on the crest of the $t$-axis, which is clearly shown in Figures~\ref{kdvfig5_3}, \ref{kdvfig5_3v} with domain $[-20,20]^2$. A bright dark solitary wave is revealed in Figures~\ref{kdvfig66}, \ref{kdvfig6v} 
  for the solution \eqref{kdvsol6} with  $F_3(t)=3/t^4$, $F_4(t)=\operatorname{sech}(t)$  and $y=x=0$.
In the end, we investigate a dark soliton solution in Figure~\ref{kdvfig7} associated with the solution \eqref{kdvsol7} with   $c_1=1$  and $y=1$. Additionally, the dark soliton is travelling over the time $t=-20$, $t=0$, $t=20$ in the Figures~\ref{kdvfig7_1}, \ref{kdvfig7_1v}, 
Figures~\ref{kdvfig7_2}, \ref{kdvfig7_2v}, 
and 
Figures~\ref{kdvfig7_3}, \ref{kdvfig7_3v}, respectively.


\section{Conservation laws of \eqref{NewKdVTren}} \label{sec6}

In this section we construct conserved vectors for the (3+1)-dimensional KdV equation \eqref{NewKdVTren}. We invoke the theorem due to Ibragimov \cite{Ibragimov2007}.
This theorem does not require a Lagrangian and can be applied to differential equations that are not  obtained from the variational principle.

For the equation \eqref{NewKdVTren}, we take the Lagrangian as
\begin{equation}
{\cal L} = v\left\lbrace 3q_{xxyt}+3q_{xxzt}-(q_{xt}-6q_{x}q_{xx}+q_{xxxx})_{yz}-2\left(q_{xx}q_{y}\right)_{xz}-2\left(q_{xx}q_{z}\right)_{xy}\right\rbrace, 
\end{equation}
where $v$ is the new dependent variable, $v=v(t,x,y,z)$ \cite{Ibragimov2007}.
The adjoint equation of   \eqref{NewKdVTren}   has the form
\begin{equation}\label{adj}
	10 q_{xx} v_{xyz}-4 q_{ xy} v_{xxz}-4 q_{xz}
	v_{xxy}+6 q_{x} v_{xxyz}-2 q_{y} v_{xxxz}-2
	q_{z} v_{xxxy}-v_{txyz}+3 v_{txxy}+3 v_{ txxz}-v_{xxxxyz}=0.
\end{equation}
Now applying the formula for conserved vectors \cite{Ibragimov2007}
	\begin{align}
	\label{conform}
	T^i =\, &\, \xi^i \mathcal{L}+W^{\sigma}\Bigg[\frac{\partial \mathcal{L} }{\partial q^{\sigma}_i}-D_j\frac{\partial \mathcal{L}}{\partial q^{\sigma}_{ij}}+D_jD_k \left(\frac{\partial \mathcal{L}}{\partial q^{\sigma}_{ijk}}\right)+\cdots\Bigg]
	+D_j (W^{\sigma})\Bigg[\frac{\partial \mathcal{L}}{\partial q^{\sigma}_{ij}}
	\nonumber\\
	&-D_k\frac{\partial\mathcal{L}}{\partial q^{\sigma}_{ijk}}+\cdots\Bigg]+D_jD_k(W^{\sigma}) \frac{\partial \mathcal{L}}{\partial q_{ijk}}+\cdots, \,\,\,  
\end{align}
where $W^\sigma$ is the Lie characteristic function  given as
$W^\sigma=\eta^\sigma-\xi^jq_j^\sigma$, we can write down the conserved vectors corresponding to all the seven symmetries \eqref{NewSymkdv} as follows:
 
	\begin{align}
	T_1^t= \,  &\frac{1}{12} v_{xz} q_{xy}
	-\frac{1}{4} v_{xx} q_{xy}
	+\frac{1}{12} q_{xz} v_{xy}
	-\frac{1}{12} v_{x} q_{xyz}
	-\frac{1}{4} q_{x} v_{xyz}
	+\frac{1}{12} v_{yz} q_{xx}
	-\frac{1}{2} v_{xz} q_{xx}
	-\frac{1}{2} v_{xy} q_{xx}
	-\frac{1}{4} q_{xz} v_{xx}  \nonumber \\
	&
	-\frac{1}{12} v_{y} q_{xxz}
	+\frac{1}{2} v_{x} q_{xxz}
	+\frac{3}{4} q_{x} v_{xxz}
	-\frac{1}{12} v_{z} q_{xxy}
	+\frac{1}{2} v_{x} q_{xxy}
	+\frac{3}{4} q_{x} v_{xxy}
	+\frac{1}{4} v q_{xxyz}
	+\frac{1}{4} v_{z} q_{xxx}
	+\frac{1}{4} v_{y} q_{xxx}  \nonumber \\
	&
	-\frac{3}{4} v  q_{xxxz}
	-\frac{3}{4} v q_{xxxy},  \nonumber \\
	T_1^x=	&3
	v_{xyz} q_{x}{}^2
	-\frac{11}{3} v_{xz} q_{xy} q_{x}
	-\frac{11}{3} q_{xz} v_{xy}	q_{x}
	+\frac{8}{3} v_{x}
	q_{xyz} q_{x}
	+\frac{4}{3}
	v_{yz} q_{xx} q_{x}
	+q_{yz} v_{xx} q_{x}
	+\frac{7}{6} v_{y} q_{xxz}q_{x}
	-\frac{3}{2} q_{y}v_{xxz} q_{x}  \nonumber \\
	&
	+\frac{7}{6}
	v_{z} q_{xxy} q_{x}
	-\frac{3}{2} q_{z} v_{xxy}q_{x}
	+v q_{xxyz}	q_{x}
	-\frac{2}{3} v_{xxxyz}
	q_{x}
	-\frac{1}{4} v_{tyz}
	q_{x}
	+\frac{3}{2} v_{txz}
	q_{x}
	+\frac{3}{2} v_{txy}
	q_{x}
	+\frac{8}{3} v_{x}
	q_{xz} q_{xy}           \nonumber \\
	&
	-2 q_{yz}
	v_{x} q_{xx}
	+\frac{2}{3}
	v_{y} q_{xz} q_{xx}
	+q_{y} v_{xz} q_{xx}
	+\frac{2}{3} v_{z} q_{xy}
	q_{xx}
	+q_{z} v_{xy}
	q_{xx}
	+\frac{1}{2} q_{y}
	q_{xz} v_{xx}
	+\frac{1}{2}
	q_{z} q_{xy} v_{xx}  
	-q_{y} v_{x} q_{xxz}  \nonumber \\
	&
	+\frac{1}{2} v q_{xy}
	q_{xxz}
	-q_{z} v_{x}
	q_{xxy}    
	+\frac{1}{2} v
	q_{xz} q_{xxy}
	-\frac{1}{5}
	v_{xxz} q_{xxy}
	-\frac{1}{5}
	q_{xxz} v_{xxy}
	+\frac{1}{5}
	v_{xx} q_{xxyz}
	+\frac{2}{5}
	q_{xx} v_{xxyz}
	-\frac{1}{2}
	v_{z} q_{y} q_{xxx}  \nonumber \\
	&
	-\frac{1}{2} q_{z} v_{y}
	q_{xxx}
	-v q_{yz}
	q_{xxx}
	-\frac{1}{5} v_{xyz}
	q_{xxx}
	-\frac{1}{15} q_{xyz}
	v_{xxx}
	-\frac{1}{2} v
	q_{y} q_{xxxz}
	+\frac{1}{5}
	v_{xy} q_{xxxz}
	+\frac{2}{15}
	q_{xy} v_{xxxz}
	-\frac{1}{2}
	v q_{z} q_{xxxy}  \nonumber \\
	&
	+\frac{1}{5} v_{xz} q_{xxxy}
	+\frac{2}{15} q_{xz} v_{xxxy}-\frac{2}{5} v_{x} q_{xxxyz}+\frac{1}{15} v_{yz} q_{xxxx}-\frac{2}{15} v_{y} q_{xxxxz}-\frac{2}{15} v_{z} q_{xxxxy}-\frac{1}{3} v q_{xxxxyz}   
	\nonumber \\
	&
	-\frac{1}{12} q_{xyz} v_{t}+\frac{1}{2} q_{xxz} v_{t}+\frac{1}{2} q_{xxy} v_{t}+\frac{1}{12} q_{xy} v_{tz}-\frac{1}{2} q_{xx} v_{tz}+\frac{1}{12} q_{xz} v_{ty}-\frac{1}{2} q_{xx} v_{ty}+\frac{1}{12} v_{yz} q_{tx}-\frac{1}{2} v_{xz} q_{tx} \nonumber \\
	&
	-\frac{1}{2} v_{xy} q_{tx}-\frac{1}{2} q_{xz} v_{tx}-\frac{1}{2} q_{xy} v_{tx}-\frac{1}{12} v_{y} q_{txz}+\frac{1}{2} v_{x} q_{txz}-\frac{1}{12} v_{z} q_{txy}+\frac{1}{2} v_{x} q_{txy}-\frac{3}{4} v q_{txyz}+\frac{1}{2} v_{z} q_{txx} \nonumber \\
	&
	+\frac{1}{2} v_{y} q_{txx}+\frac{3}{2} v q_{txxz}+\frac{3}{2} v q_{txxy},\nonumber \\
	T_1^y= \,	&\frac{3}{2} v_{xxz} q_{x}{}^2+\frac{4}{3} v_{xz} q_{xx} q_{x}-\frac{11}{6} q_{xz}
	v_{xx} q_{x}+\frac{7}{6}
	v_{x} q_{xxz} q_{x}-\frac{2}{3} v_{z} q_{xxx}
	q_{x}-\frac{1}{2} q_{z}
	v_{xxx} q_{x}-\frac{1}{2}
	v q_{xxxz} q_{x}-\frac{1}{6} v_{xxxxz} q_{x}
	\nonumber \\
	&
	-\frac{1}{4} v_{txz} q_{x}+\frac{3}{4} v_{txx} q_{x}-\frac{2}{3} v_{z} q_{xx}{}^2+\frac{2}{3} v_{x} q_{xz} q_{xx}+\frac{1}{2} q_{z}
	q_{xx} v_{xx}-\frac{1}{2}
	v q_{xx} q_{xxz}-\frac{1}{2} q_{z} v_{x}
	q_{xxx}+\frac{1}{2} v	q_{xz} q_{xxx}
		\nonumber \\
	&
	-\frac{1}{10}
	v_{xxz} q_{xxx}-\frac{1}{15}
	q_{xxz} v_{xxx}+\frac{1}{10}
	v_{xx} q_{xxxz}+\frac{2}{15}
	q_{xx} v_{xxxz}+\frac{1}{2}
	v q_{z} q_{xxxx}+\frac{1}{15} v_{xz} q_{xxxx}+\frac{1}{30} q_{xz} v_{xxxx}
	\nonumber \\
	&
	-\frac{2}{15} v_{x} q_{xxxxz}-\frac{1}{30} v_{z} q_{xxxxx}+\frac{1}{6} v q_{xxxxxz}-\frac{1}{12} q_{xxz} v_{t}+\frac{1}{4} q_{xxx} v_{t}
	+\frac{1}{12} q_{xx} v_{tz}+\frac{1}{12} v_{xz} q_{tx}-\frac{1}{4} v_{xx} q_{tx}
	\nonumber \\
	&
	+\frac{1}{12} q_{xz} v_{tx}-\frac{1}{2} q_{xx} v_{tx}-\frac{1}{12} v_{x} q_{txz}-\frac{1}{12} v_{z} q_{txx}+\frac{1}{2} v_{x} q_{txx}+\frac{1}{4} v q_{txxz}-\frac{3}{4} v q_{txxx},  \nonumber \\
	T_1^z= \, &\frac{3}{2} v_{xxy} q_{x}{}^2+\frac{4}{3} v_{xy} q_{xx} q_{x}-\frac{11}{6} q_{xy}
	v_{xx} q_{x}+\frac{7}{6}
	v_{x} q_{xxy} q_{x}-\frac{2}{3} v_{y} q_{xxx}
	q_{x}-\frac{1}{2} q_{y}
	v_{xxx} q_{x}-\frac{1}{2}
	v q_{xxxy} q_{x}-\frac{1}{6} v_{xxxxy} q_{x}
	\nonumber \\
	&
	-\frac{1}{4} v_{txy} q_{x}+\frac{3}{4} v_{txx} q_{x}-\frac{2}{3} v_{y} q_{xx}{}^2+\frac{2}{3} v_{x} q_{xy} q_{xx}+\frac{1}{2} q_{y}
	q_{xx} v_{xx}-\frac{1}{2}
	v q_{xx} q_{xxy}-\frac{1}{2} q_{y} v_{x}
	q_{xxx}+\frac{1}{2} v
	q_{xy} q_{xxx}
	\nonumber \\
	&
	-\frac{1}{10}
	v_{xxy} q_{xxx}
	-\frac{1}{15} q_{xxy} v_{xxx}
	+\frac{1}{10}v_{xx} q_{xxxy}
	+\frac{2}{15}q_{xx} v_{xxxy}
	+\frac{1}{2}v q_{y} q_{xxxx}
	+\frac{1}{15} v_{xy} q_{xxxx}
	+\frac{1}{30} q_{xy} v_{xxxx}
\nonumber \\
&
	-\frac{2}{15} v_{x} q_{xxxxy}
	-\frac{1}{30} v_{y} q_{xxxxx}
	+\frac{1}{6} v q_{xxxxxy}
	-\frac{1}{12} q_{xxy} v_{t}
	+\frac{1}{4} q_{xxx} v_{t}
	+\frac{1}{12} q_{xx} v_{ty}
	+\frac{1}{12} v_{xy} q_{tx}
	-\frac{1}{4} v_{xx} q_{tx}
	\nonumber \\
	&
	+\frac{1}{12} q_{xy} v_{tx}
	-\frac{1}{2} q_{xx} v_{tx}
	-\frac{1}{12} v_{x} q_{txy}
	-\frac{1}{12} v_{y} q_{txx}
	+\frac{1}{2} v_{x} q_{txx}
	+\frac{1}{4} v q_{txxy}
	-\frac{3}{4} v q_{txxx}; \nonumber 
\end{align}
	\begin{align}
	T_2^t= \, &-\frac{1}{12} q_{yyz} v_{x}
	+\frac{1}{2} q_{xyz} v_{x}
	+\frac{1}{2} q_{xyy} v_{x}
	+\frac{1}{12} q_{yy} v_{xz}
	+\frac{1}{12} v_{yz} q_{xy}
	-\frac{1}{2} v_{xz} q_{xy}
	+\frac{1}{12} q_{yz} v_{xy}
	-\frac{1}{2} q_{xy} v_{xy}
	-\frac{1}{12} v_{y} q_{xyz}
		\nonumber \\
	&
	-\frac{1}{4} q_{y} v_{xyz}
	-\frac{1}{12} v_{z} q_{xyy}
	+\frac{1}{4} v q_{xyyz}
	-\frac{1}{4} q_{yz} v_{xx}
	-\frac{1}{4} q_{yy} v_{xx}
	+\frac{3}{4} q_{y} v_{xxz}
	+\frac{1}{4} v_{z} q_{xxy}
	+\frac{1}{4} v_{y} q_{xxy}+\frac{3}{4} q_{y} v_{xxy}
	\nonumber \\
&
	-\frac{3}{4} v q_{xxyz}
	-\frac{3}{4} v q_{xxyy},  \nonumber \\
	T_2^x= \, &
	-\frac{3}{2} v_{xxz} q_{y}{}^2
	-\frac{5}{3} v_{xz} q_{xy} q_{y}
	-\frac{8}{3} q_{xz}
	v_{xy} q_{y}+\frac{2}{3}
	v_{x} q_{xyz} q_{y}+3
	q_{x} v_{xyz} q_{y}+\frac{7}{3} v_{yz} q_{xx}
	q_{y}+\frac{3}{2} q_{yz}
	v_{xx} q_{y}+\frac{1}{6}
	v_{y} q_{xxz} q_{y}
		\nonumber \\
	&
	-\frac{1}{3} v_{z} q_{xxy}
	q_{y}-\frac{3}{2} q_{z}
	v_{xxy} q_{y}-\frac{1}{2}
	v q_{xxyz} q_{y}-\frac{2}{3} v_{xxxyz} q_{y}-\frac{1}{4} v_{tyz} q_{y}+\frac{3}{2} v_{txz} q_{y}+\frac{3}{2} v_{txy} q_{y}+\frac{4}{3} v_{z} q_{xy}^2+q_{yyz} q_{x}
	v_{x}
		\nonumber \\
	&
	+\frac{4}{3} q_{yy}
	v_{x} q_{xz}-q_{yy}
	q_{x} v_{xz}-v_{yz}
	q_{x} q_{xy}-\frac{2}{3}
	q_{yz} v_{x} q_{xy}+\frac{4}{3} v_{y} q_{xz}
	q_{xy}-q_{yz} q_{x}
	v_{xy}+q_{z} q_{xy}
	v_{xy}+v_{y} q_{x}
	q_{xyz}
	\nonumber \\
	&
	-5 v q_{xy}
	q_{xyz}+v_{z} q_{x}
	q_{xyy}-q_{z} v_{x}
	q_{xyy}-\frac{8}{3} v
	q_{xz} q_{xyy}-3 v
	q_{x} q_{xyyz}-\frac{2}{3}
	v_{y} q_{yz} q_{xx}-\frac{2}{3} v_{z} q_{yy}
	q_{xx}+\frac{1}{3} v q_{yyz} q_{xx}
	\nonumber \\
	&
	+\frac{1}{2} q_{z}
	q_{yy} v_{xx}+\frac{1}{5}
	q_{xyyz} v_{xx}-\frac{5}{6}
	v q_{yy} q_{xxz}-\frac{1}{5} q_{xyy} v_{xxz}-\frac{1}{2} q_{z} v_{y}
	q_{xxy}+\frac{13}{6} v
	q_{yz} q_{xxy}-\frac{1}{5}
	v_{xyz} q_{xxy}-\frac{1}{5}
	q_{xyz} v_{xxy}
	\nonumber \\
	&
	+\frac{1}{5}
	v_{xy} q_{xxyz}+\frac{2}{5}
	q_{xy} v_{xxyz}+\frac{3}{2}
	v q_{z} q_{xxyy}+\frac{1}{5} v_{xz} q_{xxyy}-\frac{2}{5} v_{x} q_{xxyyz}-\frac{1}{15} q_{yyz} v_{xxx}+\frac{2}{15} q_{yy} v_{xxxz}
	\nonumber \\
	&
	+\frac{1}{15} v_{yz} q_{xxxy}+\frac{2}{15} q_{yz} v_{xxxy}-\frac{2}{15} v_{y} q_{xxxyz}-\frac{2}{15} v_{z} q_{xxxyy}+\frac{2}{3} v q_{xxxyyz}-\frac{1}{12} q_{yyz} v_{t}+\frac{1}{2} q_{xyz} v_{t}+\frac{1}{2} q_{xyy} v_{t}
	\nonumber \\
	&
	+\frac{1}{12} q_{yy} v_{tz}-\frac{1}{2} q_{xy} v_{tz}+\frac{1}{12} v_{yz} q_{ty}-\frac{1}{2} v_{xz} q_{ty}-\frac{1}{2} v_{xy} q_{ty}+\frac{1}{12} q_{yz} v_{ty}-\frac{1}{2} q_{xy} v_{ty}
		-\frac{1}{12} v_{y} q_{tyz}+\frac{1}{2} v_{x} q_{tyz}
	\nonumber \\
	&
	-\frac{1}{12} v_{z} q_{tyy}+\frac{1}{2} v_{x} q_{tyy}+\frac{1}{4} v q_{tyyz}-\frac{1}{2} q_{yz} v_{tx}-\frac{1}{2} q_{yy} v_{tx}+\frac{1}{2} v_{z} q_{txy}+\frac{1}{2} v_{y} q_{txy}-\frac{3}{2} v q_{txyz}-\frac{3}{2} v q_{txyy},
\nonumber \\
	T_2^y= \, &
	\frac{4}{3} v_{x} q_{xz}
	q_{xy}-q_{x} v_{xz}
	q_{xy}-\frac{2}{3} v_{z}
	q_{xx} q_{xy}+\frac{1}{2}
	q_{z} v_{xx} q_{xy}+\frac{19}{6} v q_{xxz}
	q_{xy}+\frac{2}{15} v_{xxxz}
	q_{xy}+\frac{1}{12} v_{tz}
	q_{xy}-\frac{1}{2} v_{tx}
	q_{xy}
	\nonumber \\
	&
	+q_{x} v_{x}
	q_{xyz}-\frac{2}{3} q_{yz}
	v_{x} q_{xx}+\frac{7}{3}
	q_{y} v_{xz} q_{xx}+\frac{7}{3} v q_{xyz}
	q_{xx}-\frac{1}{2} q_{yz}
	q_{x} v_{xx}-\frac{4}{3}
	q_{y} q_{xz} v_{xx}+\frac{1}{6} q_{y} v_{x}
	q_{xxz}+\frac{3}{2} q_{y}
	q_{x} v_{xxz}
	\nonumber \\
	&
	+\frac{1}{2}
	v_{z} q_{x} q_{xxy}-\frac{1}{2} q_{z} v_{x}
	q_{xxy}+\frac{8}{3} v
	q_{xz} q_{xxy}-\frac{1}{10}
	v_{xxz} q_{xxy}+\frac{9}{2}
	v q_{x} q_{xxyz}+\frac{1}{10} v_{xx} q_{xxyz}-\frac{7}{6} v_{z} q_{y}
	q_{xxx}
	\nonumber \\
	&
	-\frac{13}{6} v
	q_{yz} q_{xxx}-\frac{1}{2}
	q_{z} q_{y} v_{xxx}-\frac{1}{15} q_{xyz} v_{xxx}-v q_{y} q_{xxxz}-\frac{3}{2} v q_{z}
	q_{xxxy}+\frac{1}{15} v_{xz}
	q_{xxxy}-\frac{2}{15} v_{x}
	q_{xxxyz}
	\nonumber \\
	&
	+\frac{1}{30} q_{yz}
	v_{xxxx}-\frac{1}{6} q_{y}
	v_{xxxxz}
		-\frac{1}{30} v_{z}
	q_{xxxxy}-\frac{5}{6} v
	q_{xxxxyz}-\frac{1}{12} q_{xyz}
	v_{t}+\frac{1}{4} q_{xxy}
	v_{t}+\frac{1}{12} v_{xz}
	q_{ty}-\frac{1}{4} v_{xx}
	q_{ty}
	\nonumber \\
	&
	-\frac{1}{12} v_{x}
	q_{tyz}+\frac{1}{12} q_{yz}
	v_{tx}-\frac{1}{4} q_{y}
	v_{txz}-\frac{1}{12} v_{z}
	q_{txy}+\frac{1}{2} v_{x}
	q_{txy}-\frac{3}{4} v
	q_{txyz}+\frac{3}{4} q_{y}
	v_{txx}+3 v q_{txxz}+\frac{9}{4} v q_{txxy}, \nonumber
	\\
	T_2^z= \, &
	-\frac{1}{2} v_{xxx} q_{y}{}^2+\frac{7}{3} v_{xy} q_{xx} q_{y}-\frac{5}{6} q_{xy}
	v_{xx} q_{y}-\frac{1}{3}
	v_{x} q_{xxy} q_{y}+\frac{3}{2} q_{x} v_{xxy}
	q_{y}-\frac{7}{6} v_{y}
	q_{xxx} q_{y}+\frac{3}{2}
	v q_{xxxy} q_{y}-\frac{1}{6} v_{xxxxy} q_{y}
	\nonumber \\
	&
	-\frac{1}{4} v_{txy} q_{y}+\frac{3}{4} v_{txx} q_{y}+\frac{4}{3} v_{x} q_{xy}{}^2-q_{x} q_{xy}
	v_{xy}+q_{x} v_{x}
	q_{xyy}-\frac{2}{3} q_{yy}
	v_{x} q_{xx}-\frac{2}{3}
	v_{y} q_{xy} q_{xx}+\frac{1}{3} v q_{xyy}
	q_{xx}
	\nonumber \\
	&
	-\frac{1}{2} q_{yy}
	q_{x} v_{xx}+\frac{1}{2}
	v_{y} q_{x} q_{xxy}-\frac{13}{6} v q_{xy}
	q_{xxy}-\frac{1}{10} q_{xxy}
	v_{xxy}-\frac{3}{2} v
	q_{x} q_{xxyy}+\frac{1}{10}
	v_{xx} q_{xxyy}+\frac{11}{6}
	v q_{yy} q_{xxx}
	\nonumber \\
	&
	-\frac{1}{15} q_{xyy} v_{xxx}+\frac{1}{15} v_{xy} q_{xxxy}+\frac{2}{15} q_{xy} v_{xxxy}-\frac{2}{15} v_{x} q_{xxxyy}+\frac{1}{30} q_{yy} v_{xxxx}-\frac{1}{30} v_{y} q_{xxxxy}+\frac{1}{6} v q_{xxxxyy}
	\nonumber \\
	&
	-\frac{1}{12} q_{xyy} v_{t}+\frac{1}{4} q_{xxy} v_{t}+\frac{1}{12} v_{xy} q_{ty}-\frac{1}{4} v_{xx} q_{ty}+\frac{1}{12} q_{xy} v_{ty}-\frac{1}{12} v_{x} q_{tyy}+\frac{1}{12} q_{yy} v_{tx}-\frac{1}{2} q_{xy} v_{tx}-\frac{1}{12} v_{y} q_{txy}
	\nonumber \\
	&
	+\frac{1}{2} v_{x} q_{txy}+\frac{1}{4} v q_{txyy}-\frac{3}{4} v q_{txxy}; \nonumber
\end{align}

\begin{align}
	T_3^t= \, &-\frac{1}{12} q_{yzz} v_{x}+\frac{1}{2} q_{xzz} v_{x}+\frac{1}{2} q_{xyz} v_{x}+\frac{1}{12} v_{yz} q_{xz}+\frac{1}{12} q_{yz} v_{xz}-\frac{1}{2} q_{xz} v_{xz}
	-\frac{1}{12} v_{y} q_{xzz}+\frac{1}{12} q_{zz} v_{xy}-\frac{1}{2} q_{xz} v_{xy}
	\nonumber \\
	&
	-\frac{1}{12} v_{z} q_{xyz}-\frac{1}{4} q_{z} v_{xyz}+\frac{1}{4} v q_{xyzz}-\frac{1}{4} q_{zz} v_{xx}-\frac{1}{4} q_{yz} v_{xx}+\frac{1}{4} v_{z} q_{xxz}+\frac{1}{4} v_{y} q_{xxz}+\frac{3}{4} q_{z} v_{xxz}-\frac{3}{4} v q_{xxzz}
	\nonumber \\
	&
	+\frac{3}{4} q_{z} v_{xxy}-\frac{3}{4} v q_{xxyz}, \nonumber
	\\
	T_3^x= \, &
	-\frac{3}{2} v_{xxy} q_{z}{}^2-\frac{8}{3} v_{xz} q_{xy} q_{z}-\frac{5}{3} q_{xz}
	v_{xy} q_{z}+\frac{2}{3}
	v_{x} q_{xyz} q_{z}+3
	q_{x} v_{xyz} q_{z}+\frac{7}{3} v_{yz} q_{xx}
	q_{z}+\frac{3}{2} q_{yz}
	v_{xx} q_{z}-\frac{1}{3}
	v_{y} q_{xxz} q_{z}
	\nonumber \\
	&
	-\frac{3}{2} q_{y} v_{xxz}
	q_{z}+\frac{1}{6} v_{z}
	q_{xxy} q_{z}-\frac{1}{2}
	v q_{xxyz} q_{z}-\frac{2}{3} v_{xxxyz} q_{z}-\frac{1}{4} v_{tyz} q_{z}+\frac{3}{2} v_{txz} q_{z}+\frac{3}{2} v_{txy} q_{z}+\frac{4}{3} v_{y} q_{xz}^2+q_{yzz} q_{x}
	v_{x}
	\nonumber \\
	&
	-v_{yz} q_{x}
	q_{xz}-\frac{2}{3} q_{yz}
	v_{x} q_{xz}-q_{yz}
	q_{x} v_{xz}+q_{y}
	q_{xz} v_{xz}+v_{y}
	q_{x} q_{xzz}-q_{y}
	v_{x} q_{xzz}+\frac{4}{3}
	q_{zz} v_{x} q_{xy}+\frac{4}{3} v_{z} q_{xz}
	q_{xy}
	\nonumber \\
	&
	-\frac{8}{3} v q_{xzz} q_{xy}-q_{zz} q_{x} v_{xy}+v_{z} q_{x}
	q_{xyz}-5 v q_{xz}
	q_{xyz}-3 v q_{x}
	q_{xyzz}-\frac{2}{3} q_{zz}
	v_{y} q_{xx}-\frac{2}{3}
	v_{z} q_{yz} q_{xx}+\frac{1}{3} v q_{yzz}
	q_{xx}
	\nonumber \\
	&
	+\frac{1}{2} q_{zz}
	q_{y} v_{xx}+\frac{1}{5}
	q_{xyzz} v_{xx}-\frac{1}{2}
	v_{z} q_{y} q_{xxz}+\frac{13}{6} v q_{yz}
	q_{xxz}-\frac{1}{5} v_{xyz}
	q_{xxz}-\frac{1}{5} q_{xyz}
	v_{xxz}+\frac{3}{2} v
	q_{y} q_{xxzz}+\frac{1}{5}
	v_{xy} q_{xxzz}
	\nonumber \\
	&
	-\frac{5}{6}
	v q_{zz} q_{xxy}-\frac{1}{5} q_{xzz} v_{xxy}+\frac{1}{5} v_{xz} q_{xxyz}+\frac{2}{5} q_{xz} v_{xxyz}-\frac{2}{5} v_{x} q_{xxyzz}-\frac{1}{15} q_{yzz} v_{xxx}+\frac{1}{15} v_{yz} q_{xxxz}
	\nonumber \\
	&
	+\frac{2}{15} q_{yz} v_{xxxz}-\frac{2}{15} v_{y} q_{xxxzz}
	+\frac{2}{15} q_{zz} v_{xxxy}-\frac{2}{15} v_{z} q_{xxxyz}+\frac{2}{3} v q_{xxxyzz}-\frac{1}{12} q_{yzz} v_{t}+\frac{1}{2} q_{xzz} v_{t}+\frac{1}{2} q_{xyz} v_{t}
	\nonumber \\
	&
	+\frac{1}{12} v_{yz} q_{tz}-\frac{1}{2} v_{xz} q_{tz}-\frac{1}{2} v_{xy} q_{tz}
	+\frac{1}{12} q_{yz} v_{tz}-\frac{1}{2} q_{xz} v_{tz}
	-\frac{1}{12} v_{y} q_{tzz}+\frac{1}{2} v_{x} q_{tzz}+\frac{1}{12} q_{zz} v_{ty}-\frac{1}{2} q_{xz} v_{ty}
	\nonumber \\
	&
	-\frac{1}{12} v_{z} q_{tyz}+\frac{1}{2} v_{x} q_{tyz}+\frac{1}{4} v q_{tyzz}-\frac{1}{2} q_{zz} v_{tx}-\frac{1}{2} q_{yz} v_{tx}+\frac{1}{2} v_{z} q_{txz}+\frac{1}{2} v_{y} q_{txz}-\frac{3}{2} v q_{txzz}-\frac{3}{2} v q_{txyz},  \nonumber
	\\
	T_3^y= \, &
	-\frac{1}{2} v_{xxx} q_{z}{}^2+\frac{7}{3} v_{xz} q_{xx} q_{z}-\frac{5}{6} q_{xz}
	v_{xx} q_{z}-\frac{1}{3}
	v_{x} q_{xxz} q_{z}+\frac{3}{2} q_{x} v_{xxz}
	q_{z}-\frac{7}{6} v_{z}
	q_{xxx} q_{z}+\frac{3}{2}
	v q_{xxxz} q_{z}-\frac{1}{6} v_{xxxxz} q_{z}
	\nonumber \\
	&
	-\frac{1}{4} v_{txz} q_{z}+\frac{3}{4} v_{txx} q_{z}+\frac{4}{3} v_{x} q_{xz}{}^2-q_{x} q_{xz}
	v_{xz}+q_{x} v_{x}
	q_{xzz}-\frac{2}{3} q_{zz}
	v_{x} q_{xx}-\frac{2}{3}
	v_{z} q_{xz} q_{xx}+\frac{1}{3} v q_{xzz}
	q_{xx}
	\nonumber \\
	&
	-\frac{1}{2} q_{zz}
	q_{x} v_{xx}+\frac{1}{2}
	v_{z} q_{x} q_{xxz}-\frac{13}{6} v q_{xz}
	q_{xxz}-\frac{1}{10} q_{xxz}
	v_{xxz}-\frac{3}{2} v
	q_{x} q_{xxzz}+\frac{1}{10}
	v_{xx} q_{xxzz}+\frac{11}{6}
	v q_{zz} q_{xxx}
	\nonumber \\
	&
	-\frac{1}{15} q_{xzz} v_{xxx}+\frac{1}{15} v_{xz} q_{xxxz}+\frac{2}{15} q_{xz} v_{xxxz}-\frac{2}{15} v_{x} q_{xxxzz}+\frac{1}{30} q_{zz} v_{xxxx}-\frac{1}{30} v_{z} q_{xxxxz}+\frac{1}{6} v q_{xxxxzz}
	\nonumber \\
	&
	-\frac{1}{12} q_{xzz} v_{t}+\frac{1}{4} q_{xxz} v_{t}+\frac{1}{12} v_{xz} q_{tz}-\frac{1}{4} v_{xx} q_{tz}+\frac{1}{12} q_{xz} v_{tz}-\frac{1}{12} v_{x} q_{tzz}+\frac{1}{12} q_{zz} v_{tx}-\frac{1}{2} q_{xz} v_{tx}-\frac{1}{12} v_{z} q_{txz}
	\nonumber \\
	&
	+\frac{1}{2} v_{x} q_{txz}+\frac{1}{4} v q_{txzz}-\frac{3}{4} v q_{txxz}, \nonumber
	\\
	T_3^z= \, &
	\frac{4}{3} v_{x} q_{xz}
	q_{xy}-\frac{4}{3} q_{z}
	v_{xx} q_{xy}+\frac{8}{3}
	v q_{xxz} q_{xy}-q_{x} q_{xz} v_{xy}+q_{x} v_{x} q_{xyz}-\frac{2}{3} q_{yz} v_{x}
	q_{xx}-\frac{2}{3} v_{y}
	q_{xz} q_{xx}+\frac{7}{3}
	q_{z} v_{xy} q_{xx}
	\nonumber \\
	&
	+\frac{7}{3} v q_{xyz}
	q_{xx}-\frac{1}{2} q_{yz}
	q_{x} v_{xx}+\frac{1}{2}
	q_{y} q_{xz} v_{xx}+\frac{1}{2} v_{y} q_{x}
	q_{xxz}-\frac{1}{2} q_{y}
	v_{x} q_{xxz}+\frac{1}{6}
	q_{z} v_{x} q_{xxy}+\frac{19}{6} v q_{xz}
	q_{xxy}+\frac{3}{2} q_{z}
	q_{x} v_{xxy}
	\nonumber \\
	&
	-\frac{1}{10}
	q_{xxz} v_{xxy}+\frac{9}{2}
	v q_{x} q_{xxyz}+\frac{1}{10} v_{xx} q_{xxyz}-\frac{7}{6} q_{z} v_{y}
	q_{xxx}-\frac{13}{6} v
	q_{yz} q_{xxx}-\frac{1}{2}
	q_{z} q_{y} v_{xxx}-\frac{1}{15} q_{xyz} v_{xxx}
	\nonumber \\
	&
	-\frac{3}{2} v q_{y}
	q_{xxxz}+\frac{1}{15} v_{xy}
	q_{xxxz}-v q_{z}
	q_{xxxy}+\frac{2}{15} q_{xz}
	v_{xxxy}-\frac{2}{15} v_{x}
	q_{xxxyz}+\frac{1}{30} q_{yz}
	v_{xxxx}-\frac{1}{30} v_{y}
	q_{xxxxz}
	\nonumber \\
	&
	-\frac{1}{6} q_{z}
	v_{xxxxy}-\frac{5}{6} v
	q_{xxxxyz}-\frac{1}{12} q_{xyz}
	v_{t}+\frac{1}{4} q_{xxz}
	v_{t}+\frac{1}{12} v_{xy}
	q_{tz}-\frac{1}{4} v_{xx}
	q_{tz}+\frac{1}{12} q_{xz}
	v_{ty}-\frac{1}{12} v_{x}
	q_{tyz}
	\nonumber \\
	&
	+\frac{1}{12} q_{yz}
	v_{tx}-\frac{1}{2} q_{xz}
	v_{tx}-\frac{1}{12} v_{y}
	q_{txz}+\frac{1}{2} v_{x}
	q_{txz}-\frac{1}{4} q_{z}
	v_{txy}-\frac{3}{4} v
	q_{txyz}+\frac{3}{4} q_{z}
	v_{txx}+\frac{9}{4} v
	q_{txxz}+3 v q_{txxy}; \nonumber
\end{align}

\begin{align}
	T_4^t= \, & 2 v q_{xyz} q_{xx}+4 v q_{xy} q_{xxz}+4 v q_{xz} q_{xxy}+6 v q_{x} q_{xxyz}-4 v q_{yz} q_{xxx}-2 v q_{y} q_{xxxz}-2 v q_{z} q_{xxxy}-v q_{xxxxyz}
	\nonumber \\
	&
	-\frac{1}{4}
	v_{xyz} q_{t}+\frac{3}{4}
	v_{xxz} q_{t}+\frac{3}{4}
	v_{xxy} q_{t}+\frac{1}{12}
	v_{xy} q_{tz}-\frac{1}{4}
	v_{xx} q_{tz}+\frac{1}{12}
	v_{xz} q_{ty}-\frac{1}{4}
	v_{xx} q_{ty}-\frac{1}{12}
	v_{x} q_{tyz}+\frac{1}{12}
	v_{yz} q_{tx}
	\nonumber \\
	&
	-\frac{1}{2}
	v_{xz} q_{tx}-\frac{1}{2}
	v_{xy} q_{tx}-\frac{1}{12}
	v_{y} q_{txz}+\frac{1}{2}
	v_{x} q_{txz}-\frac{1}{12}
	v_{z} q_{txy}+\frac{1}{2}
	v_{x} q_{txy}-\frac{3}{4}
	v q_{txyz}+\frac{1}{4}
	v_{z} q_{txx}+\frac{1}{4}
	v_{y} q_{txx}
	\nonumber \\
	&
	+\frac{9}{4}
	v q_{txxz}+\frac{9}{4}
	v q_{txxy},  \nonumber
	\\
	T_4^x=&
	-\frac{8}{3}
	v_{xz} q_{xy} q_{t}-\frac{8}{3} q_{xz} v_{xy}
	q_{t}+\frac{5}{3} v_{x}
	q_{xyz} q_{t}+3 q_{x}
	v_{xyz} q_{t}+\frac{7}{3}
	v_{yz} q_{xx} q_{t}+q_{yz} v_{xx} q_{t}+\frac{1}{6} v_{y} q_{xxz}
	q_{t}-\frac{3}{2} q_{y}
	v_{xxz} q_{t}
	\nonumber \\
	&
	+\frac{1}{6}
	v_{z} q_{xxy} q_{t}-\frac{3}{2} q_{z} v_{xxy}
	q_{t}-2 v q_{xxyz}
	q_{t}-\frac{2}{3} v_{xxxyz}
	q_{t}-\frac{1}{4} v_{tyz}
	q_{t}+\frac{3}{2} v_{txz}
	q_{t}+\frac{3}{2} v_{txy}
	q_{t}+\frac{4}{3} v_{x}
	q_{xy} q_{tz}-q_{x}
	v_{xy} q_{tz}
	\nonumber \\
	&
	-\frac{2}{3}
	v_{y} q_{xx} q_{tz}+\frac{1}{2} q_{y} v_{xx}
	q_{tz}-\frac{5}{6} v q_{xxy} q_{tz}+\frac{2}{15} v_{xxxy} q_{tz}+\frac{4}{3} v_{x}
	q_{xz} q_{ty}-q_{x}
	v_{xz} q_{ty}-\frac{2}{3}
	v_{z} q_{xx} q_{ty}+\frac{1}{2} q_{z} v_{xx}
	q_{ty}
	\nonumber \\
	&
	-\frac{5}{6} v q_{xxz} q_{ty}+\frac{2}{15} v_{xxxz} q_{ty}+\frac{1}{12} v_{tz} q_{ty}+\frac{1}{12} q_{tz}
	v_{ty}+q_{x} v_{x}
	q_{tyz}+\frac{1}{3} v
	q_{xx} q_{tyz}-\frac{1}{15}
	v_{xxx} q_{tyz}-\frac{1}{12}
	v_{t} q_{tyz}
	\nonumber \\
	&
	-v_{yz}
	q_{x} q_{tx}-2 q_{yz}
	v_{x} q_{tx}+\frac{4}{3}
	v_{y} q_{xz} q_{tx}+q_{y} v_{xz} q_{tx}+\frac{4}{3} v_{z} q_{xy}
	q_{tx}+q_{z} v_{xy}
	q_{tx}-\frac{7}{3} v q_{xyz} q_{tx}+\frac{2}{5} v_{xxyz} q_{tx}
	\nonumber \\
	&
	-\frac{1}{2} v_{tz} q_{tx}-\frac{1}{2} v_{ty}
	q_{tx}-\frac{1}{2} q_{tz}
	v_{tx}-\frac{1}{2} q_{ty}
	v_{tx}+v_{y} q_{x}
	q_{txz}-q_{y} v_{x}
	q_{txz}-\frac{8}{3} v
	q_{xy} q_{txz}-\frac{1}{5}
	v_{xxy} q_{txz}+\frac{1}{2}
	v_{t} q_{txz}
	\nonumber \\
	&
	+v_{z}
	q_{x} q_{txy}-q_{z}
	v_{x} q_{txy}-\frac{8}{3}
	v q_{xz} q_{txy}-\frac{1}{5} v_{xxz} q_{txy}+\frac{1}{2} v_{t} q_{txy}-3 v q_{x} q_{txyz}+\frac{1}{5} v_{xx} q_{txyz}-\frac{1}{2} v_{z} q_{y}
	q_{txx}
	\nonumber \\
	&
	-\frac{1}{2} q_{z}
	v_{y} q_{txx}+3 v
	q_{yz} q_{txx}-\frac{1}{5}
	v_{xyz} q_{txx}+\frac{3}{2}
	v q_{y} q_{txxz}+\frac{1}{5} v_{xy} q_{txxz}+\frac{3}{2} v q_{z}
	q_{txxy}+\frac{1}{5} v_{xz}
	q_{txxy}-\frac{2}{5} v_{x}
	q_{txxyz}
	\nonumber \\
	&
	+\frac{1}{15} v_{yz}
	q_{txxx}-\frac{2}{15} v_{y}
	q_{txxxz}-\frac{2}{15} v_{z}
	q_{txxxy}+\frac{2}{3} v
	q_{txxxyz}+\frac{1}{12} v_{yz}
	q_{tt}-\frac{1}{2} v_{xz}
	q_{tt}-\frac{1}{2} v_{xy}
	q_{tt}-\frac{1}{12} v_{y}
	q_{ttz}
	\nonumber \\
	&
	+\frac{1}{2} v_{x}
	q_{ttz}-\frac{1}{12} v_{z}
	q_{tty}+\frac{1}{2} v_{x}
	q_{tty}+\frac{1}{4} v
	q_{ttyz}+\frac{1}{2} v_{z}
	q_{ttx}+\frac{1}{2} v_{y}
	q_{ttx}-\frac{3}{2} v
	q_{ttxz}-\frac{3}{2} v
	q_{ttxy},   \nonumber
	\\
	T_4^y= \, &
	\frac{7}{3} v_{xz}
	q_{xx} q_{t}-\frac{4}{3}
	q_{xz} v_{xx} q_{t}+\frac{1}{6} v_{x} q_{xxz}
	q_{t}+\frac{3}{2} q_{x}
	v_{xxz} q_{t}-\frac{7}{6}
	v_{z} q_{xxx} q_{t}-\frac{1}{2} q_{z} v_{xxx}
	q_{t}+v q_{xxxz}
	q_{t}-\frac{1}{6} v_{xxxxz}
	q_{t}
	\nonumber \\
	&
	-\frac{1}{4} v_{txz}
	q_{t}+\frac{3}{4} v_{txx}
	q_{t}-\frac{2}{3} v_{x}
	q_{xx} q_{tz}-\frac{1}{2}
	q_{x} v_{xx} q_{tz}+\frac{11}{6} v q_{xxx}
	q_{tz}+\frac{1}{30} v_{xxxx}
	q_{tz}+\frac{4}{3} v_{x}
	q_{xz} q_{tx}-q_{x}
	v_{xz} q_{tx}
	\nonumber \\
	&
	-\frac{2}{3}
	v_{z} q_{xx} q_{tx}+\frac{1}{2} q_{z} v_{xx}
	q_{tx}-\frac{5}{6} v q_{xxz} q_{tx}+\frac{2}{15} v_{xxxz} q_{tx}+\frac{1}{12} v_{tz} q_{tx}+\frac{1}{12} q_{tz}
	v_{tx}-\frac{1}{2} q_{tx}
	v_{tx}+q_{x} v_{x}
	q_{txz}
	\nonumber \\
	&
	+\frac{1}{3} v
	q_{xx} q_{txz}-\frac{1}{15}
	v_{xxx} q_{txz}-\frac{1}{12}
	v_{t} q_{txz}+\frac{1}{2}
	v_{z} q_{x} q_{txx}-\frac{1}{2} q_{z} v_{x}
	q_{txx}-\frac{4}{3} v
	q_{xz} q_{txx}-\frac{1}{10}
	v_{xxz} q_{txx}+\frac{1}{4}
	v_{t} q_{txx}
	\nonumber \\
	&
	-\frac{3}{2}
	v q_{x} q_{txxz}+\frac{1}{10} v_{xx} q_{txxz}+\frac{1}{2} v q_{z}
	q_{txxx}+\frac{1}{15} v_{xz}
	q_{txxx}-\frac{2}{15} v_{x}
	q_{txxxz}-\frac{1}{30} v_{z}
	q_{txxxx}+\frac{1}{6} v
	q_{txxxxz}
	\nonumber \\
	&
	+\frac{1}{12} v_{xz}
	q_{tt}-\frac{1}{4} v_{xx}
	q_{tt}-\frac{1}{12} v_{x}
	q_{ttz}-\frac{1}{12} v_{z}
	q_{ttx}+\frac{1}{2} v_{x}
	q_{ttx}+\frac{1}{4} v
	q_{ttxz}-\frac{3}{4} v
	q_{ttxx},   \nonumber
	\\
	T_4^z= \, &
	\frac{7}{3} v_{xy}
	q_{xx} q_{t}-\frac{4}{3}
	q_{xy} v_{xx} q_{t}+\frac{1}{6} v_{x} q_{xxy}
	q_{t}+\frac{3}{2} q_{x}
	v_{xxy} q_{t}-\frac{7}{6}
	v_{y} q_{xxx} q_{t}-\frac{1}{2} q_{y} v_{xxx}
	q_{t}+v q_{xxxy}
	q_{t}-\frac{1}{6} v_{xxxxy}
	q_{t}
	\nonumber \\
	&
	-\frac{1}{4} v_{txy}
	q_{t}+\frac{3}{4} v_{txx}
	q_{t}-\frac{2}{3} v_{x}
	q_{xx} q_{ty}-\frac{1}{2}
	q_{x} v_{xx} q_{ty}+\frac{11}{6} v q_{xxx}
	q_{ty}+\frac{1}{30} v_{xxxx}
	q_{ty}+\frac{4}{3} v_{x}
	q_{xy} q_{tx}-q_{x}
	v_{xy} q_{tx}
	\nonumber \\
	&
	-\frac{2}{3}
	v_{y} q_{xx} q_{tx}+\frac{1}{2} q_{y} v_{xx}
	q_{tx}-\frac{5}{6} v q_{xxy} q_{tx}+\frac{2}{15} v_{xxxy} q_{tx}+\frac{1}{12} v_{ty} q_{tx}+\frac{1}{12} q_{ty}
	v_{tx}-\frac{1}{2} q_{tx}
	v_{tx}+q_{x} v_{x}
	q_{txy}
	\nonumber \\
	&
	+\frac{1}{3} v
	q_{xx} q_{txy}-\frac{1}{15}
	v_{xxx} q_{txy}-\frac{1}{12}
	v_{t} q_{txy}+\frac{1}{2}
	v_{y} q_{x} q_{txx}-\frac{1}{2} q_{y} v_{x}
	q_{txx}-\frac{4}{3} v
	q_{xy} q_{txx}-\frac{1}{10}
	v_{xxy} q_{txx}+\frac{1}{4}
	v_{t} q_{txx}
	\nonumber \\
	&
	-\frac{3}{2}
	v q_{x} q_{txxy}+\frac{1}{10} v_{xx} q_{txxy}+\frac{1}{2} v q_{y}
	q_{txxx}+\frac{1}{15} v_{xy}
	q_{txxx}-\frac{2}{15} v_{x}
	q_{txxxy}-\frac{1}{30} v_{y}
	q_{txxxx}+\frac{1}{6} v
	q_{txxxxy}
	\nonumber \\
	&
	+\frac{1}{12} v_{xy}
	q_{tt}-\frac{1}{4} v_{xx}
	q_{tt}-\frac{1}{12} v_{x}
	q_{tty}-\frac{1}{12} v_{y}
	q_{ttx}+\frac{1}{2} v_{x}
	q_{ttx}+\frac{1}{4} v
	q_{ttxy}-\frac{3}{4} v
	q_{ttxx}; \nonumber
\end{align}

\begin{align}
	T_5^t= \, & \frac{1}{4} t v_{xyz}-\frac{3}{4} t
	v_{xxz}-\frac{3}{4} t v_{xxy},  \nonumber
	\\
	T_5^x= \, &
	-\frac{7}{3} t q_{xx} v_{yz}-\frac{v_{yz}}{12}+\frac{v_{xz}}{2}+\frac{8}{3} t v_{xz} q_{xy}+\frac{8}{3} t q_{xz} v_{xy}+\frac{v_{xy}}{2}-\frac{5}{3} t
	v_{x} q_{xyz}-3 t q_{x} v_{xyz}-t q_{yz} v_{xx}-\frac{1}{6} t v_{y} q_{xxz}
	\nonumber \\
	&
	+\frac{3}{2} t q_{y} v_{xxz}-\frac{1}{6} t v_{z} q_{xxy}+\frac{3}{2} t q_{z} v_{xxy}+2 t v q_{xxyz}+\frac{2}{3} t v_{xxxyz}+\frac{1}{4} t v_{tyz}-\frac{3}{2} t v_{txz}-\frac{3}{2}
	t v_{txy},  \nonumber
	\\
	T_5^y= \, &
	-\frac{7}{3} t q_{xx}
	v_{xz}-\frac{v_{xz}}{12}+\frac{4}{3} t q_{xz}
	v_{xx}+\frac{v_{xx}}{4}-\frac{1}{6} t v_{x} q_{xxz}-\frac{3}{2} t q_{x} v_{xxz}+\frac{7}{6} t v_{z} q_{xxx}+\frac{1}{2} t q_{z} v_{xxx}-t v q_{xxxz}
	\nonumber \\
	&
	+\frac{1}{6}
	t v_{xxxxz}+\frac{1}{4} t v_{txz}-\frac{3}{4} t v_{txx}, \nonumber
	\\
	T_5^z= \, &
	-\frac{7}{3}
	t q_{xx} v_{xy}-\frac{v_{xy}}{12}+\frac{4}{3} t
	q_{xy} v_{xx}+\frac{v_{xx}}{4}-\frac{1}{6} t v_{x} q_{xxy}-\frac{3}{2} t q_{x} v_{xxy}+\frac{7}{6} t v_{y} q_{xxx}+\frac{1}{2} t q_{y} v_{xxx}-t v q_{xxxy}
	\nonumber \\
	&
	+\frac{1}{6}
	t v_{xxxxy}+\frac{1}{4} t v_{txy}-\frac{3}{4} t v_{txx}; \nonumber
\end{align}
\begin{align}
	T_6^t= \, & \frac{1}{12} t q_{xx} v_{yz}+\frac{v_{yz}}{72}+\frac{v_{xz}}{24}+\frac{1}{12} t v_{xz}
	q_{xy}+\frac{1}{12} t q_{xz}
	v_{xy}+\frac{v_{xy}}{24}-\frac{1}{12} t v_{x}
	q_{xyz}-\frac{1}{24} x v_{xyz}-\frac{3}{8} y v_{xyz}-\frac{3}{8}
	z v_{xyz}
	\nonumber \\
	&
	-\frac{1}{4} t q_{x}
	v_{xyz}-\frac{1}{2} t v_{xz}
	q_{xx}-\frac{1}{2} t v_{xy}
	q_{xx}-\frac{1}{4} t q_{xz}
	v_{xx}-\frac{1}{4} t q_{xy}
	v_{xx}-\frac{3 v_{xx}}{4}-\frac{1}{12} t v_{y} q_{xxz}+\frac{1}{2} t v_{x} q_{xxz}+\frac{1}{8} x v_{xxz}
	\nonumber \\
	&
	+\frac{9}{8}
	y v_{xxz}+\frac{9}{8} z v_{xxz}+\frac{3}{4} t q_{x} v_{xxz}-\frac{1}{12} t v_{z} q_{xxy}+\frac{1}{2} t v_{x} q_{xxy}+\frac{1}{8} x v_{xxy}+\frac{9}{8}
	y v_{xxy}+\frac{9}{8} z v_{xxy}+\frac{3}{4} t q_{x} v_{xxy}
	\nonumber \\
	&
	+\frac{1}{4} t v q_{xxyz}+\frac{1}{4} t v_{z} q_{xxx}+\frac{1}{4} t v_{y} q_{xxx}-\frac{3}{4} t v q_{xxxz}-\frac{3}{4} t v q_{xxxy}, \nonumber
	\end{align}
\begin{align}
	T_6^x= \, &
	3 t v_{xyz} q_{x}{}^2-\frac{1}{12} v_{yz} q_{x}-2 v_{xz} q_{x}-\frac{11}{3}
	t v_{xz} q_{xy} q_{x}-\frac{11}{3} t q_{xz} v_{xy} q_{x}-2 v_{xy} q_{x}+\frac{8}{3} t v_{x} q_{xyz}
	q_{x}+\frac{1}{2} x v_{xyz}
	q_{x}
	\nonumber \\
	&
	+\frac{9}{2} y v_{xyz}
	q_{x}+\frac{9}{2} z v_{xyz}
	q_{x}+\frac{4}{3} t v_{yz}
	q_{xx} q_{x}+t q_{yz}
	v_{xx} q_{x}+\frac{7}{6} t
	v_{y} q_{xxz} q_{x}-\frac{3}{2} t q_{y} v_{xxz}
	q_{x}+\frac{7}{6} t v_{z}
	q_{xxy} q_{x}
	\nonumber \\
	&
	-\frac{3}{2} t
	q_{z} v_{xxy} q_{x}+t
	v q_{xxyz} q_{x}-\frac{2}{3} t v_{xxxyz} q_{x}-\frac{1}{4} t v_{tyz} q_{x}+\frac{3}{2} t v_{txz} q_{x}+\frac{3}{2} t v_{txy} q_{x}-\frac{1}{3} q_{yz} v_{x}+\frac{5}{36} v_{y} q_{xz}
	\nonumber \\
	&
	+\frac{5}{2} v_{x} q_{xz}+\frac{1}{6} q_{y} v_{xz}+\frac{5}{36} v_{z} q_{xy}+\frac{5}{2} v_{x} q_{xy}+\frac{8}{3} t v_{x} q_{xz}
	q_{xy}-\frac{4}{9} x v_{xz}
	q_{xy}-4 y v_{xz} q_{xy}-4 z v_{xz} q_{xy}+\frac{1}{6} q_{z} v_{xy}
	\nonumber \\
	&
	-\frac{4}{9} x q_{xz} v_{xy}-4 y q_{xz} v_{xy}-4 z
	q_{xz} v_{xy}-\frac{5}{36}
	v q_{xyz}+\frac{5}{18} x
	v_{x} q_{xyz}+\frac{5}{2} y
	v_{x} q_{xyz}+\frac{5}{2} z
	v_{x} q_{xyz}-\frac{1}{2}
	v_{z} q_{xx}
	-\frac{1}{2}
	v_{y} q_{xx}
	\nonumber \\
	&
	+\frac{7}{18} x
	v_{yz} q_{xx}+\frac{7}{2} y
	v_{yz} q_{xx}+\frac{7}{2} z
	v_{yz} q_{xx}-2 t q_{yz} v_{x} q_{xx}+\frac{2}{3} t
	v_{y} q_{xz} q_{xx}+t
	q_{y} v_{xz} q_{xx}+\frac{2}{3} t v_{z} q_{xy}
	q_{xx}+t q_{z} v_{xy}
	q_{xx}
	\nonumber \\
	&
	+\frac{3}{4} q_{z}
	v_{xx}+\frac{3}{4} q_{y}
	v_{xx}+\frac{1}{6} x q_{yz}
	v_{xx}+\frac{3}{2} y q_{yz}
	v_{xx}+\frac{3}{2} z q_{yz}
	v_{xx}+\frac{1}{2} t q_{y}
	q_{xz} v_{xx}+\frac{1}{2} t
	q_{z} q_{xy} v_{xx}-\frac{11}{4} v q_{xxz}
	\nonumber \\
	&
	+\frac{1}{36} x v_{y} q_{xxz}+\frac{1}{4} y v_{y} q_{xxz}+\frac{1}{4} z v_{y} q_{xxz}-t q_{y} v_{x} q_{xxz}+\frac{1}{2} t v q_{xy}
	q_{xxz}-\frac{1}{4} x q_{y}
	v_{xxz}-\frac{9}{4} y q_{y}
	v_{xxz}-\frac{9}{4} z q_{y}
	v_{xxz}
	\nonumber \\
	&
	-\frac{11}{4} v
	q_{xxy}
	+\frac{1}{36} x v_{z}
	q_{xxy}+\frac{1}{4} y v_{z}
	q_{xxy}+\frac{1}{4} z v_{z}
	q_{xxy}-t q_{z} v_{x}
	q_{xxy}+\frac{1}{2} t v
	q_{xz} q_{xxy}-\frac{1}{5} t
	v_{xxz} q_{xxy}-\frac{1}{4} x
	q_{z} v_{xxy}
	\nonumber \\
	&
	-\frac{9}{4} y
	q_{z} v_{xxy}-\frac{9}{4} z
	q_{z} v_{xxy}-\frac{1}{5} t
	q_{xxz} v_{xxy}-\frac{1}{3} x
	v q_{xxyz}-3 y v
	q_{xxyz}-3 z v q_{xxyz}+\frac{1}{5} t v_{xx} q_{xxyz}+\frac{2}{5} t q_{xx} v_{xxyz}
	\nonumber \\
	&
	+\frac{1}{15} v_{xxyz}-\frac{1}{2}
	t v_{z} q_{y} q_{xxx}-\frac{1}{2} t q_{z} v_{y}
	q_{xxx}-t v q_{yz}
	q_{xxx}-\frac{1}{5} t v_{xyz}
	q_{xxx}-\frac{1}{15} t q_{xyz}
	v_{xxx}-\frac{1}{2} t v
	q_{y} q_{xxxz}
	\nonumber \\
	&
	+\frac{1}{5} t
	v_{xy} q_{xxxz}+\frac{2}{15} t
	q_{xy} v_{xxxz}+\frac{1}{5}
	v_{xxxz}-\frac{1}{2} t v
	q_{z} q_{xxxy}+\frac{1}{5} t
	v_{xz} q_{xxxy}+\frac{2}{15} t
	q_{xz} v_{xxxy}+\frac{1}{5}
	v_{xxxy}-\frac{2}{5} t v_{x}
	q_{xxxyz}
	\nonumber \\
	&
	-\frac{1}{9} x v_{xxxyz}-y v_{xxxyz}-z v_{xxxyz}+\frac{1}{15} t v_{yz} q_{xxxx}-\frac{2}{15} t v_{y} q_{xxxxz}-\frac{2}{15} t v_{z} q_{xxxxy}-\frac{1}{3} t v q_{xxxxyz}-\frac{1}{12} t q_{xyz}
	v_{t}
	\nonumber \\
	&
	+\frac{1}{2} t q_{xxz}
	v_{t}+\frac{1}{2} t q_{xxy}
	v_{t}+\frac{1}{12} t q_{xy}
	v_{tz}-\frac{1}{2} t q_{xx}
	v_{tz}+\frac{v_{tz}}{24}+\frac{1}{12} t q_{xz}
	v_{ty}-\frac{1}{2} t q_{xx}
	v_{ty}+\frac{v_{ty}}{24}-\frac{1}{24} x v_{tyz}-\frac{3}{8} y v_{tyz}
	\nonumber \\
	&
	-\frac{3}{8}
	z v_{tyz}+\frac{1}{12} t v_{yz}
	q_{tx}-\frac{1}{2} t v_{xz}
	q_{tx}-\frac{1}{2} t v_{xy}
	q_{tx}-\frac{1}{2} t q_{xz}
	v_{tx}-\frac{1}{2} t q_{xy}
	v_{tx}-\frac{3 v_{tx}}{2}-\frac{1}{12} t v_{y} q_{txz}+\frac{1}{2} t v_{x} q_{txz}
	\nonumber \\
	&
	+\frac{1}{4} x v_{txz}+\frac{9}{4}
	y v_{txz}+\frac{9}{4} z v_{txz}-\frac{1}{12} t v_{z} q_{txy}+\frac{1}{2} t v_{x} q_{txy}+\frac{1}{4} x v_{txy}+\frac{9}{4}
	y v_{txy}+\frac{9}{4} z v_{txy}-\frac{3}{4} t v q_{txyz}+\frac{1}{2} t v_{z} q_{txx}
	\nonumber \\
	&
	+\frac{1}{2} t v_{y} q_{txx}+\frac{3}{2} t v q_{txxz}+\frac{3}{2} t v q_{txxy}, 
	\nonumber
\end{align}
\begin{align}
	T_6^y= \, &
	\frac{3}{2} t v_{xxz} q_{x}{}^2-\frac{1}{12} v_{xz} q_{x}+\frac{4}{3} t v_{xz} q_{xx}
	q_{x}-\frac{11}{6} t q_{xz}
	v_{xx} q_{x}-v_{xx}
	q_{x}+\frac{7}{6} t v_{x}
	q_{xxz} q_{x}+\frac{1}{4} x
	v_{xxz} q_{x}+\frac{9}{4} y
	v_{xxz} q_{x}
	\nonumber \\
	&
	+\frac{9}{4} z
	v_{xxz} q_{x}-\frac{2}{3} t
	v_{z} q_{xxx} q_{x}-\frac{1}{2} t q_{z} v_{xxx}
	q_{x}-\frac{1}{2} t v
	q_{xxxz} q_{x}-\frac{1}{6} t
	v_{xxxxz} q_{x}-\frac{1}{4} t
	v_{txz} q_{x}+\frac{3}{4} t
	v_{txx} q_{x}-\frac{2}{3} t
	v_{z} q_{xx}^2
	\nonumber \\
	&
	+\frac{5}{36}
	v_{x} q_{xz}-\frac{7}{36}
	v_{z} q_{xx}-\frac{1}{2}
	v_{x} q_{xx}+\frac{2}{3} t
	v_{x} q_{xz} q_{xx}+\frac{7}{18} x v_{xz} q_{xx}+\frac{7}{2} y v_{xz} q_{xx}+\frac{7}{2} z v_{xz} q_{xx}+\frac{1}{12} q_{z} v_{xx}
	\nonumber \\
	&
	-\frac{2}{9} x q_{xz} v_{xx}-2 y q_{xz} v_{xx}-2 z
	q_{xz} v_{xx}+\frac{1}{2} t
	q_{z} q_{xx} v_{xx}+\frac{1}{9} v q_{xxz}+\frac{1}{36} x v_{x} q_{xxz}+\frac{1}{4} y v_{x} q_{xxz}+\frac{1}{4} z v_{x} q_{xxz}
	\nonumber \\
	&
	-\frac{1}{2} t v q_{xx}
	q_{xxz}+2 v q_{xxx}-\frac{7}{36} x v_{z} q_{xxx}-\frac{7}{4} y v_{z} q_{xxx}-\frac{7}{4} z v_{z} q_{xxx}-\frac{1}{2} t q_{z} v_{x}
	q_{xxx}+\frac{1}{2} t v
	q_{xz} q_{xxx}-\frac{1}{10} t
	v_{xxz} q_{xxx}
	\nonumber \\
	&
	-\frac{1}{12} x
	q_{z} v_{xxx}-\frac{3}{4} y
	q_{z} v_{xxx}-\frac{3}{4} z
	q_{z} v_{xxx}-\frac{1}{15} t
	q_{xxz} v_{xxx}+\frac{1}{6} x
	v q_{xxxz}+\frac{3}{2} y
	v q_{xxxz}+\frac{3}{2} z
	v q_{xxxz}+\frac{1}{10} t
	v_{xx} q_{xxxz}
	\nonumber \\
	&
	+\frac{2}{15} t
	q_{xx} v_{xxxz}+\frac{1}{45}
	v_{xxxz}+\frac{1}{2} t v
	q_{z} q_{xxxx}+\frac{1}{15} t
	v_{xz} q_{xxxx}+\frac{1}{30} t
	q_{xz} v_{xxxx}+\frac{1}{20}
	v_{xxxx}-\frac{2}{15} t v_{x}
	q_{xxxxz}
	\nonumber \\
	&
	-\frac{1}{36} x v_{xxxxz}-\frac{1}{4} y v_{xxxxz}-\frac{1}{4} z v_{xxxxz}-\frac{1}{30} t v_{z} q_{xxxxx}+\frac{1}{6} t v q_{xxxxxz}-\frac{1}{12} t q_{xxz}
	v_{t}+\frac{1}{4} t q_{xxx}
	v_{t}+\frac{1}{12} t q_{xx}
	v_{tz}
		\nonumber \\
	&
	+\frac{1}{72}v_{tz}+\frac{1}{12} t v_{xz}
	q_{tx}-\frac{1}{4} t v_{xx}
	q_{tx}+\frac{1}{12} t q_{xz}
	v_{tx}-\frac{1}{2} t q_{xx}
	v_{tx}+\frac{v_{tx}}{24}-\frac{1}{12} t v_{x}
	q_{txz}-\frac{1}{24} x v_{txz}-\frac{3}{8} y v_{txz}-\frac{3}{8}
	z v_{txz}
		\nonumber \\
	&
	-\frac{1}{12} t v_{z}
	q_{txx}+\frac{1}{2} t v_{x}
	q_{txx}+\frac{1}{8} x v_{txx}+\frac{9}{8} y v_{txx}+\frac{9}{8}
	z v_{txx}+\frac{1}{4} t v
	q_{txxz}-\frac{3}{4} t v
	q_{txxx},  \nonumber
\end{align}
\begin{align}
	T_6^z=  \, &
	\frac{3}{2} t v_{xxy}
	q_{x}{}^2-\frac{1}{12} v_{xy}
	q_{x}+\frac{4}{3} t v_{xy}
	q_{xx} q_{x}-\frac{11}{6} t
	q_{xy} v_{xx} q_{x}-v_{xx} q_{x}+\frac{7}{6} t
	v_{x} q_{xxy} q_{x}+\frac{1}{4} x v_{xxy} q_{x}+\frac{9}{4} y v_{xxy} q_{x}
\nonumber \\
	&
	+\frac{9}{4} z v_{xxy} q_{x}-\frac{2}{3} t v_{y} q_{xxx}
	q_{x}-\frac{1}{2} t q_{y}
	v_{xxx} q_{x}-\frac{1}{2} t
	v q_{xxxy} q_{x}-\frac{1}{6} t v_{xxxxy} q_{x}-\frac{1}{4} t v_{txy} q_{x}+\frac{3}{4} t v_{txx} q_{x}-\frac{2}{3} t v_{y} q_{xx}^2
			\nonumber \\
	&
	+\frac{5}{36} v_{x} q_{xy}-\frac{7}{36} v_{y} q_{xx}-\frac{1}{2} v_{x} q_{xx}+\frac{2}{3} t v_{x} q_{xy}
	q_{xx}+\frac{7}{18} x v_{xy}
	q_{xx}+\frac{7}{2} y v_{xy}
	q_{xx}+\frac{7}{2} z v_{xy}
	q_{xx}+\frac{1}{12} q_{y}
	v_{xx}
			\nonumber \\
	&
	-\frac{2}{9} x q_{xy}
	v_{xx}-2 y q_{xy} v_{xx}-2 z q_{xy} v_{xx}+\frac{1}{2} t q_{y} q_{xx}
	v_{xx}+\frac{1}{9} v q_{xxy}+\frac{1}{36} x v_{x} q_{xxy}+\frac{1}{4} y v_{x} q_{xxy}+\frac{1}{4} z v_{x} q_{xxy}
			\nonumber \\
	&
	-\frac{1}{2} t v q_{xx}
	q_{xxy}+2 v q_{xxx}-\frac{7}{36} x v_{y} q_{xxx}-\frac{7}{4} y v_{y} q_{xxx}-\frac{7}{4} z v_{y} q_{xxx}-\frac{1}{2} t q_{y} v_{x}
	q_{xxx}+\frac{1}{2} t v
	q_{xy} q_{xxx}-\frac{1}{10} t
	v_{xxy} q_{xxx}
			\nonumber \\
	&
	-\frac{1}{12} x
	q_{y} v_{xxx}-\frac{3}{4} y
	q_{y} v_{xxx}-\frac{3}{4} z
	q_{y} v_{xxx}-\frac{1}{15} t
	q_{xxy} v_{xxx}+\frac{1}{6} x
	v q_{xxxy}+\frac{3}{2} y
	v q_{xxxy}+\frac{3}{2} z
	v q_{xxxy}+\frac{1}{10} t
	v_{xx} q_{xxxy}
			\nonumber \\
	&
	+\frac{2}{15} t
	q_{xx} v_{xxxy}+\frac{1}{45}
	v_{xxxy}+\frac{1}{2} t v
	q_{y} q_{xxxx}+\frac{1}{15} t
	v_{xy} q_{xxxx}+\frac{1}{30} t
	q_{xy} v_{xxxx}+\frac{1}{20}
	v_{xxxx}-\frac{2}{15} t v_{x}
	q_{xxxxy}
			\nonumber \\
	&
	-\frac{1}{36} x v_{xxxxy}-\frac{1}{4} y v_{xxxxy}-\frac{1}{4} z v_{xxxxy}-\frac{1}{30} t v_{y} q_{xxxxx}+\frac{1}{6} t v q_{xxxxxy}-\frac{1}{12} t q_{xxy}
	v_{t}+\frac{1}{4} t q_{xxx}
	v_{t}+\frac{1}{12} t q_{xx}
	v_{ty}
			\nonumber \\
	&
	+\frac{1}{72}v_{ty}+\frac{1}{12} t v_{xy}
	q_{tx}-\frac{1}{4} t v_{xx}
	q_{tx}+\frac{1}{12} t q_{xy}
	v_{tx}-\frac{1}{2} t q_{xx}
	v_{tx}+\frac{v_{tx}}{24}-\frac{1}{12} t v_{x}
	q_{txy}-\frac{1}{24} x v_{txy}-\frac{3}{8} y v_{txy}-\frac{3}{8}
	z v_{txy}
			\nonumber \\
	&
	-\frac{1}{12} t v_{y}
	q_{txx}+\frac{1}{2} t v_{x}
	q_{txx}+\frac{1}{8} x v_{txx}+\frac{9}{8} y v_{txx}+\frac{9}{8}
	z v_{txx}+\frac{1}{4} t v
	q_{txxy}-\frac{3}{4} t v
	q_{txxx}; \nonumber 
\end{align}
\begin{align}
	T_7^t=&\frac{1}{6} v_{yz} q_{x}
	-v_{xz} q_{x}-v_{xy} q_{x}
	-\frac{1}{4} x v_{xyz} q_{x}
	+\frac{3}{4} x v_{xxz} q_{x}
	+\frac{3}{4} x v_{xxy} q_{x}
	+18 t v q_{xxyz} q_{x}
	-\frac{1}{4} q_{yz}
	v_{x}-\frac{1}{12} z q_{yzz}
	v_{x}  -\frac{1}{4} v_{y}
	q_{xz}\nonumber
	\\
	&
	-\frac{1}{12} y q_{yyz}
	v_{x}+\frac{1}{12} z v_{yz}
	q_{xz}+\frac{3}{2} v_{x}
	q_{xz}+\frac{1}{6} q_{y}
	v_{xz}+\frac{1}{12} z q_{yz}
	v_{xz}+\frac{1}{12} y q_{yy}-\frac{1}{2} z q_{xz}
	v_{xz}
	-\frac{1}{12} z v_{y}
	q_{xzz}
	v_{xz}   	+v q_{xyz}                     \nonumber
	\\
	&
	+\frac{1}{2} z v_{x}
	q_{xzz}
	-\frac{1}{4} v_{z}
	q_{xy}
	+\frac{1}{12} y v_{yz}
	q_{xy}
	+\frac{3}{2} v_{x}
	q_{xy}
	+\frac{1}{12} x v_{xz}
	q_{xy}-\frac{1}{2} y v_{xz}
	q_{xy}        +\frac{1}{6} q_{z}
	v_{xy}+\frac{1}{12} z q_{zz}
	v_{xy}   -\frac{1}{2} y q_{xy}
	v_{xy}               \nonumber \\      
	&+\frac{1}{12} y q_{yz}
	v_{xy}+\frac{1}{12} x q_{xz}
	v_{xy}-\frac{1}{2} z q_{xz}
	v_{xy}         
	-\frac{1}{12} z v_{z} q_{xyz}
	-\frac{1}{12} y v_{y} q_{xyz}
	-\frac{1}{12} x v_{x} q_{xyz}
	+\frac{1}{2} y v_{x} q_{xyz}
	+\frac{1}{2} z v_{x} q_{xyz} \nonumber \\
	&
	-\frac{1}{4} q v_{xyz}
	-\frac{1}{4} z q_{z} v_{xyz}
	-\frac{1}{4} y q_{y} v_{xyz}
	+\frac{1}{4} z v q_{xyzz}
	-\frac{1}{12} y v_{z} q_{xyy}
	+\frac{1}{2} y v_{x} q_{xyy}
	+\frac{1}{4} y v q_{xyyz}
	+\frac{3}{4} v_{z} q_{xx}
	+\frac{3}{4} v_{y} q_{xx}
	 \nonumber \\
	&
	+\frac{1}{12} x v_{yz} q_{xx}
	-\frac{1}{2} x v_{xz} q_{xx}
	-\frac{1}{2} x v_{xy} q_{xx}
	+6 t v q_{xyz} q_{xx}
	-\frac{1}{2} q_{z} v_{xx}
	-\frac{1}{4} z q_{zz} v_{xx}
	-\frac{1}{2} q_{y} v_{xx}
	-\frac{1}{4} y q_{yz} v_{xx}
	-\frac{1}{4} z q_{yz} v_{xx}
	 \nonumber \\
	&
	-\frac{1}{4} y q_{yy} v_{xx}
	-\frac{1}{4} x q_{xz} v_{xx}
	-\frac{1}{4} x q_{xy} v_{xx}
	-3 v q_{xxz}
	+\frac{1}{4} z	v_{z} q_{xxz}
	-\frac{1}{12} x	v_{y} q_{xxz}
	+\frac{1}{4} z	v_{y} q_{xxz}
	+\frac{1}{2} x	v_{x} q_{xxz}
		+\frac{3}{4}q v_{xxz}
	 \nonumber \\
	&
	+12 t v	q_{xy} q_{xxz}
	+\frac{3}{4} z	q_{z} v_{xxz}
	+\frac{3}{4} y	q_{y} v_{xxz}
	-\frac{3}{4} z	v q_{xxzz}
	-3 v	q_{xxy}
	-\frac{1}{12} x v_{z}	q_{xxy}
	+\frac{1}{4} y v_{z}	q_{xxy}
	+\frac{1}{4} y v_{y}	q_{xxy}
	+\frac{1}{2} x v_{x}	q_{xxy} \nonumber \\
	&
	+12 t v q_{xz}	q_{xxy}
	+\frac{3}{4} q	v_{xxy}
	+\frac{3}{4} z q_{z}	v_{xxy}
	+\frac{3}{4} y q_{y}	v_{xxy}
	+\frac{1}{4} x v	q_{xxyz}
	-\frac{3}{4} y v	q_{xxyz}
	-\frac{3}{4} z v	q_{xxyz}
	-\frac{3}{4} y v	q_{xxyy}
	+\frac{1}{4} x v_{z}	q_{xxx} \nonumber \\
	&
	+\frac{1}{4} x v_{y}	q_{xxx}
	-12 t v q_{yz}	q_{xxx}
	-\frac{3}{4} x v	q_{xxxz}
	-6 t v q_{y}	q_{xxxz}
	-\frac{3}{4} x v	q_{xxxy}
	-6 t v q_{z}
	q_{xxxy}-3 t v q_{xxxxyz}-\frac{3}{4} t v_{xyz}
	q_{t} \nonumber \\
	&+\frac{9}{4} t v_{xxz}
	q_{t}+\frac{9}{4} t v_{xxy}
	q_{t}+\frac{1}{4} t v_{xy}
	q_{tz}-\frac{3}{4} t v_{xx}
	q_{tz}+\frac{1}{4} t v_{xz}
	q_{ty}-\frac{3}{4} t v_{xx}
	q_{ty}-\frac{1}{4} t v_{x}
	q_{tyz}+\frac{1}{4} t v_{yz}
	q_{tx}-\frac{3}{2} t v_{xz}
	q_{tx} \nonumber \\
	&-\frac{3}{2} t v_{xy}
	q_{tx}-\frac{1}{4} t v_{y}
	q_{txz}+\frac{3}{2} t v_{x}
	q_{txz}-\frac{1}{4} t v_{z}
	q_{txy}+\frac{3}{2} t v_{x}
	q_{txy}-\frac{9}{4} t v
	q_{txyz}+\frac{3}{4} t v_{z}
	q_{txx}+\frac{3}{4} t v_{y}
	q_{txx}+\frac{27}{4} t v
	q_{txxz} \nonumber \\
	&+\frac{27}{4} t v
	q_{txxy},\nonumber
\end{align}
\begin{align}
T_7^x=\, &-\frac{3}{2} z v_{xxy}
q_{z}^2-\frac{1}{3} v_{x}
q_{xy} q_{z}-\frac{8}{3} z
v_{xz} q_{xy} q_{z}-\frac{5}{3} z q_{xz} v_{xy}
q_{z}+y q_{xy} v_{xy}
q_{z}+\frac{2}{3} z v_{x}
q_{xyz} q_{z}+3 z q_{x} v_{xyz} q_{z}-v_{tx} q_{z}
	\nonumber \\
&
-y v_{x} q_{xyy} q_{z}-\frac{17}{6}
v_{y} q_{xx} q_{z}+\frac{7}{3} z v_{yz} q_{xx}
q_{z}+x v_{xy} q_{xx}
q_{z}+2 q_{y} v_{xx}
q_{z}+\frac{3}{2} z q_{yz}
v_{xx} q_{z}+\frac{1}{2} y
q_{yy} v_{xx} q_{z}+\frac{1}{6} v_{ty} q_{z}
	\nonumber \\
&
+\frac{1}{2} x q_{xy} v_{xx}
q_{z}-\frac{1}{3} z v_{y}
q_{xxz} q_{z}-\frac{3}{2} z
q_{y} v_{xxz} q_{z}+\frac{13}{3} v q_{xxy}
q_{z}+\frac{1}{6} z v_{z}
q_{xxy} q_{z}-\frac{1}{2} y
v_{y} q_{xxy} q_{z}-x
v_{x} q_{xxy} q_{z}-2
v_{yz} q_{x}^2
	\nonumber \\
&
-\frac{3}{2} q v_{xxy}
q_{z}
-\frac{3}{2} y q_{y}
v_{xxy} q_{z}-\frac{3}{2} x
q_{x} v_{xxy} q_{z}-\frac{1}{2} z v q_{xxyz}
q_{z}+\frac{3}{2} y v
q_{xxyy} q_{z}-\frac{1}{2} x
v_{y} q_{xxx} q_{z}-\frac{1}{2} x v q_{xxxy}
q_{z}
	\nonumber \\
&
+\frac{4}{15} v_{xxxy}
q_{z}-\frac{2}{3} z v_{xxxyz}
q_{z}-\frac{9}{2} t v_{xxy}
q_{t} q_{z}+\frac{3}{2} t
v_{xx} q_{ty} q_{z}-\frac{1}{4} z v_{tyz} q_{z}+3 t v_{xy} q_{tx}
q_{z}
+\frac{3}{2} z v_{txz} q_{z}-3 t v_{x} q_{txy}
q_{z}	\nonumber \\
&+\frac{3}{2} z v_{txy}
q_{z}-\frac{3}{2} t v_{y}
q_{txx} q_{z}+\frac{9}{2} t
v q_{txxy} q_{z}+\frac{4}{3} z
v_{y} q_{xz}^2+\frac{4}{3} y
v_{z} q_{xy}^2-q_{yz} q_{x} v_{x}+z q_{yzz} q_{x} v_{x}+y q_{yyz} q_{x} v_{x}+\frac{17}{3}
v_{y} q_{x} q_{xz}
\nonumber \\
&
-z
v_{yz} q_{x} q_{xz}-\frac{1}{3} q_{y} v_{x}
q_{xz}-\frac{2}{3} z q_{yz}
v_{x} q_{xz}+\frac{4}{3} y
q_{yy} v_{x} q_{xz}-z
q_{yz} q_{x} v_{xz}-y
q_{yy} q_{x} v_{xz}+z
q_{y} q_{xz} v_{xz}+z
v_{y} q_{x} q_{xzz}\nonumber \\
&-z
q_{y} v_{x} q_{xzz}+\frac{17}{3} v_{z} q_{x}
q_{xy}-y v_{yz} q_{x}
q_{xy}+\frac{4}{3} z q_{zz}
v_{x} q_{xy}-\frac{2}{3} y
q_{yz} v_{x} q_{xy}-16
v q_{xz} q_{xy}+\frac{4}{3} z v_{z} q_{xz}
q_{xy}+\frac{4}{3} y v_{y}
q_{xz} q_{xy}\nonumber \\
&+\frac{8}{3} x
v_{x} q_{xz} q_{xy}-\frac{8}{3} q v_{xz}
q_{xy}-\frac{5}{3} y q_{y}
v_{xz} q_{xy}-\frac{11}{3} x
q_{x} v_{xz} q_{xy}-\frac{8}{3} z v q_{xzz}
q_{xy}-z q_{zz} q_{x}
v_{xy}-y q_{yz} q_{x}
v_{xy}-\frac{8}{3} q q_{xz} v_{xy}\nonumber \\
&-\frac{8}{3} y q_{y}
q_{xz} v_{xy}-\frac{11}{3} x
q_{x} q_{xz} v_{xy}-\frac{50}{3} v q_{x}
q_{xyz}+z v_{z} q_{x}
q_{xyz}+y v_{y} q_{x}
q_{xyz}+\frac{5}{3} q
v_{x} q_{xyz}+\frac{2}{3} y
q_{y} v_{x} q_{xyz}\nonumber \\
&+\frac{8}{3} x q_{x} v_{x}
q_{xyz}-5 z v q_{xz}
q_{xyz}-5 y v q_{xy}
q_{xyz}+3 x q_{x}^2
v_{xyz}+3 q q_{x}
v_{xyz}+3 y q_{y} q_{x} v_{xyz}-3 z v q_{x} q_{xyzz}+y v_{z} q_{x} q_{xyy}\nonumber \\
&-\frac{8}{3} y v
q_{xz} q_{xyy}-3 y v
q_{x} q_{xyyz}-\frac{17}{6}
v_{z} q_{y} q_{xx}-\frac{2}{3} z q_{zz} v_{y}
q_{xx}+10 v q_{yz}
q_{xx}-\frac{2}{3} z v_{z}
q_{yz} q_{xx}-\frac{2}{3} y
v_{y} q_{yz} q_{xx}\nonumber \\
&+\frac{7}{3} q v_{yz}
q_{xx}+\frac{7}{3} y q_{y}
v_{yz} q_{xx}+\frac{1}{3} z
v q_{yzz} q_{xx}-\frac{2}{3} y v_{z} q_{yy}
q_{xx}+\frac{1}{3} y v
q_{yyz} q_{xx}+\frac{4}{3} x
v_{yz} q_{x} q_{xx}-2
x q_{yz} v_{x} q_{xx}\nonumber \\
&+\frac{2}{3} x v_{y} q_{xz}
q_{xx}+x q_{y} v_{xz}
q_{xx}+\frac{2}{3} x v_{z}
q_{xy} q_{xx}-\frac{3}{5}
v_{xyz} q_{xx}+\frac{1}{2} z
q_{zz} q_{y} v_{xx}+q q_{yz} v_{xx}+\frac{3}{2} y q_{y} q_{yz}
v_{xx}+x q_{yz} q_{x}
v_{xx}\nonumber \\
&+\frac{1}{2} x q_{y}
q_{xz} v_{xx}+\frac{4}{5}
q_{xyz} v_{xx}+\frac{1}{5} z
q_{xyzz} v_{xx}+\frac{1}{5} y
q_{xyyz} v_{xx}+\frac{13}{3}
v q_{y} q_{xxz}-\frac{1}{2} z v_{z} q_{y}
q_{xxz}+\frac{1}{6} q
v_{y} q_{xxz}\nonumber \\
&+\frac{1}{6} y
q_{y} v_{y} q_{xxz}+\frac{13}{6} z v q_{yz}
q_{xxz}-\frac{5}{6} y v
q_{yy} q_{xxz}+\frac{7}{6} x
v_{y} q_{x} q_{xxz}-x
q_{y} v_{x} q_{xxz}+\frac{1}{2} x v q_{xy}
q_{xxz}+\frac{4}{5} v_{xy}
q_{xxz}\nonumber \\
&-\frac{1}{5} z v_{xyz}
q_{xxz}-\frac{3}{2} y q_{y}^2
v_{xxz}-\frac{3}{2} q
q_{y} v_{xxz}-\frac{3}{2} x
q_{y} q_{x} v_{xxz}-\frac{3}{5} q_{xy} v_{xxz}-\frac{1}{5} z q_{xyz} v_{xxz}-\frac{1}{5} y q_{xyy} v_{xxz}+\frac{3}{2} z v q_{y}
q_{xxzz}\nonumber \\
&+\frac{1}{5} z v_{xy}
q_{xxzz}+\frac{1}{6} q
v_{z} q_{xxy}-\frac{5}{6} z
v q_{zz} q_{xxy}-\frac{1}{3} y v_{z} q_{y}
q_{xxy}+\frac{13}{6} y v
q_{yz} q_{xxy}+\frac{7}{6} x
v_{z} q_{x} q_{xxy}+\frac{1}{2} x v q_{xz}
q_{xxy}\nonumber \\
&+\frac{4}{5} v_{xz}
q_{xxy}-\frac{1}{5} y v_{xyz}
q_{xxy}-\frac{1}{5} x v_{xxz}
q_{xxy}-\frac{3}{5} q_{xz}
v_{xxy}-\frac{1}{5} z q_{xzz}
v_{xxy}-\frac{1}{5} y q_{xyz}
v_{xxy}-\frac{1}{5} x q_{xxz}
v_{xxy}-2 q v
q_{xxyz}
\nonumber \\
&-\frac{1}{2} y v
q_{y} q_{xxyz}+x v
q_{x} q_{xxyz}-2 v_{x}
q_{xxyz}+\frac{1}{5} z v_{xz}
q_{xxyz}+\frac{1}{5} y v_{xy}
q_{xxyz}+\frac{1}{5} x v_{xx}
q_{xxyz}+\frac{4}{5} q_{x}
v_{xxyz}+\frac{2}{5} z q_{xz}
v_{xxyz}
\nonumber \\
&+\frac{2}{5} y q_{xy}
v_{xxyz}+\frac{2}{5} x q_{xx}
v_{xxyz}-\frac{2}{5} z v_{x}
q_{xxyzz}+\frac{1}{5} y v_{xz}
q_{xxyy}-\frac{2}{5} y v_{x}
q_{xxyyz}-\frac{1}{2} x v_{z}
q_{y} q_{xxx}-x v
q_{yz} q_{xxx}\nonumber \\
&+\frac{4}{15}
v_{yz} q_{xxx}-\frac{1}{5} x
v_{xyz} q_{xxx}-\frac{1}{5}
q_{yz} v_{xxx}-\frac{1}{15} z
q_{yzz} v_{xxx}-\frac{1}{15} y
q_{yyz} v_{xxx}-\frac{1}{15} x
q_{xyz} v_{xxx}-\frac{1}{2} x
v q_{y} q_{xxxz}\nonumber \\
&-\frac{2}{3} v_{y} q_{xxxz}+\frac{1}{15} z v_{yz} q_{xxxz}+\frac{1}{5} x v_{xy} q_{xxxz}+\frac{4}{15} q_{y} v_{xxxz}+\frac{2}{15} z q_{yz} v_{xxxz}+\frac{2}{15} y q_{yy} v_{xxxz}+\frac{2}{15} x q_{xy} v_{xxxz}
\nonumber \\
&
-\frac{2}{15} z v_{y} q_{xxxzz}-\frac{2}{3} v_{z} q_{xxxy}+\frac{1}{15} y v_{yz} q_{xxxy}+\frac{1}{5} x v_{xz} q_{xxxy}+\frac{2}{15} z q_{zz} v_{xxxy}+\frac{2}{15} y q_{yz} v_{xxxy}+\frac{2}{15} x q_{xz} v_{xxxy}
\nonumber \\
&
+4 v q_{xxxyz}-\frac{2}{15} z v_{z} q_{xxxyz}-\frac{2}{15} y v_{y} q_{xxxyz}-\frac{2}{5} x v_{x} q_{xxxyz}-\frac{2}{3} q v_{xxxyz}-\frac{2}{3} y q_{y} v_{xxxyz}-\frac{2}{3} x q_{x} v_{xxxyz}
\nonumber \\
&+\frac{2}{3} z v q_{xxxyzz}-\frac{2}{15} y v_{z} q_{xxxyy}+\frac{2}{3} y v q_{xxxyyz}+\frac{1}{15} x v_{yz}
q_{xxxx}-\frac{2}{15} x v_{y}
q_{xxxxz}-\frac{2}{15} x v_{z}
q_{xxxxy}-\frac{1}{3} x v
q_{xxxxyz}
\nonumber \\
&+\frac{1}{3} v_{yz}
q_{t}-2 v_{xz} q_{t}-8
t v_{xz} q_{xy} q_{t}-8 t q_{xz} v_{xy}
q_{t}-2 v_{xy} q_{t}+5
t v_{x} q_{xyz} q_{t}+9 t q_{x} v_{xyz}
q_{t}+7 t v_{yz} q_{xx} q_{t}+3 t q_{yz}
v_{xx} q_{t}
\nonumber \\
&
+\frac{1}{2} t
v_{y} q_{xxz} q_{t}-\frac{9}{2} t q_{y} v_{xxz}
q_{t}+\frac{1}{2} t v_{z}
q_{xxy} q_{t}-6 t v
q_{xxyz} q_{t}-2 t v_{xxxyz} q_{t}-\frac{1}{4} q_{yz} v_{t}-\frac{1}{12} z q_{yzz} v_{t}-\frac{1}{12} y q_{yyz} v_{t}
\nonumber \\
&
+\frac{3}{2} q_{xz}
v_{t}+\frac{1}{2} z q_{xzz}
v_{t}+\frac{3}{2} q_{xy}
v_{t}-\frac{1}{12} x q_{xyz}
v_{t}+\frac{1}{2} y q_{xyz}
v_{t}+\frac{1}{2} z q_{xyz}
v_{t}+\frac{1}{2} y q_{xyy}
v_{t}+\frac{1}{2} x q_{xxz}
v_{t}+\frac{1}{2} x q_{xxy}
v_{t}\nonumber \\
&-\frac{5}{12} v_{y}
q_{tz}+\frac{1}{12} z v_{yz}
q_{tz}+\frac{5}{2} v_{x}
q_{tz}-\frac{1}{2} z v_{xz}
q_{tz}+4 t v_{x} q_{xy} q_{tz}-\frac{1}{2} z v_{xy} q_{tz}-3 t q_{x}
v_{xy} q_{tz}-2 t v_{y} q_{xx} q_{tz}\nonumber \\
&+\frac{3}{2} t
q_{y} v_{xx} q_{tz}-\frac{5}{2} t v q_{xxy}
q_{tz}+\frac{2}{5} t v_{xxxy}
q_{tz}+\frac{1}{6} q_{y}
v_{tz}+\frac{1}{12} z q_{yz}
v_{tz}+\frac{1}{12} y q_{yy}
v_{tz}-q_{x} v_{tz}-\frac{1}{2} z q_{xz} v_{tz}\nonumber \\
&+\frac{1}{12} x q_{xy} v_{tz}
-\frac{1}{2} y q_{xy} v_{tz}
-\frac{1}{2} x q_{xx} v_{tz}
-\frac{1}{12} z v_{y} q_{tzz}
+\frac{1}{2} z v_{x} q_{tzz}
-\frac{5}{12} v_{z} q_{ty}
+\frac{1}{12} y v_{yz} q_{ty}
+\frac{5}{2} v_{x} q_{ty}
+4 t v_{x} q_{xz} q_{ty}
\nonumber\\
&
-\frac{1}{2} y v_{xz} q_{ty}-3 t q_{x} v_{xz}
q_{ty}-\frac{1}{2} y v_{xy}
q_{ty}-2 t v_{z} q_{xx} q_{ty}-\frac{5}{2} t v
q_{xxz} q_{ty}+\frac{2}{5} t
v_{xxxz} q_{ty}+\frac{1}{4} t
v_{tz} q_{ty}+\frac{1}{12} z
q_{zz} v_{ty}
\nonumber\\
&
+\frac{1}{12} y
q_{yz} v_{ty}-q_{x}
v_{ty}+\frac{1}{12} x q_{xz}
v_{ty}-\frac{1}{2} z q_{xz}
v_{ty}-\frac{1}{2} y q_{xy}
v_{ty}-\frac{1}{2} x q_{xx}
v_{ty}+\frac{1}{4} t q_{tz}
v_{ty}+\frac{3}{2} v q_{tyz}-\frac{1}{12} z v_{z} q_{tyz}
\nonumber\\
&
-\frac{1}{12} y v_{y} q_{tyz}+\frac{1}{2} y v_{x} q_{tyz}+\frac{1}{2} z v_{x} q_{tyz}+3 t q_{x} v_{x}
q_{tyz}+t v q_{xx}
q_{tyz}-\frac{1}{5} t v_{xxx}
q_{tyz}-\frac{1}{4} t v_{t}
q_{tyz}-\frac{1}{4} q
v_{tyz}-\frac{1}{4} y q_{y}
v_{tyz}
\nonumber\\
&
-\frac{1}{4} x q_{x}
v_{tyz}-\frac{3}{4} t q_{t}
v_{tyz}+\frac{1}{4} z v
q_{tyzz}-\frac{1}{12} y v_{z}
q_{tyy}+\frac{1}{2} y v_{x}
q_{tyy}+\frac{1}{4} y v
q_{tyyz}+\frac{5}{2} v_{z}
q_{tx}+\frac{5}{2} v_{y}
q_{tx}+\frac{1}{12} x v_{yz}
q_{tx}
\nonumber\\
&
-3 t v_{yz} q_{x} q_{tx}-6 t q_{yz}
v_{x} q_{tx}+4 t v_{y}
q_{xz} q_{tx}-\frac{1}{2} x
v_{xz} q_{tx}+3 t q_{y} v_{xz} q_{tx}+4 t
v_{z} q_{xy} q_{tx}-\frac{1}{2} x v_{xy} q_{tx}-7 t v q_{xyz} q_{tx}
\nonumber\\
&
+\frac{6}{5} t v_{xxyz} q_{tx}-\frac{3}{2} t v_{tz} q_{tx}-\frac{3}{2} t v_{ty} q_{tx}-\frac{1}{2} z q_{zz} v_{tx}-q_{y} v_{tx}-\frac{1}{2} y
q_{yz} v_{tx}-\frac{1}{2} z
q_{yz} v_{tx}-\frac{1}{2} y
q_{yy} v_{tx}-\frac{1}{2} x
q_{xz} v_{tx}
\nonumber\\
&
-\frac{1}{2} x
q_{xy} v_{tx}-\frac{3}{2} t
q_{tz} v_{tx}-\frac{3}{2} t
q_{ty} v_{tx}-9 v
q_{txz}+\frac{1}{2} z v_{z}
q_{txz}-\frac{1}{12} x v_{y}
q_{txz}+\frac{1}{2} z v_{y}
q_{txz}+3 t v_{y} q_{x} q_{txz}+\frac{1}{2} x v_{x}
q_{txz}
\nonumber\\
&
-3 t q_{y} v_{x} q_{txz}-8 t v q_{xy} q_{txz}-\frac{3}{5} t v_{xxy} q_{txz}+\frac{3}{2} t v_{t} q_{txz}+\frac{3}{2} q
v_{txz}+\frac{3}{2} y q_{y}
v_{txz}+\frac{3}{2} x q_{x}
v_{txz}+\frac{9}{2} t q_{t}
v_{txz}-\frac{3}{2} z v
q_{txzz}
\nonumber\\
&
-9 v q_{txy}-\frac{1}{12} x v_{z} q_{txy}+\frac{1}{2} y v_{z} q_{txy}+\frac{1}{2} y v_{y} q_{txy}+3 t v_{z} q_{x}
q_{txy}+\frac{1}{2} x v_{x}
q_{txy}-8 t v q_{xz}
q_{txy}-\frac{3}{5} t v_{xxz}
q_{txy}
\nonumber\\
&
+\frac{3}{2} t v_{t}
q_{txy}+\frac{3}{2} q
v_{txy}+\frac{3}{2} y q_{y}
v_{txy}+\frac{3}{2} x q_{x}
v_{txy}+\frac{9}{2} t q_{t}
v_{txy}-\frac{3}{4} x v
q_{txyz}-\frac{3}{2} y v
q_{txyz}-\frac{3}{2} z v
q_{txyz}-9 t v q_{x}
q_{txyz}
\nonumber\\
&
+\frac{3}{5} t v_{xx}
q_{txyz}-\frac{3}{2} y v
q_{txyy}+\frac{1}{2} x v_{z}
q_{txx}-\frac{3}{2} t v_{z}
q_{y} q_{txx}+\frac{1}{2} x
v_{y} q_{txx}+9 t v
q_{yz} q_{txx}-\frac{3}{5} t
v_{xyz} q_{txx}+\frac{3}{2} x
v q_{txxz}
\nonumber\\
&
+\frac{9}{2} t
v q_{y} q_{txxz}+\frac{3}{5} t v_{xy} q_{txxz}+\frac{3}{2} x v q_{txxy}+\frac{3}{5} t v_{xz} q_{txxy}-\frac{6}{5} t v_{x} q_{txxyz}+\frac{1}{5} t v_{yz} q_{txxx}-\frac{2}{5} t v_{y} q_{txxxz}-\frac{2}{5} t v_{z} q_{txxxy}
\nonumber\\
&
+2 t v q_{txxxyz}+\frac{1}{4} t v_{yz} q_{tt}-\frac{3}{2} t v_{xz} q_{tt}-\frac{3}{2} t v_{xy} q_{tt}-\frac{1}{4} t v_{y} q_{ttz}+\frac{3}{2} t v_{x} q_{ttz}-\frac{1}{4} t v_{z} q_{tty}+\frac{3}{2} t v_{x} q_{tty}+\frac{3}{4} t v q_{ttyz}
\nonumber\\
&
+\frac{3}{2} t v_{z} q_{ttx}+\frac{3}{2} t v_{y} q_{ttx}-\frac{9}{2} t v q_{ttxz}-\frac{9}{2} t v q_{ttxy},\nonumber
\end{align}
\begin{align}
	T_7^y=\, &-\frac{1}{2} z v_{xxx} q_{z}^2-\frac{17}{6} v_{x} q_{xx} q_{z}+\frac{7}{3} z v_{xz}
	q_{xx} q_{z}-\frac{5}{6} z
	q_{xz} v_{xx} q_{z}+\frac{1}{2} y q_{xy} v_{xx}
	q_{z}+\frac{1}{2} x q_{xx}
	v_{xx} q_{z}-\frac{1}{3} z
	v_{x} q_{xxz} q_{z}
	\nonumber\\
	&
	+\frac{3}{2} z q_{x} v_{xxz}
	q_{z}-\frac{1}{2} y v_{x}
	q_{xxy} q_{z}+\frac{17}{3}
	v q_{xxx} q_{z}-\frac{7}{6} z v_{z} q_{xxx}
	q_{z}-\frac{1}{2} x v_{x}
	q_{xxx} q_{z}-\frac{1}{2}
	q v_{xxx} q_{z}-\frac{1}{2} y q_{y} v_{xxx}
	q_{z}
	\nonumber\\
	&
	-\frac{1}{2} x q_{x}
	v_{xxx} q_{z}+\frac{3}{2} z
	v q_{xxxz} q_{z}-\frac{3}{2} y v q_{xxxy}
	q_{z}+\frac{1}{2} x v
	q_{xxxx} q_{z}+\frac{1}{15}
	v_{xxxx} q_{z}-\frac{1}{6} z
	v_{xxxxz} q_{z}-\frac{3}{2} t
	v_{xxx} q_{t} q_{z}
	\nonumber\\
	&
	+\frac{3}{2} t v_{xx} q_{tx}
	q_{z}+\frac{1}{6} v_{tx}
	q_{z}-\frac{1}{4} z v_{txz}
	q_{z}-\frac{3}{2} t v_{x}
	q_{txx} q_{z}+\frac{3}{4} z
	v_{txx} q_{z}+\frac{3}{2} t
	v q_{txxx} q_{z}+\frac{4}{3} z v_{x} q_{xz}^2-\frac{2}{3} x v_{z} q_{xx}^2
	\nonumber\\
	&
	+\frac{17}{3} q_{x} v_{x} q_{xz}-2 q_{x}^2
	v_{xz}-z q_{x} q_{xz}
	v_{xz}+z q_{x} v_{x}
	q_{xzz}+\frac{4}{3} y v_{x}
	q_{xz} q_{xy}-y q_{x}
	v_{xz} q_{xy}+y q_{x}
	v_{x} q_{xyz}+\frac{1}{6}
	v_{z} q_{x} q_{xx}
	\nonumber\\
	&
	-\frac{2}{3} z q_{zz} v_{x}	q_{xx}
	-\frac{2}{3} y q_{yz}v_{x} q_{xx}
	-3 v q_{xz} q_{xx}
	-\frac{2}{3} z	v_{z} q_{xz} q_{xx}
	+\frac{2}{3} x v_{x} q_{xz}	q_{xx}
	+\frac{7}{3} q v_{xz} q_{xx}
	+\frac{7}{3} y q_{y} v_{xz} q_{xx}
	\nonumber\\
	&
	+\frac{4}{3} x	q_{x} v_{xz} q_{xx}
	+\frac{1}{3} z v q_{xzz} q_{xx}
	-\frac{2}{3} y v_{z} q_{xy} q_{xx}
	+\frac{7}{3} y	v q_{xyz} q_{xx}
	-\frac{1}{2} z q_{zz} q_{x}	v_{xx}
	-\frac{1}{2} y q_{yz} 	q_{x} v_{xx}
	-\frac{4}{3}q q_{xz} v_{xx}
	\nonumber\\
	&
	-\frac{4}{3} y q_{y} q_{xz}
	v_{xx}-\frac{11}{6} x q_{x}
	q_{xz} v_{xx}-\frac{23}{3}
	v q_{x} q_{xxz}+\frac{1}{2} z v_{z} q_{x}
	q_{xxz}+\frac{1}{6} q
	v_{x} q_{xxz}+\frac{1}{6} y
	q_{y} v_{x} q_{xxz}+\frac{7}{6} x q_{x} v_{x}
	q_{xxz}
	\nonumber\\
	&
	-\frac{13}{6} z v
	q_{xz} q_{xxz}+\frac{19}{6} y
	v q_{xy} q_{xxz}-\frac{1}{2} x v q_{xx}
	q_{xxz}+\frac{2}{5} v_{xx}
	q_{xxz}+\frac{3}{2} x q_{x}^2
	v_{xxz}+\frac{3}{2} q
	q_{x} v_{xxz}+\frac{3}{2} y
	q_{y} q_{x} v_{xxz}
	\nonumber\\
	&
	-\frac{3}{10} q_{xx} v_{xxz}-\frac{1}{10} z q_{xxz} v_{xxz}-\frac{3}{2} z v q_{x}
	q_{xxzz}+\frac{1}{10} z v_{xx}
	q_{xxzz}+\frac{1}{2} y v_{z}
	q_{x} q_{xxy}+\frac{8}{3} y
	v q_{xz} q_{xxy}-\frac{1}{10} y v_{xxz} q_{xxy}
	\nonumber\\
	&
	+\frac{9}{2} y v q_{x}
	q_{xxyz}+\frac{1}{10} y v_{xx}
	q_{xxyz}-\frac{7}{6} q
	v_{z} q_{xxx}+\frac{11}{6} z
	v q_{zz} q_{xxx}-\frac{7}{6} y v_{z} q_{y}
	q_{xxx}-\frac{13}{6} y v
	q_{yz} q_{xxx}-\frac{2}{3} x
	v_{z} q_{x} q_{xxx}
	\nonumber\\
	&
	+\frac{1}{2} x v q_{xz}
	q_{xxx}+\frac{4}{15} v_{xz}
	q_{xxx}-\frac{1}{10} x v_{xxz}
	q_{xxx}-\frac{1}{5} q_{xz}
	v_{xxx}-\frac{1}{15} z q_{xzz}
	v_{xxx}-\frac{1}{15} y q_{xyz}
	v_{xxx}-\frac{1}{15} x q_{xxz}
	v_{xxx}
	\nonumber\\
	&
	+q v q_{xxxz}-y v q_{y} q_{xxxz}-\frac{1}{2} x v q_{x}
	q_{xxxz}-\frac{2}{3} v_{x}
	q_{xxxz}+\frac{1}{15} z v_{xz}
	q_{xxxz}+\frac{1}{10} x v_{xx}
	q_{xxxz}+\frac{4}{15} q_{x}
	v_{xxxz}+\frac{2}{15} z q_{xz}
	v_{xxxz}
	\nonumber\\
	&
	+\frac{2}{15} y q_{xy}
	v_{xxxz}+\frac{2}{15} x q_{xx}
	v_{xxxz}-\frac{2}{15} z v_{x}
	q_{xxxzz}+\frac{1}{15} y v_{xz}
	q_{xxxy}-\frac{2}{15} y v_{x}
	q_{xxxyz}-\frac{1}{6} v_{z}
	q_{xxxx}+\frac{1}{15} x v_{xz}
	q_{xxxx}
	\nonumber\\
	&
	+\frac{1}{30} z q_{zz}
	v_{xxxx}+\frac{1}{30} y q_{yz}
	v_{xxxx}+\frac{1}{30} x q_{xz}
	v_{xxxx}+v q_{xxxxz}
	-\frac{1}{30} z v_{z} q_{xxxxz}
	-\frac{2}{15} x v_{x} q_{xxxxz}
	-\frac{1}{6} q v_{xxxxz}
	 \nonumber\\
	&
	-\frac{1}{6} y q_{y} v_{xxxxz}
	-\frac{1}{6} x q_{x} v_{xxxxz}
	+\frac{1}{6} z v q_{xxxxzz}
	-\frac{1}{30} y v_{z} q_{xxxxy}
	-\frac{5}{6} y v q_{xxxxyz}
	-\frac{1}{30} x v_{z} q_{xxxxx}
	+\frac{1}{6} x v q_{xxxxxz}
	\nonumber\\
	&
		+\frac{1}{3} v_{xz} q_{t}
	+7 t v_{xz} q_{xx}	q_{t}
	-4 t q_{xz} v_{xx} q_{t}
	-v_{xx} q_{t}
	+\frac{1}{2} t v_{x} q_{xxz}q_{t}
	+\frac{9}{2} t q_{x}	v_{xxz} q_{t}
	-\frac{7}{2} t
	v_{z} q_{xxx} q_{t}
	+3	t v q_{xxxz} q_{t}
	\nonumber\\
	&
	-\frac{1}{2} t v_{xxxxz} q_{t}
	-\frac{1}{4} q_{xz} v_{t}
	-\frac{1}{12} z q_{xzz} v_{t}
	-\frac{1}{12} y q_{xyz} v_{t}
	+\frac{3}{4} q_{xx} v_{t}
	-\frac{1}{12} x q_{xxz} v_{t}
	+\frac{1}{4} z q_{xxz} v_{t}
	+\frac{1}{4} y q_{xxy} v_{t}
	+\frac{1}{4} x q_{xxx} v_{t}
	\nonumber\\
	&
	-\frac{5}{12} v_{x} q_{tz}
	+\frac{1}{12} z v_{xz} q_{tz}
	-2 t v_{x} q_{xx}	q_{tz}
	-\frac{1}{4} z v_{xx}	q_{tz}
	-\frac{3}{2} t q_{x}
	v_{xx} q_{tz}+\frac{11}{2} t
	v q_{xxx} q_{tz}+\frac{1}{10} t v_{xxxx} q_{tz}+\frac{1}{6} q_{x} v_{tz}
	\nonumber\\
	&
	+\frac{1}{12} z q_{xz} v_{tz}+\frac{1}{12} y q_{xy} v_{tz}+\frac{1}{12} x q_{xx} v_{tz}-\frac{1}{12} z v_{x} q_{tzz}+\frac{1}{12} y v_{xz} q_{ty}-\frac{1}{4} y v_{xx} q_{ty}-\frac{1}{12} y v_{x} q_{tyz}-\frac{5}{12} v_{z} q_{tx}
	\nonumber\\
	&
	+\frac{5}{2} v_{x} q_{tx}+4
	t v_{x} q_{xz} q_{tx}+\frac{1}{12} x v_{xz} q_{tx}-3 t q_{x} v_{xz}
	q_{tx}-2 t v_{z} q_{xx} q_{tx}-\frac{1}{4} x v_{xx} q_{tx}-\frac{5}{2} t v
	q_{xxz} q_{tx}+\frac{2}{5} t
	v_{xxxz} q_{tx}
	\nonumber\\
	&
	+\frac{1}{4} t
	v_{tz} q_{tx}+\frac{1}{12} z
	q_{zz} v_{tx}+\frac{1}{12} y
	q_{yz} v_{tx}-q_{x}
	v_{tx}+\frac{1}{12} x q_{xz}
	v_{tx}-\frac{1}{2} z q_{xz}
	v_{tx}-\frac{1}{2} y q_{xy}
	v_{tx}-\frac{1}{2} x q_{xx}
	v_{tx}+\frac{1}{4} t q_{tz}
	v_{tx}
	\nonumber\\
	&
	-\frac{3}{2} t q_{tx}
	v_{tx}+\frac{3}{2} v q_{txz}-\frac{1}{12} z v_{z} q_{txz}-\frac{1}{12} x v_{x} q_{txz}+\frac{1}{2} z v_{x} q_{txz}+3 t q_{x} v_{x}
	q_{txz}+t v q_{xx}
	q_{txz}-\frac{1}{5} t v_{xxx}
	q_{txz}
	\nonumber\\
	&
	-\frac{1}{4} t v_{t}
	q_{txz}-\frac{1}{4} q
	v_{txz}-\frac{1}{4} y q_{y}
	v_{txz}-\frac{1}{4} x q_{x}
	v_{txz}-\frac{3}{4} t q_{t}
	v_{txz}+\frac{1}{4} z v
	q_{txzz}-\frac{1}{12} y v_{z}
	q_{txy}+\frac{1}{2} y v_{x}
	q_{txy}-\frac{3}{4} y v
	q_{txyz}
	\nonumber\\
	&
	-\frac{9}{2} v
	q_{txx}-\frac{1}{12} x v_{z}
	q_{txx}+\frac{3}{2} t v_{z}
	q_{x} q_{txx}+\frac{1}{2} x
	v_{x} q_{txx}-4 t v
	q_{xz} q_{txx}-\frac{3}{10} t
	v_{xxz} q_{txx}+\frac{3}{4} t
	v_{t} q_{txx}+\frac{3}{4}
	q v_{txx}
	\nonumber\\
	&
	+\frac{3}{4} y
	q_{y} v_{txx}+\frac{3}{4} x
	q_{x} v_{txx}+\frac{9}{4} t
	q_{t} v_{txx}+\frac{1}{4} x
	v q_{txxz}+3 y v
	q_{txxz}-\frac{3}{4} z v
	q_{txxz}-\frac{9}{2} t v
	q_{x} q_{txxz}+\frac{3}{10} t
	v_{xx} q_{txxz}
	\nonumber\\
	&
	+\frac{9}{4} y
	v q_{txxy}-\frac{3}{4} x
	v q_{txxx}+\frac{1}{5} t
	v_{xz} q_{txxx}-\frac{2}{5} t
	v_{x} q_{txxxz}-\frac{1}{10} t
	v_{z} q_{txxxx}+\frac{1}{2} t
	v q_{txxxxz}+\frac{1}{4} t
	v_{xz} q_{tt}-\frac{3}{4} t
	v_{xx} q_{tt}
	\nonumber\\
	&
	-\frac{1}{4} t
	v_{x} q_{ttz}-\frac{1}{4} t
	v_{z} q_{ttx}+\frac{3}{2} t
	v_{x} q_{ttx}+\frac{3}{4} t
	v q_{ttxz}-\frac{9}{4} t
	v q_{ttxx},\nonumber
\end{align}

\begin{align}
T_7^z=\,& -\frac{1}{2} y
v_{xxx} q_{y}^2-\frac{17}{6}
v_{x} q_{xx} q_{y}+\frac{7}{3} y v_{xy} q_{xx}
q_{y}+\frac{1}{2} z q_{xz}
v_{xx} q_{y}-\frac{5}{6} y
q_{xy} v_{xx} q_{y}+\frac{1}{2} x q_{xx} v_{xx}
q_{y}-\frac{1}{2} z v_{x}
q_{xxz} q_{y}	\nonumber\\
&-\frac{1}{3} y
v_{x} q_{xxy} q_{y}+\frac{3}{2} y q_{x} v_{xxy}
q_{y}+\frac{17}{3} v q_{xxx} q_{y}-\frac{7}{6} y v_{y}
q_{xxx} q_{y}-\frac{1}{2} x
v_{x} q_{xxx} q_{y}-\frac{1}{2} q v_{xxx}
q_{y}-\frac{1}{2} z q_{z}
v_{xxx} q_{y}	\nonumber\\
&-\frac{1}{2} x
q_{x} v_{xxx} q_{y}-\frac{3}{2} z v q_{xxxz}
q_{y}+\frac{3}{2} y v
q_{xxxy} q_{y}+\frac{1}{2} x
v q_{xxxx} q_{y}+\frac{1}{15} v_{xxxx} q_{y}-\frac{1}{6} y v_{xxxxy} q_{y}-\frac{3}{2} t v_{xxx} q_{t}q_{y}	
\nonumber\\
&
+\frac{3}{2} t v_{xx}
q_{tx} q_{y}+\frac{1}{6}
v_{tx} q_{y}-\frac{1}{4} y
v_{txy} q_{y}-\frac{3}{2} t
v_{x} q_{txx} q_{y}+\frac{3}{4} y v_{txx} q_{y}+\frac{3}{2} t v q_{txxx}
q_{y}+\frac{4}{3} y v_{x}
q_{xy}^2-\frac{2}{3} x v_{y}
q_{xx}^2
	\nonumber\\
&
+\frac{17}{3} q_{x}
v_{x} q_{xy}+\frac{4}{3} z
v_{x} q_{xz} q_{xy}-2
q_{x}^2 v_{xy}-z q_{x} q_{xz} v_{xy}-y q_{x} q_{xy} v_{xy}+z q_{x} v_{x} q_{xyz}+y q_{x} v_{x} q_{xyy}+\frac{1}{6}
v_{y} q_{x} q_{xx}
	\nonumber\\
&
-\frac{2}{3} z q_{yz} v_{x}
q_{xx}-\frac{2}{3} y q_{yy}
v_{x} q_{xx}-\frac{2}{3} z
v_{y} q_{xz} q_{xx}-3
v q_{xy} q_{xx}-\frac{2}{3} y v_{y} q_{xy}
q_{xx}+\frac{2}{3} x v_{x}
q_{xy} q_{xx}+\frac{7}{3}
q v_{xy} q_{xx}	
\nonumber\\
&
+\frac{7}{3} z q_{z} v_{xy}
q_{xx}+\frac{4}{3} x q_{x}
v_{xy} q_{xx}+\frac{7}{3} z
v q_{xyz} q_{xx}+\frac{1}{3} y v q_{xyy}
q_{xx}-\frac{1}{2} z q_{yz}
q_{x} v_{xx}-\frac{1}{2} y
q_{yy} q_{x} v_{xx}-\frac{4}{3} q q_{xy}
v_{xx}
	\nonumber\\
&
-\frac{4}{3} z q_{z}
q_{xy} v_{xx}-\frac{11}{6} x
q_{x} q_{xy} v_{xx}+\frac{1}{2} z v_{y} q_{x}
q_{xxz}+\frac{8}{3} z v
q_{xy} q_{xxz}-\frac{23}{3}
v q_{x} q_{xxy}+\frac{1}{2} y v_{y} q_{x}
q_{xxy}+\frac{1}{6} q
v_{x} q_{xxy}	\nonumber\\
&+\frac{1}{6} z
q_{z} v_{x} q_{xxy}+\frac{7}{6} x q_{x} v_{x}
q_{xxy}+\frac{19}{6} z v
q_{xz} q_{xxy}-\frac{13}{6} y
v q_{xy} q_{xxy}-\frac{1}{2} x v q_{xx}
q_{xxy}+\frac{2}{5} v_{xx}
q_{xxy}+\frac{3}{2} x q_{x}^2
v_{xxy}
	\nonumber\\
&
+\frac{3}{2} q
q_{x} v_{xxy}+\frac{3}{2} z
q_{z} q_{x} v_{xxy}-\frac{3}{10} q_{xx} v_{xxy}-\frac{1}{10} z q_{xxz} v_{xxy}-\frac{1}{10} y q_{xxy} v_{xxy}+\frac{9}{2} z v q_{x}
q_{xxyz}+\frac{1}{10} z v_{xx}
q_{xxyz}
	\nonumber\\
&
-\frac{3}{2} y v
q_{x} q_{xxyy}+\frac{1}{10} y
v_{xx} q_{xxyy}-\frac{7}{6}
q v_{y} q_{xxx}-\frac{7}{6} z q_{z} v_{y}
q_{xxx}-\frac{13}{6} z v
q_{yz} q_{xxx}+\frac{11}{6} y
v q_{yy} q_{xxx}-\frac{2}{3} x v_{y} q_{x}
q_{xxx}
	\nonumber\\
&
+\frac{1}{2} x v
q_{xy} q_{xxx}+\frac{4}{15}
v_{xy} q_{xxx}-\frac{1}{10} x
v_{xxy} q_{xxx}-\frac{1}{5}
q_{xy} v_{xxx}-\frac{1}{15} z
q_{xyz} v_{xxx}-\frac{1}{15} y
q_{xyy} v_{xxx}-\frac{1}{15} x
q_{xxy} v_{xxx}
	\nonumber\\
&
+\frac{1}{15} z
v_{xy} q_{xxxz}+q
v q_{xxxy}-z v
q_{z} q_{xxxy}-\frac{1}{2} x
v q_{x} q_{xxxy}-\frac{2}{3} v_{x} q_{xxxy}+\frac{1}{15} y v_{xy} q_{xxxy}+\frac{1}{10} x v_{xx} q_{xxxy}
	\nonumber\\
&
+\frac{4}{15} q_{x} v_{xxxy}+\frac{2}{15} z q_{xz} v_{xxxy}+\frac{2}{15} y q_{xy} v_{xxxy}+\frac{2}{15} x q_{xx} v_{xxxy}-\frac{2}{15} z v_{x} q_{xxxyz}-\frac{2}{15} y v_{x} q_{xxxyy}-\frac{1}{6} v_{y} q_{xxxx}
	\nonumber\\
&
+\frac{1}{15} x v_{xy} q_{xxxx}+\frac{1}{30} z q_{yz} v_{xxxx}+\frac{1}{30} y q_{yy} v_{xxxx}+\frac{1}{30} x q_{xy} v_{xxxx}-\frac{1}{30} z v_{y} q_{xxxxz}+v q_{xxxxy}-\frac{1}{30} y v_{y} q_{xxxxy}
	\nonumber\\
&
-\frac{2}{15} x v_{x} q_{xxxxy}-\frac{1}{6} q v_{xxxxy}-\frac{1}{6} z q_{z} v_{xxxxy}-\frac{1}{6} x q_{x} v_{xxxxy}-\frac{5}{6} z v q_{xxxxyz}+\frac{1}{6} y v q_{xxxxyy}-\frac{1}{30} x v_{y} q_{xxxxx}
	\nonumber\\
&
+\frac{1}{6} x v q_{xxxxxy}+\frac{1}{3} v_{xy} q_{t}+7 t v_{xy} q_{xx}
q_{t}-4 t q_{xy} v_{xx} q_{t}-v_{xx} q_{t}+\frac{1}{2} t v_{x} q_{xxy}
q_{t}+\frac{9}{2} t q_{x}
v_{xxy} q_{t}-\frac{7}{2} t
v_{y} q_{xxx} q_{t}
	\nonumber\\
&
+3
t v q_{xxxy} q_{t}-\frac{1}{2} t v_{xxxxy} q_{t}-\frac{1}{4} q_{xy} v_{t}-\frac{1}{12} z q_{xyz} v_{t}-\frac{1}{12} y q_{xyy} v_{t}+\frac{3}{4} q_{xx} v_{t}+\frac{1}{4} z q_{xxz} v_{t}-\frac{1}{12} x q_{xxy} v_{t}
	\nonumber\\
&
+\frac{1}{4} y q_{xxy} v_{t}+\frac{1}{4} x q_{xxx} v_{t}+\frac{1}{12} z v_{xy} q_{tz}-\frac{1}{4} z v_{xx} q_{tz}-\frac{5}{12} v_{x} q_{ty}+\frac{1}{12} y v_{xy} q_{ty}-2 t v_{x} q_{xx}
q_{ty}-\frac{1}{4} y v_{xx}
q_{ty}
	\nonumber\\
&
-\frac{3}{2} t q_{x}
v_{xx} q_{ty}+\frac{11}{2} t
v q_{xxx} q_{ty}+\frac{1}{10} t v_{xxxx} q_{ty}+\frac{1}{6} q_{x} v_{ty}+\frac{1}{12} z q_{xz} v_{ty}+\frac{1}{12} y q_{xy} v_{ty}+\frac{1}{12} x q_{xx} v_{ty}-\frac{1}{12} z v_{x} q_{tyz}
	\nonumber\\
&
-\frac{1}{12} y v_{x} q_{tyy}-\frac{5}{12} v_{y} q_{tx}+\frac{5}{2} v_{x} q_{tx}+4
t v_{x} q_{xy} q_{tx}+\frac{1}{12} x v_{xy} q_{tx}-3 t q_{x} v_{xy}
q_{tx}-2 t v_{y} q_{xx} q_{tx}-\frac{1}{4} x v_{xx} q_{tx}
	\nonumber\\
&
-\frac{5}{2} t v
q_{xxy} q_{tx}+\frac{2}{5} t
v_{xxxy} q_{tx}+\frac{1}{4} t
v_{ty} q_{tx}+\frac{1}{12} z
q_{yz} v_{tx}+\frac{1}{12} y
q_{yy} v_{tx}-q_{x}
v_{tx}-\frac{1}{2} z q_{xz}
v_{tx}+\frac{1}{12} x q_{xy}
v_{tx}
	\nonumber\\
&
-\frac{1}{2} y q_{xy}
v_{tx}-\frac{1}{2} x q_{xx}
v_{tx}+\frac{1}{4} t q_{ty}
v_{tx}-\frac{3}{2} t q_{tx}
v_{tx}-\frac{1}{12} z v_{y}
q_{txz}+\frac{1}{2} z v_{x}
q_{txz}+\frac{3}{2} v
q_{txy}-\frac{1}{12} y v_{y}
q_{txy}-\frac{1}{12} x v_{x}
q_{txy}
	\nonumber\\
&
+\frac{1}{2} y v_{x}
q_{txy}+3 t q_{x} v_{x} q_{txy}+t v q_{xx}
q_{txy}-\frac{1}{5} t v_{xxx}
q_{txy}-\frac{1}{4} t v_{t}
q_{txy}-\frac{1}{4} q
v_{txy}-\frac{1}{4} z q_{z}
v_{txy}-\frac{1}{4} x q_{x}
v_{txy}-\frac{3}{4} t q_{t}
v_{txy}
	\nonumber\\
&
-\frac{3}{4} z v
q_{txyz}+\frac{1}{4} y v
q_{txyy}-\frac{9}{2} v
q_{txx}-\frac{1}{12} x v_{y}
q_{txx}+\frac{3}{2} t v_{y}
q_{x} q_{txx}+\frac{1}{2} x
v_{x} q_{txx}-4 t v
q_{xy} q_{txx}-\frac{3}{10} t
v_{xxy} q_{txx}
	\nonumber\\
&
+\frac{3}{4} t
v_{t} q_{txx}+\frac{3}{4}
q v_{txx}+\frac{3}{4} z
q_{z} v_{txx}+\frac{3}{4} x
q_{x} v_{txx}+\frac{9}{4} t
q_{t} v_{txx}+\frac{9}{4} z
v q_{txxz}+\frac{1}{4} x
v q_{txxy}-\frac{3}{4} y
v q_{txxy}+3 z v
q_{txxy}
	\nonumber\\
&
-\frac{9}{2} t v
q_{x} q_{txxy}+\frac{3}{10} t
v_{xx} q_{txxy}-\frac{3}{4} x
v q_{txxx}+\frac{1}{5} t
v_{xy} q_{txxx}-\frac{2}{5} t
v_{x} q_{txxxy}-\frac{1}{10} t
v_{y} q_{txxxx}+\frac{1}{2} t
v q_{txxxxy}+\frac{1}{4} t
v_{xy} q_{tt}
	\nonumber\\
&
-\frac{3}{4} t
v_{xx} q_{tt}-\frac{1}{4} t
v_{x} q_{tty}-\frac{1}{4} t
v_{y} q_{ttx}+\frac{3}{2} t
v_{x} q_{ttx}+\frac{3}{4} t
v q_{ttxy}-\frac{9}{4} t
v q_{ttxx}.\nonumber
\end{align}

\section{Conclusion} \label{sec7}

In this paper we reported the construction of a new $(3+1)$-dimensional Korteweg-de Vries (KdV) equation. The construction was done by utilizing the KdV's recursion operator.  
Two equations were extracted and with elemental computation steps, Eq.\ref{NewKdV} was formulated. This equation was then transformed to Eq.\eqref{NewKdVTren} so as to avoid the integral in the equation.
Lie symmetry analysis was applied to  Eq.\eqref{NewKdVTren} which resulted in a $7$-dimensional Lie algebra $L_7$ of point symmetries. Optimal system of one-dimensional  Lie subalgebras was constructed and exact solutions were computed.
These exact solutions were then sketched as 3D and 2D plots which showed distinct propagations of solitary wave solutions such as breather, periodic, bell shape, and others.
Finally, seven conserved vectors were derived by invoking Ibragimov's method. These included the energy conservation law coming from time-symmetry, and momentum conservation
laws acquired the spatial-symmetries.

\section*{Data Availability Statement}
Data sharing not applicable to this article as no datasets were generated or analysed during the current study.

\section*{Conflict of interest}
 The authors declare that they have no con- flict of interest.

 \section*{Ethical approval}
 Not applicable

\bibliographystyle{ieeetr}
\bibliography{Thesis}

   \end{document}